\documentclass{aa}

\usepackage{graphicx}
\usepackage{color}
\usepackage{array}
\usepackage[compatibility=false]{caption}
\usepackage{txfonts}
\usepackage[citecolor=blue,colorlinks=true,linkcolor=blue,urlcolor=blue]{hyperref}
\makeatletter
\renewcommand*\aa@pageof{, page \thepage{} of \pageref*{LastPage}}
\makeatother

\def\erosita{eROSITA\ }

\def\erobubbles{\erosita bubbles\ }

\begin{document}

\title{The SRG/eROSITA diffuse soft X-ray background}
\subtitle{II.~spectra and morphology of the eROSITA bubbles in the western Galactic hemisphere}

\author{Michael~C.~H.~Yeung\inst{1}            
       \and
       Martin~G.~F.~Mayer \inst{2}
       \and
       Andy Strong \inst{1}
       \and
       Michael~J.~Freyberg\inst{1}
       \and
       Gabriele~Ponti\inst{3,1,4}
       \and
       Konrad~Dennerl\inst{1}
       \and
       Junjie~Mao\inst{5}
       \and
       Manami~Sasaki\inst{2}
       \and
       Xueying~Zheng\inst{1}
       \and
       Jeremy~S.~Sanders\inst{1}
       \and
       Yi~Zhang\inst{1,6}           
       \and
       Jiejia~Liu\inst{5}
       \and
       Liyi~Gu\inst{7,8,9,10}
       \and
       Werner~Becker\inst{1}
       \and
       Frank~Haberl\inst{1}
       \and
       Teng~Liu\inst{11,12}
       \and
       Andrea~Merloni\inst{1}
       \and
       Peter~Predehl\inst{1}
      }

\institute{Max-Planck-Institut f\"ur extraterrestrische Physik, Giessenbachstra{\ss}e 1, 85748 Garching, Germany\\
            \email{myeung@mpe.mpg.de}
     \and Dr. Karl Remeis-Sternwarte and Erlangen Centre for Astroparticle Physics, Friedrich-Alexander Universität Erlangen-Nürnberg, Sternwartstra{\ss}e 7, 96049 Bamberg, Germany\\
     \email{mgf.mayer@fau.de}
     \and INAF-Osservatorio Astronomico di Brera, Via E. Bianchi 46, 23807 Merate (LC), Italy
     \and Como Lake centre for AstroPhysics (CLAP), DiSAT, Università dell’Insubria, via Valleggio 11, 22100 Como, Italy
     \and Department of Astronomy, Tsinghua University, Beijing 100084, China
     \and Max-Planck-Institut für Astrophysik, Karl-Schwarzschild-Stra{\ss}e 1, 85748 Garching, Germany
     \and SRON Space Research Organisation Netherlands, Niels Bohrweg 4, 2333 CA Leiden, The Netherlands
     \and Leiden Observatory, Leiden University, PO Box 9513, 2300 RA Leiden, The Netherlands
     \and RIKEN High Energy Astrophysics Laboratory, 2-1 Hirosawa, Wako, Saitama 351-0198, Japan
     \and Department of Physics, Tokyo University of Science, 1-3 Kagurazaka, Shinjuku-ku, Tokyo 162-8601, Japan
     \and Department of Astronomy, University of Science and Technology of China, Hefei 230026, China
    \and School of Astronomy and Space Science, University of Science and Technology of China, Hefei 230026, China
         } 

\date{Submitted 21 November 2025; Accepted YYY}

\abstract
{The eROSITA bubbles (eRObub) were discovered in 2020 in the first {\it SRG}/eROSITA All-Sky Survey, and are among the most extended structures in the X-ray sky. In projection, the eRObub encloses the $\gamma$-ray Fermi bubbles (FB), and the northern eRObub appears to be bounded by the north polar spur (NPS).
}
{Using eROSITA all-sky maps and spatially resolved spectra, we aim to infer the three-dimensional structure and measure the hot gas properties of the eRObub.}
{We fit spectra binned to a constant signal-to-noise ratio (S/N) and high-S/N spectra from custom regions to examine gas properties in more detail. We fit the morphology of eRObub with a parametrised geometrical model that describes a blast wave propagating into an idealised Galactic halo from the centre.}
{We found the interior of the western eRObub is best characterised by two emission components with relatively uniform temperatures: a hotter component at $kT=0.60\pm0.02\,$keV, and a colder one at $kT=0.21^{+0.03}_{-0.01}$\,keV,  where the latter's emission measure is about five times higher on average.    Our spectra suggest sub-solar abundances ($Z=0.2\pm0.1\,Z_{\rm \odot}$), consistent with expectations for the Galactic halo, while we find no conclusive evidence for $\alpha$-element enhancement.
In contrast, the NPS exhibits higher abundances ($Z>0.5\,Z_{\rm \odot}$), which, at face value, disfavours a common origin. We spectrally confirm an apparent cool shell at $kT\sim0.18$--$0.2\,$keV surrounding the northern eRObub, assuming collisional ionisation equilibrium.
We found no noticeable difference in X-ray emission in regions overlapping with the FB. 
Our geometrical model suggests that the horizontal size of both eRObub is well-constrained (semi-minor axis $\sim 6\,$kpc), but their vertical extent is uncertain, as the observed X-ray emission is almost insensitive to the existence and location of a bubble cap.
Additionally, a tilt ($\sim 30\degr$) towards $l\sim 220\degr$ is needed to reproduce the projected image of the northern eRObub, whereas the southern bubble requires little tilt.}
{}


\titlerunning{\erobubbles}
\authorrunning{M.~C.~H.~Yeung et al.}
\maketitle
%

\section{Introduction}
One of the most impactful early discoveries of the {\it SRG}/eROSITA telescope \citep{Merloni12,Predehl21,Sunyaev21} was that of the two giant X-ray emitting shells visible in its all-sky map, subsequently dubbed the eROSITA bubbles (eRObub) \citep{Predehl20}. These structures appear to be approximately symmetric about the Galactic centre (GC), with each bubble extending over around $80^{\circ}\times80^{\circ}$ in the sky, and emitting most prominently at energies in the range $0.6$--$1.0\,\mathrm{keV}$ \citep{Predehl20}. 
The eRObub bear strong similarities to structures seen at other wavelengths. Most prominently, this includes the Fermi bubbles (FB), which are nonthermally emitting symmetric $\gamma$-ray structures visible around $1\,\mathrm{GeV}$ \citep{Su10, Ackermann14}, extending over $55^{\circ}\times45^{\circ}$, located `inside' the eRObub in projection. Furthermore, polarised radio and microwave emissions are coincident with the north polar spur (NPS) \citep{HanburyBrown60,Egger95, Crocker15}, an X-ray bright arc which appears to mark the whole eastern edge and extend to the north-western side of the eRObub. 

The morphology of the eRObub suggests a vertical extent on scales comparable to the entire Galaxy, assuming their symmetry about the GC, which indicates an energy budget of $\sim 10^{56}\,\mathrm{erg}$ necessary for their inflation \citep{Kataoka18,Predehl20}. Possible sources for the necessary power ($P\sim 10^{41}$--$10^{43}\,\mathrm{erg\,s^{-1}}$) are: first, a short-term ($\sim 10^5\,\mathrm{yr}$) outburst or jet activity of the GC black hole close to its Eddington luminosity several $10^6\,\mathrm{yr}$ ago may reproduce the observed morphology well \citep{Yang22}. In this scenario, the jet drove a shock into the circumgalactic medium (CGM) of the Milky Way, heating the swept-up material to X-ray-emitting temperatures and producing a shell akin to the \mbox{eRObub}. The outflow, along with its leptonic cosmic-ray content \citep{Ackermann14}, is compressed by the reverse shock and separated from the shocked CGM by a contact discontinuity \citep{Predehl20}. The inverse-Compton emission from the confined outflow is responsible for the gamma-ray bright FB. 
Alternatively, the bubbles could have been fuelled by past nuclear starburst activity in our Galaxy \citep[e.g.,][]{Nguyen22}, which would lead to large energy input from stellar winds and supernovae. Collimation by the pressure of the surrounding medium would produce bi-conical or bubble-like outflows on timescales similar to the former scenario. Scenarios involving energy injections from tidal-disruption events have also been proposed \citep{Cheng11,Scheffler25}.

Even though the bipolar morphology of both eRObub and FB likely indicates a physical extent on Galactic scales, other evidence suggests that the NPS could be located very nearby ($\sim 100\,\mathrm{pc}$), most recently based on multiwavelength polarisation information \citep{Panopoulou21}. Such a low distance seems to be contradicted by X-ray absorption measurements \citep{Lallement16} and by a lack of radio dispersion \citep{Koljonen24} toward the NPS. However, the question regarding the location of the individual features persists. This concerns, in particular, whether all features are intrinsically physically connected or only superimposed along the line of sight. In order to address this question, a measurement of the physical parameters of the hot gas in multiple locations in the eRObub would be crucial to test for their internal consistency. 

Much like the matter of the distance, many of the physical properties of the eRObub have only been probed in pencil-beam observations using a variety of instruments and approaches. For instance, the collisionally heated X-ray emitting plasma appears to have a typical temperature of $0.3\,\mathrm{keV}$ \citep{Kataoka13, Kataoka2015, Ursino16}, even though temperature substructure, non-equilibrium ionisation (NEI), or multiple emitting components may complicate the matter \citep{Yamamoto22, Gupta23}, as is also evidenced by a putative cool shell observed in eROSITA line ratio maps \citep{line-ratio}.
The metal abundance and $\alpha$-element enrichment in the bubbles could hold the key to distinguishing the origin of the bubbles \citep{Inoue15}. The hot gas metallicity has often been assumed to be subsolar at $\sim0.2\,Z_{\odot}$ \citep[e.g.][]{Kataoka2015} under a shock-heated CGM scenario. However, this contrasts with findings of enhanced relative abundances of neon and magnesium \citep{Ursino16, Gupta23},  which may be interpreted as a signature of a stellar origin of the bubbles. 
Finally, it is unclear whether there is any discernible X-ray signature of the compressed outflow predicted inside the FB \citep{Yang22}, apart from a tentative edge detection in ROSAT band ratios \citep{Su10}.

This work is the second in a series of papers on the diffuse soft X-ray background observed with eROSITA, with the previous work focusing on the properties of the local hot bubble (LHB) \citep{Yeung2024}. The target of this paper is the quantitative study of the physical properties of the eRObub, obtained via the analysis of their X-ray morphology and the spatially resolved spectroscopy of the diffuse X-ray emission in the first all-sky survey of {\it SRG}/eROSITA (eRASS1). 
The main questions we study include the three-dimensional geometry giving rise to the observed morphology of the bubbles, the physical mechanisms contributing to the detected X-ray emission, the temperature and density of the emitting material, and the relative abundances of light $\alpha$-elements and Fe. 
Based on these properties, we quantify the energetics and shock velocities needed to inflate the bubbles, discuss how the NPS, the FB, and the putative cool shell fit into the global picture of the bubbles in terms of distance and brightness, and test previous findings of non-solar abundance ratios in the context of different scenarios for the bubbles. 
Our paper is structured as follows: we describe the underlying data in Sect.~\ref{Data}, and present the complementary spectral and geometric analyses in Sects.~\ref{Spectra} and ~\ref{Morph}, respectively. In Sect.~\ref{Discussion}, we discuss our findings in the context of the physical properties of the eRObub, and we summarise our results in Sect.~\ref{Summary}.

\begin{figure}[htbp]
    \centering
    \includegraphics[width=\linewidth]{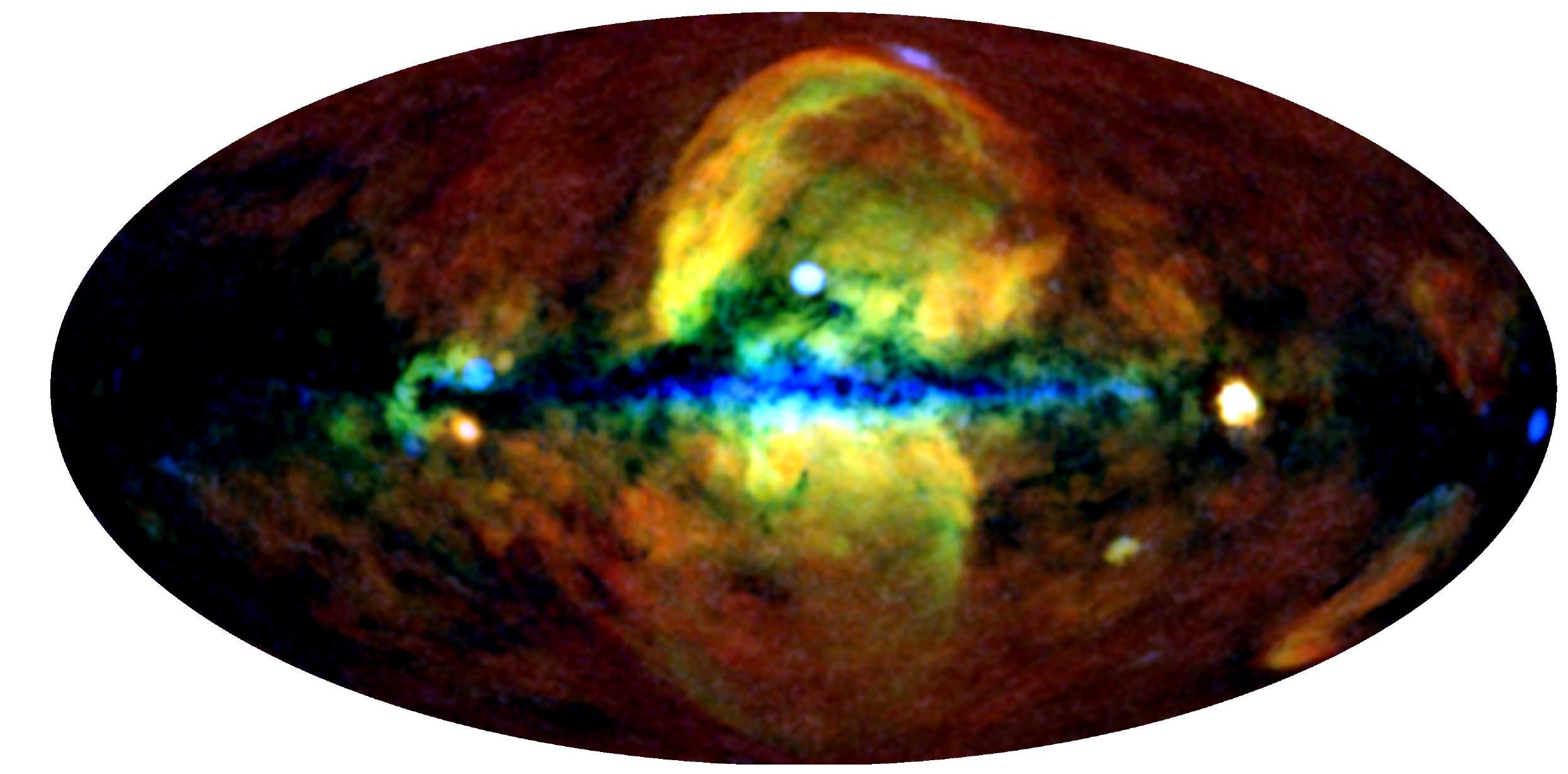}
    \caption{Multi-band view of eROSITA bubbles in the $0.3$--$0.6$, $0.6$--$1.0$, $1.0$--$2.3\,\rm keV$ bands \citep[as in Fig.~1 of][]{Predehl20}, after point source removal. The individual bands were smoothed with a Gaussian kernel of $0\fdg5$ and are displayed on a logarithmic scale.}
    \label{fig:bubble_rgb}
\end{figure}

\section{Data} \label{Data}
This section outlines the data products used for the analysis described in this paper. Since our work focuses on morphological and spectral analysis of the X-ray emission of the eRObub, our primary data products are based on publicly available eROSITA all-sky survey data.

\subsection{eROSITA Imaging data}
In order to gain a full impression of the morphology of the \mbox{eRObub}, including its potential asymmetries, it is necessary to use an all-sky X-ray map reflecting its diffuse X-ray emission. Details on this reconstruction and point-like source subtraction are given in Appendix~\ref{sec:ero_recons}. However, briefly, we used the eRASS1 broad-band half-sky maps \citep{Zheng2024_broad} to calibrate an all-sky emission map in different energy bands, reconstructed from the eRObub discovery paper \citep{Predehl20}. 
The maps in the different energy bands were then cleaned of point-like sources to isolate the diffuse emission component for subsequent morphological analysis (Sect.~\ref{Morph}), which is displayed in Fig.~\ref{fig:bubble_rgb}. Note that the brightest point-like sources (e.g., Sco X-1) appear extended due to their stray light extending over multiple degrees, preventing full masking of their emission. We would like to emphasise that all all-sky maps shown in this work are based entirely on publicly available resources.

\subsection{Spectroscopic data}
This work builds on the methods and results of the half-sky spectral analysis conducted by \citet{Yeung2024}, which decomposed the spectra of diffuse X-ray emission across the sky into its individual components, namely emission from the LHB, CGM, the extragalactic X-ray background (CXB), and the eRObub. This work additionally incorporates an unresolved stellar contribution, which is recently found to dominate the 0.7\,keV emission \citep{Ponti26}.
For this analysis, the eRASS1 data of the western Galactic hemisphere were used. In addition, regions around bright compact sources \citep[see][]{Merloni24} were excluded from spectral extraction regions. While a deeper data set would be available with multiple cumulative eRASSs, using only the first survey has the advantage of negligible heliospheric charge exchange emission \citep{Ponti2023,Yeung23,Dennerl26}, reducing the number of necessary model components. 
In order to break the temperature degeneracy of the LHB and CGM components, the eRASS1 data were supplemented with photometry derived from the ROSAT all-sky survey \citep{Yeung2024}, which has a larger effective area than eROSITA below the carbon K-edge (284\,eV) \citep{Snowden97}.

\section{Spectral analysis} \label{Spectra}
We divide the spectral analysis into two parts. 
The first part makes use of large (median of $139$\,deg$^{2}$, spanning a range of $25$--$670$\,deg$^{2}$), manually defined regions to target interesting areas of the eRObub, including the NPS and the FB, and to accumulate maximum photon statistics to study their spectra in detail. The high S/N provides the greatest chance to measure abundances and identify processes beyond, or in addition to, the simplest single-temperature plasma in collisional ionisation equilibrium (CIE), such as non-equilibrium effects. The second part builds upon the first, but uses spectra extracted from smaller regions (5--10\,deg$^{2}$) defined by a uniform ${\rm S/N = 80}$ criterion, and aims to produce a finer temperature and EM map of the eRObub.

\subsection{Spectra from large regions} \label{subsec:large_reg}
Defining larger, targeted spectral extraction regions could dissect the eRObub into distinct, representative features while providing higher S/N for spectral fitting, at the cost of somewhat subjective region definition. We also note here that while using all data from all five eRASSs would increase the S/N, the heliospheric solar wind charge exchange emission has been shown to increase monotonically with the solar cycle and cannot be ignored in later eRASSs, unlike in eRASS1 \citep{Ponti2023,Yeung23, Dennerl26}. A more robust spectral-fitting setup that accounts for the time domain is needed to extract reliable information from spectra accumulated across all four completed eRASSs (approximately doubling the S/N), which we leave for future work.

\subsubsection{Extraction regions} \label{subsubsec:reg}
\begin{figure}[htbp]
    \centering
    \hspace*{-0.5cm}\includegraphics[width=0.5\textwidth]{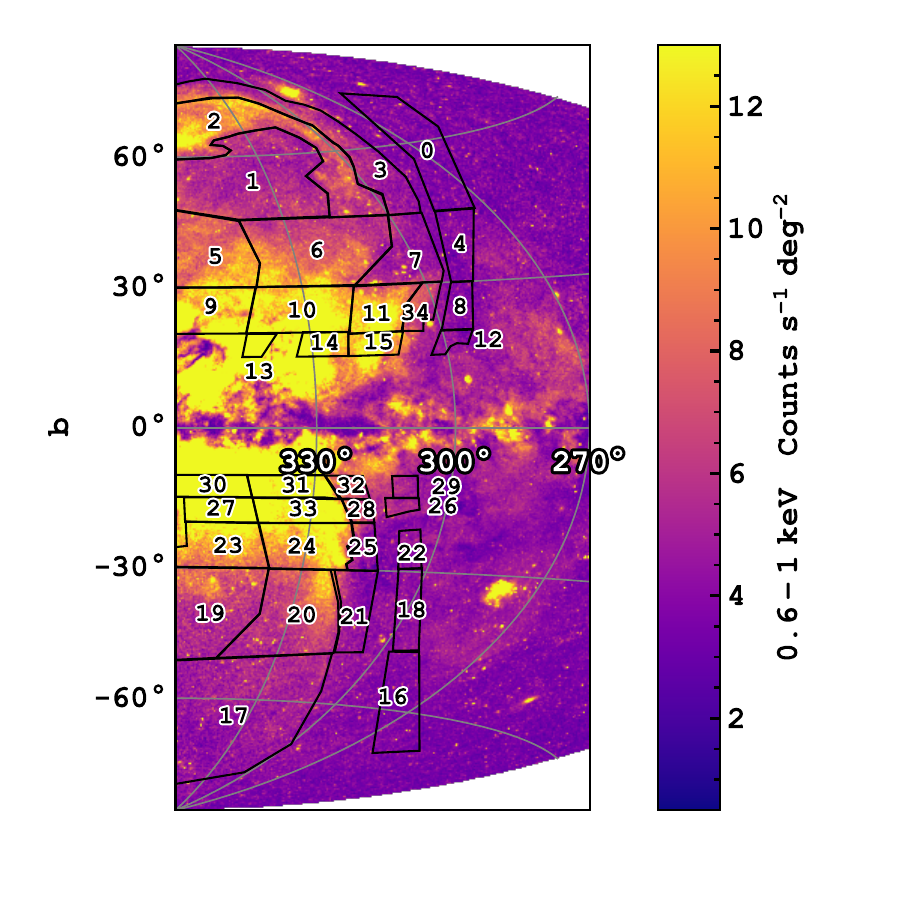}
    \caption{eRASS1 surface brightness map in the $0.6\text{--}1.0\,$keV band, overlaid with the outlines and numbering of our large spectral extraction regions. Reg~0, 4, 8, 12, 16, 18, 22, 26 and 29 are the background regions.}
    \label{fig:reg_num}
\end{figure}

Figure~\ref{fig:reg_num} presents our large extraction regions in relation to the eRASS1 $0.6\text{--}1.0$\,keV surface brightness map \citep{Zheng2024_broad}. 
We defined the extraction regions, which can be divided into four groups: background, cool shell, eRObub interior and FB.
An important point is to assign each `source' spectrum a background region within the same latitude range. This helped obtain good fits, probably because the soft X-ray background (SXRB) has a larger gradient along the latitudes than the longitudes. 
Regions (reg) 0, 4, 8, 12, 16, 18, 22, 26, and 29 are background regions that should represent the general background level at each latitude range.
Reg~3, 7, 21, 25, 28, 32 and 34 capture much of the cool shell, drawn primarily following the cool shell contour in the \ion{O}{VIII}/\ion{O}{VII} line-ratio map in \citet{line-ratio}.
Reg~2 covers the remaining part of the NPS, which is bright and shows a higher line ratio than the cool shell.
Reg~5, 9, 19, 23, 27, and 30 are the FB regions within the outline specified in \citet{Ackermann14}. The remaining regions are within the eRObub interior.
All the regions were confined to a specific latitudinal range, such that within each latitude interval, it is possible to obtain a comparison of the four spectral groups. Areas close to the Galactic plane ($-10\degr \lesssim b\lesssim15\degr$) were not included in the analysis to avoid confusion from complex line-of-sight structures during spectral fitting.

We used all events of valid patterns (singles, doubles, triples and quadruples) from the telescope modules (TMs) with the on-chip filter (TM1, 2, 3, 4, 6 or abbreviated as the fictitious TM8) in eRASS1 to construct the spectra.
The sources in the eRASS1 source catalogue \citep{Merloni24} with fluxes $F_{0.5-2.0\,\text{keV}}>10^{-12}\,{\rm erg\,s^{-1}\,cm^{-2}}$ were masked to ensure an almost uniform CXB slope and normalisation in all regions. This threshold modifies the CXB photon index from the conventional value of 1.45 to 1.7, as tested over a large fraction of the western Galactic hemisphere away from the Galactic plane \citep{Yeung2024}. We also used the latest \texttt{c030} eSASS pipeline configuration, which had multiple improvements compared to the previous \texttt{c020} configuration, including correction for the $\sim 5\%$ `missing flux' problem \citep{Brunner22,Merloni24}.

\subsubsection{Preparation for spectral fitting} \label{subsubsec:spec_fit}
\begin{figure*}[htbp]
\centering
\includegraphics[width=0.49\textwidth]{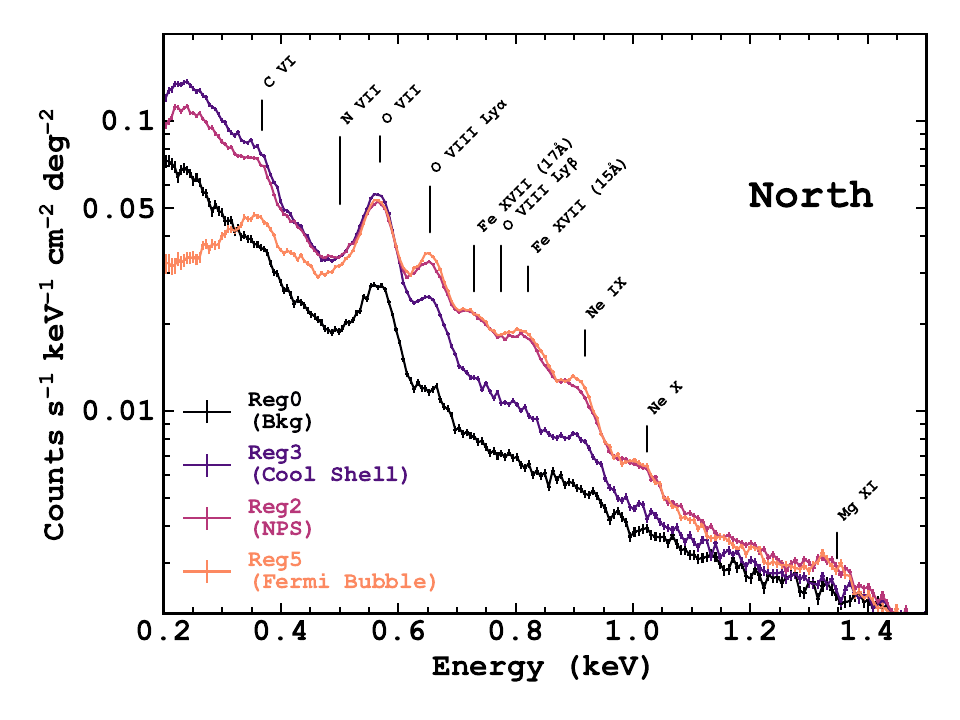}
\includegraphics[width=0.49\textwidth]{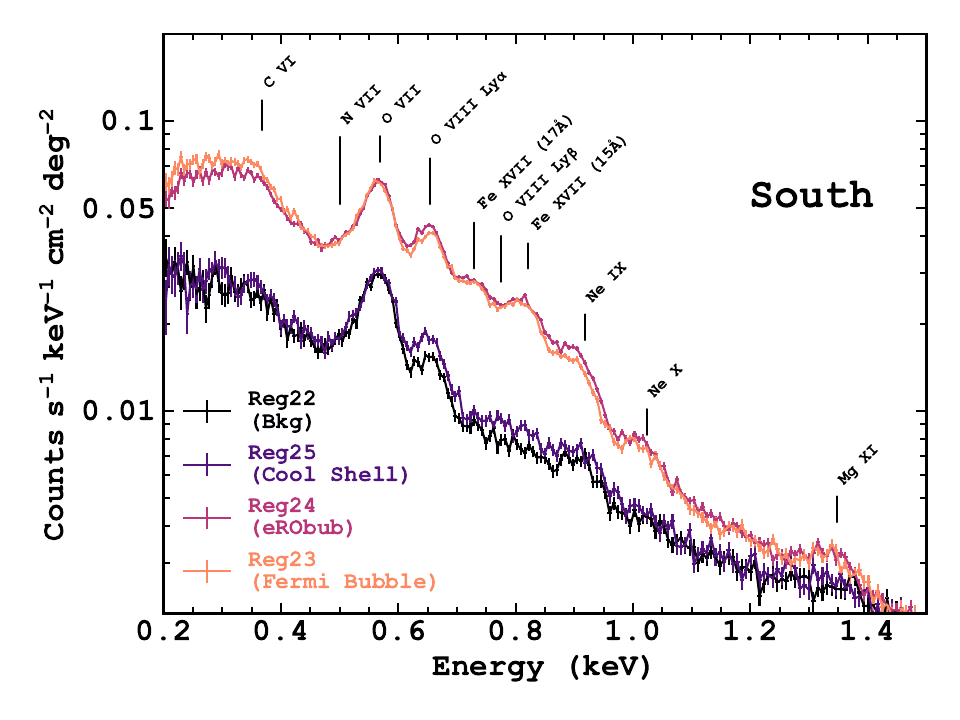}
\caption{Overlay of selected spectra (with error bars) from the northern (\textit{left}) and the southern (\textit{right}) regions. The variations redward of the \ion{O}{VII} line are mainly driven by differences in foreground absorption.}
\label{fig:overlay}
\end{figure*}

The error bars in the spectra from the large regions are extremely small, as shown in Fig.~\ref{fig:overlay}, thanks to the excellent photon statistics we obtained from large sky areas. The better statistics also raise the calibration accuracy we require. In fact, we discovered systematic residuals around emission lines, hinting at slightly inaccurate energy resolution in the ground-calibrated RMF utilised by the standard pipeline. Therefore, we used an improved response matrix file (RMF) and ancillary response file (ARF) to fit each region, which are calibrated from the isolated neutron star RX~J1856.5-3754 and the SNR 1E~0102-72.3, the latter using the IACHEC model \citep{Plucinsky17}. A finer input energy grid in the RMF was also needed to properly fold the model spectra into observed space to avoid numerical systematics when discretising a model spectrum from 1024 input energy bins into 1024 observed channels. We found that a factor of four increase in the density of the input energy grid (4096 bins into 1024 observed channels) can effectively eliminate this problem. In practice, we extracted one spectrum, one ARF from each region from the eSASS task \texttt{srctool} in the standard \texttt{c030} pipeline, then interpolated the extracted ARF into a finer grid, compatible with the new RMF. In addition to a finer energy grid, the improved ARF features improvements on the filter transmission calibration due to possibly more carbon in the optical blocking filters than the specification. This correction was applied together with the interpolation step.

We used \texttt{SPEX} version 3.08.1 \citep{SPEX, SPEX3.08} to analyse spectra from the large regions. We found \texttt{SPEX} \texttt{cie} model in version 3.08.1 seems to reproduce our spectra better than a few versions of \texttt{AtomDB/apec} \citep{apec}. We will return to this point in Sect.~\ref{subsec:plasma_code}. The spectra from TM1, 2, 3, 4 and 6 were summed before fitting. An important aspect when modelling X-ray diffuse emission is the non-X-ray background, which can be isolated by the data taken when the filter wheel was in the \texttt{CLOSED} (FWC) position \citep{Freyberg20,Yeung23}. We now use a complete set of FWC data accumulated since eROSITA's launch to create an empirical instrumental background model, and in the \texttt{c030} pipeline configuration for self-consistency with the data. The empirical model for the FWC data is created from a known list of fluorescence lines and a continuum described by a 24-order spline. For all regions, the FWC model is fixed in spectral shape, but is allowed to vary in normalisation to fit the $5$--$9$\,keV part of the spectrum, where the X-ray background is negligible. The FWC model normalisation factor was then fixed in subsequent analysis. We used the proto-Sun abundance table in \citet{Lodders2003} as our Solar abundance reference.

We followed Sect.~2 of \citet{Yeung2024} to use ROSAT R1 and R2 band fluxes, in addition to eROSITA data, to constrain the models below $0.3\,$keV.

\subsubsection{Background models} \label{subsubsec:bkg_reg}
We defined a series of background regions sampling the respective latitude ranges of the eRObub regions (see Fig.~\ref{fig:reg_num}). We modelled the emissions from the background regions as the sum of the LHB, CGM, CXB and stellar contribution, where all the components except for the LHB are absorbed by a single or two layers of neutral ISM in solar abundance modelled by the \texttt{hot} model in \texttt{SPEX}, with temperature fixed at $10^{-3}$\,eV.
We estimated the total Galactic $N_{\rm H}$ following the treatment described in \citet{Yeung2024}, which was predominantly based on \ion{H}{I} map from \citet{HI4PI} with a correction accounting for the molecular phase. The total Galactic $N_{\rm H}$ served as the initial guess for the variable $N_{\rm H}$ used to attenuate the emissions from the CGM, CXB and the stellar component in the fitting process.
This prescription of the soft X-ray background components, except for the stellar contribution, is the standard template for many studies for more than at least two decades \citep[e.g.][]{Kuntz00,Yoshino09,Henley13,Henley15,Wulf19,Kaaret20,Ringuette21,Bluem22,Ponti2023,Pan24,Qu24,Yeung2024,Gupta25}. Recent analyses of the general soft X-ray background by HaloSat, Suzaku and eROSITA, have consistently found deficit near $0.7$--$1.0$\,keV, and could be readily accounted for by an additional CIE component at $\sim0.7$\,keV.

The LHB and CGM components were modelled using the \texttt{cie} model, with the former having abundance fixed to solar, while the CGM can have variable abundances for C, O, Ne, and Fe; other elements heavier than He are coupled to the O abundance. H and He are fixed to solar.
Allowing C, O, Fe and Ne to have variable abundances may appear as a complicated model. In fact, recent work has shown that assuming the CGM has $\sim$\,0.1 to 0.3 solar metallicities is sufficient to reproduce the SXRB spectra. We refer the reader to Appendix~\ref{app:CGM_abund} for more details about the necessity of this model complexity when fitting our dataset.

The CXB component was described by a simple power-law model, with the photon index fixed at 1.7.

As mentioned, a $kT\sim0.7$\,keV CIE component could be used to explain the enhancement near $0.7$--$1.0$\,keV in the SXRB spectra, which is commonly interpreted as a hotter component of the CGM. In this work, the major change is replacing this $0.7$\,keV component with a stellar component. This change is based on the recent work by \citet{Ponti26}, who reports that low-mass stars with spectral types F to M dominate the emission from the 0.7\,keV component. They present a strong correlation between the MW's projected stellar mass profile and the $0.7\,$keV component's EM profile. Utilising the strong correlation, they compute the ratio of these two quantities, which can be understood physically as the average X-ray luminosity per solar mass. They show that this specific luminosity is consistent with measurements from the 10\,pc volume-complete sample of low-mass stars, thereby associating the 0.7\,keV component with coronal emission from low-mass stars.

In this work, we assume their interpretation is correct and reverse the process: we took the best-fit model (their 2T-VAPEC model) to the stacked X-ray spectrum of the 10\,pc low-mass star sample \citep{zheng25}, assuming this is reasonably representative of the average X-ray emissions from stars in the MW, and scaled this model according to the MW stellar mass model \citep{Hunter24} to yield an estimated spectral component from unresolved stars in each large region. \citet{zheng25} extract stacked spectra separately for FGK and M stars, and we followed \citet{Ponti26} to scale the two spectral type groups following the initial mass function by \citet[][$\rm FGK=47\%$ and $\rm M=53\%$ in mass]{Kirkpatrick24}. With this relative mass fraction, we then scaled the combined FGKM stars model ($L_*(E)/M_\sun$; in luminosity per solar mass of FGKM stars) with the 3-dimensional MW stellar mass model ($\rho_{\rm MW}(\Vec{r})$), while considering the inverse-square drop-off in flux. More precisely, our model stellar spectrum can be written as
\begin{align}
    F_*(E) &=\frac{L_*(E)}{M_\sun}\int{\frac{\rho_{\rm MW}(\Vec{r})}{4\pi r^2} \mathrm{d}V}\\
           &= \frac{L_*(E)}{M_\sun}\int{\frac{\rho_{\rm MW}(\Vec{r})}{4\pi r^2} r^2 \mathrm{d}r \mathrm{d}\Omega}\\
           &= \frac{L_*(E)\Omega}{M_\sun}\int{\frac{\rho_{\rm MW}(\Vec{r})}{4\pi} \mathrm{d}r},
\end{align}
where $\Omega$ is the solid angle of the region. One caveat is that the MW stellar mass model also contains stars with spectral types earlier than F. However, we simplify the situation by computing the integral assuming FGKM stars contribute to all stellar mass. There is no free parameter from the stellar component, as the spectral shape is determined by \citet{zheng25} and its normalisation is determined by the MW mass model \citep{Hunter24}.

\begin{figure*}[htbp]
    \centering
    \includegraphics[width=0.49\textwidth]{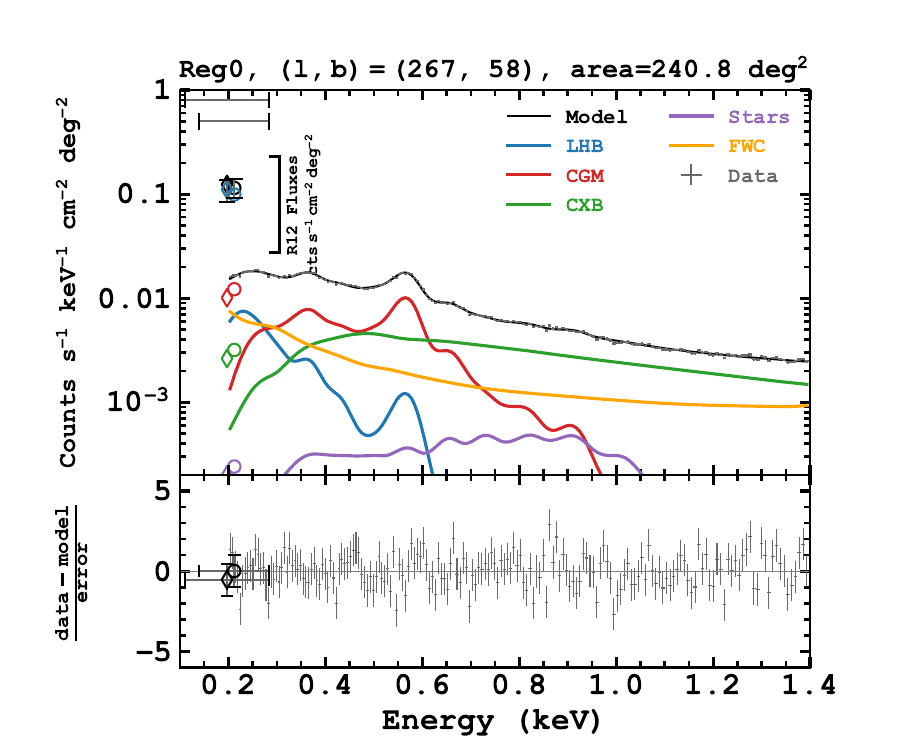}
    \includegraphics[width=0.49\textwidth]{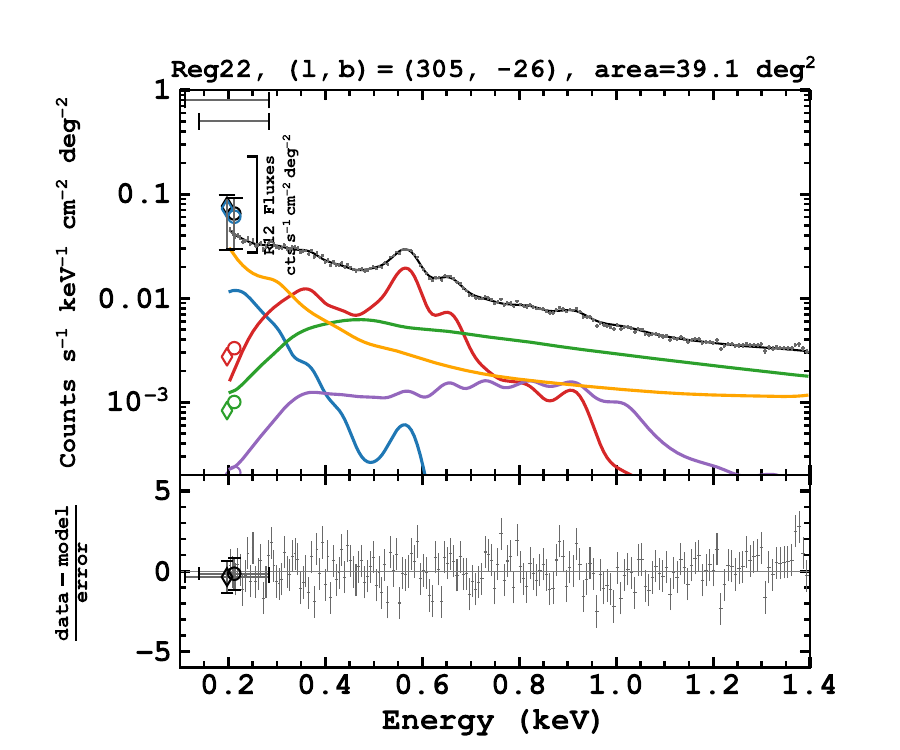}
    \caption{Example of the spectra, models and residuals of the two background regions (north: reg~0; south: reg~22), overlaid with the underlying model components. The two horizontal error bars at the top left corner of each panel indicate the bandwidth of the R1 and R2 bands.}
    \label{fig:bkg_spec}
\end{figure*}

We show an example spectrum and model of a background region in each hemisphere in Fig.~\ref{fig:bkg_spec}.
The underlying components described above are plotted, with the corresponding model parameters listed in Table~\ref{tab:bkg_param}. We obtained excellent fits to all background regions, as evidenced by $\chi^2/$dof values consistently close to 1. Most parameters are within the expectations from the X-ray background spectral analysis in \citet{Yeung2024}. 
The CXB normalisation from the background regions ($(3.56\pm0.12)\times10^{-3}\,{\rm ph\,s^{-1}\,cm^{-2}\,keV^{-1}\,deg^{-2}}$ at $1\,\rm keV$) is fully consistent with the measured values from cleaner extragalactic sky at high latitudes $\vert b \vert > 30\degr$, which is $3.54^{+0.24}_{-0.17}\times10^{-3}\,{\rm ph\,s^{-1}\,cm^{-2}\,keV^{-1}\,deg^{-2}}$ at $1\,\rm keV$ \citep{Yeung2024}.
The mean LHB and CGM temperatures, $kT_{\rm LHB}=82\pm6$\,eV and $kT_{\rm CGM}=164\pm6\,$eV, are slightly lower compared to \citet{Yeung2024}, driven by the replacement of the 0.7\,keV COR component with the stellar component which has their coronal emissions described by multiple CIE temperatures as reported in \citet{zheng25}, most notably at $\sim0.3\,$keV. Interestingly, the abundances of $\alpha$-elements, C, O and Ne, are consistently higher than Fe, which might reflect a non-negligible $\alpha$-enrichment from core-collapse supernovae into the CGM.

The best-fit temperatures of the LHB and CGM and elemental abundances of the CGM were taken and fixed to fit the spectra within the eRObub of the same latitude range. This helps to limit the number of free parameters in the fits when the focus is on the eRObub. However, we allow the normalisations (or EM) of the LHB, CGM and CXB to vary during the fits to account for potential variations in the fore- and background in the eRObub regions.

\begin{table*}[htbp]
\renewcommand{\arraystretch}{1.2}
    \caption{Parameters of the background regions.
    }
    \centering
    {\small
    \begin{tabular}{c|cccccc}
    \hline\hline
Background region & $l$ & $b$ & $kT_{\rm LHB}$ & ${\rm EM_{LHB}}$ & $kT_{\rm CGM}$ & ${\rm EM_{CGM}}$ \\
 & (\degr) & (\degr) & (eV) & ($10^{-3}\,{\rm cm^{-6}\,pc}$) & (eV) & ($10^{-2}\,{\rm cm^{-6}\,pc}$) \\\hline
$0$ & $267$ & $58$ & $89\pm4$ & $4.20^{+0.25}_{-0.33}$ & $149^{+1}_{-2}$ & $7.00^{+0.55}_{-0.53}$ \\ 
$4$ & $286$ & $38$ & $83^{+8}_{-6}$ & $2.55^{+0.71}_{-0.67}$ & $156\pm1$ & $2.56^{+0.53}_{-0.51}$ \\ 
$8$ & $294$ & $25$ & $74^{+10}_{-7}$ & $2.21^{+0.78}_{-0.74}$ & $163^{+2}_{-1}$ & $0.21^{+0.05}_{-0.04}$ \\ 
$12$ & $299$ & $18$ & $86^{+90}_{-10}$ & $1.54^{+0.62}_{-0.75}$ & $167^{+8}_{-2}$ & $0.14^{+0.04}_{-0.07}$ \\ 
$16$ & $283$ & $-57$ & $88^{+6}_{-0}$ & $5.91^{+0.24}_{-0.60}$ & $164^{+3}_{-1}$ & $3.62^{+0.30}_{-0.41}$ \\ 
$18$ & $299$ & $-38$ & $86^{+11}_{-2}$ & $4.25^{+0.34}_{-1.01}$ & $170^{+2}_{-1}$ & $1.53^{+0.39}_{-0.60}$ \\ 
$22$ & $305$ & $-26$ & $79^{+8}_{-6}$ & $2.67^{+0.77}_{-0.73}$ & $163^{+2}_{-1}$ & $2.48^{+0.75}_{-0.73}$ \\ 
$26$ & $310$ & $-16$ & $78^{+11}_{-7}$ & $2.40^{+0.93}_{-0.83}$ & $170\pm2$ & $4.85^{+0.91}_{-0.88}$ \\ 
$29$ & $310$ & $-12$ & $75^{+5}_{-4}$ & $3.07^{+0.62}_{-0.58}$ & $172\pm2$ & $2.90^{+0.96}_{-0.91}$ \\ 

\hline
    \end{tabular}
    
\vspace*{0.2cm}\begin{tabular}{c|ccccc}
\hline\hline
Background region & $Z_{\rm C}$ & $Z_{\rm O}$ & $Z_{\rm Ne}$ & $Z_{\rm Fe}$ & ${\rm CXB_{norm}}$ \\
 & ($Z_{\rm C, \odot}$) & $(Z_{\rm O, \odot})$ & $(Z_{\rm Ne, \odot})$ & $(Z_{\rm Fe, \odot}$) & ($10^{-3}\,{\rm ph\,s^{-1}\,cm^{-2}\,deg^{-2}}$) \\ \hline
 $0$ & $0.16\pm0.02$ & $0.09\pm0.01$ & $0.16\pm0.02$ & $0.07\pm0.01$ & $3.67^{+0.02}_{-0.03}$ \\ 
$4$ & $0.58^{+0.18}_{-0.12}$ & $0.29^{+0.07}_{-0.05}$ & $0.31^{+0.09}_{-0.07}$ & $0.09^{+0.03}_{-0.02}$ & $3.46\pm0.03$ \\ 
$8$ & $8.48^{+1.52}_{-1.71}$ & $3.51^{+0.85}_{-0.63}$ & $3.20^{+1.05}_{-0.81}$ & $1$\tablefootmark{($a$)} & $3.39\pm0.04$ \\ 
$12$ & $15.74^{+12.25}_{-4.74}$ & $6.02^{+3.98}_{-1.34}$ & $5.00^{+4.29}_{-1.47}$ & $1$\tablefootmark{($a$)} & $3.08^{+0.06}_{-0.05}$ \\ 
$16$ & $0.14\pm0.03$ & $0.12\pm0.01$ & $0.14\pm0.02$ & $0.07\pm0.01$ & $3.58\pm0.02$ \\ 
$18$ & $0.91^{+0.51}_{-0.25}$ & $0.51^{+0.31}_{-0.10}$ & $0.80^{+0.48}_{-0.16}$ & $0.20^{+0.08}_{-0.04}$ & $3.34^{+0.03}_{-0.04}$ \\ 
$22$ & $0.67^{+0.34}_{-0.21}$ & $0.35^{+0.14}_{-0.08}$ & $0.58^{+0.24}_{-0.14}$ & $0.12^{+0.06}_{-0.03}$ & $3.55\pm0.04$ \\ 
$26$ & $0.44^{+0.16}_{-0.12}$ & $0.23^{+0.05}_{-0.04}$ & $0.23^{+0.06}_{-0.05}$ & $0.07\pm0.02$ & $3.44\pm0.06$ \\ 
$29$ & $0.54^{+0.36}_{-0.22}$ & $0.32^{+0.15}_{-0.08}$ & $0.25^{+0.13}_{-0.08}$ & $0.08^{+0.04}_{-0.03}$ & $3.42\pm0.06$ \\ 
\hline
\end{tabular}

\vspace*{0.2cm}\begin{tabular}{c|ccccccc}
\hline\hline
Background region & $\log\left(\frac{N_{\rm H}}{{\rm cm^{-2}}}\right)$ & $\chi^2/{\rm dof}$ & dof\tablefootmark{($b$)} & Bkg of reg\tablefootmark{($c$)} & Sky area & ${\rm FGKM_{norm}}$ \\
 &  &  &  &  & (${\rm deg^2}$) & ($10^3\,M_\odot\,{\rm cm^{-2}}$) \\\hline
$0$ & $20.58^{+0.02}_{-0.03}$ & $1.07$ & $175$ & 1, 2, 3 & $240.78$ & $6.88$ \\ 
$4$ & $20.48^{+0.03}_{-0.04}$ & $1.06$ & $175$ & 5, 6, 7 & $98.78$ & $4.15$ \\ 
$8$ & $20.62\pm0.03$ & $1.11$ & $176$ & 9, 10, 11, 34 & $57.63$ & $3.84$ \\ 
$12$ & $20.67^{+0.04}_{-0.06}$ & $1.28$ & $176$ & 13, 14, 15 & $28.37$ & $2.65$ \\ 
$16$ & $20.40\pm0.03$ & $1.09$ & $175$ & 17 & $171.48$ & $5.29$ \\ 
$18$ & $20.64^{+0.02}_{-0.05}$ & $1.25$ & $175$ & 19, 20, 21 & $90.99$ & $4.38$ \\ 
$22$ & $20.77\pm0.03$ & $1.01$ & $175$ & 23, 24, 25 & $39.09$ & $2.94$ \\ 
$26$ & $20.72\pm0.03$ & $1.07$ & $175$ & 27, 28, 33 & $23.33$ & $2.96$ \\ 
$29$ & $20.83\pm0.04$ & $1.15$ & $175$ & 30, 31, 32 & $25.39$ & $4.30$ \\ 

\hline
\end{tabular}
}
    \label{tab:bkg_param}
    \tablefoot{
    \tablefoottext{a}{The absolute metallicity (relative to H) cannot be constrained. Hence, we set the reference element to be Fe, such that the Fe/H ratio follows Solar abundance. The reported abundances and the associated uncertainties of other elements are therefore relative to Fe.}
    \tablefoottext{b}{Degrees of freedom.}
    \tablefoottext{c}{The eRObub regions that the background best-fit parameters (only $kT_{\rm LHB}$, $kT_{\rm CGM}$ and $Z_{\rm CGM}$) were applied to.}
    }
\end{table*}

\subsubsection{Evidence of a temperature transition within the Bubbles} \label{kT_jump}
The spectra of the eRObub are fairly complex. This complexity can be appreciated even without any spectral fitting, but by overlaying spectra of large regions within the bubbles. The overlays are shown separately for selected regions within the northern and southern eRObub in Fig.~\ref{fig:overlay}, folded with the response in surface brightness units. The dominant emission lines are labelled. We show a direct hemispheric comparison in Appendix~\ref{app:NS_com}.
Much of the difference in the spectral shape redwards of $\sim 0.5$\,keV is mainly driven by the difference in foreground absorption.
Qualitatively, the eRObub regions (including the NPS and the FB) are enhanced in \ion{Fe}{XVII}, \ion{Ne}{IX} and \ion{Ne}{X} lines compared to the background regions.
These lines alone are strong evidence of a temperature jump in the eRObub compared to the background. The \ion{O}{VIII} Ly$\alpha$ line is also enhanced in most regions within the eRObub. However, it is difficult at this stage to draw any conclusions as the absorption column density also modifies the \ion{O}{VIII}/\ion{O}{VII} ratio.

We draw attention to the cool shell region in the north (reg~3). Despite its higher EM than the background, its Fe and Ne emission lines are weaker than in other eRObub interior regions, suggesting a cooler temperature than the interior. Instead, it shows a similar \ion{O}{VIII}/\ion{O}{VII} ratio compared to the background (reg~0), potentially indicating a similar temperature as the CGM, keeping in mind that absorption could alter this ratio.
In the opposite hemisphere, the cool shell region (reg~25) is much less pronounced when compared to the background (reg~22), only showing a slight enhancement in the 0.6--1.0\,keV range, which could easily be the difference from the unresolved stellar contribution or the ISM (Sect.~\ref{subsubsec:bkg_reg} and \citet{Ponti26}). We found a stronger cool shell signature at lower latitudes ($-20\degr\lesssim b \lesssim -15\degr$; reg~32 versus reg~29, not shown); however, we could not rule out that it is caused by the steep rise expected from the stellar contribution or the ISM towards the Galactic plane.

\begin{figure}[htbp]
    \centering
    \includegraphics[width=0.49\textwidth]{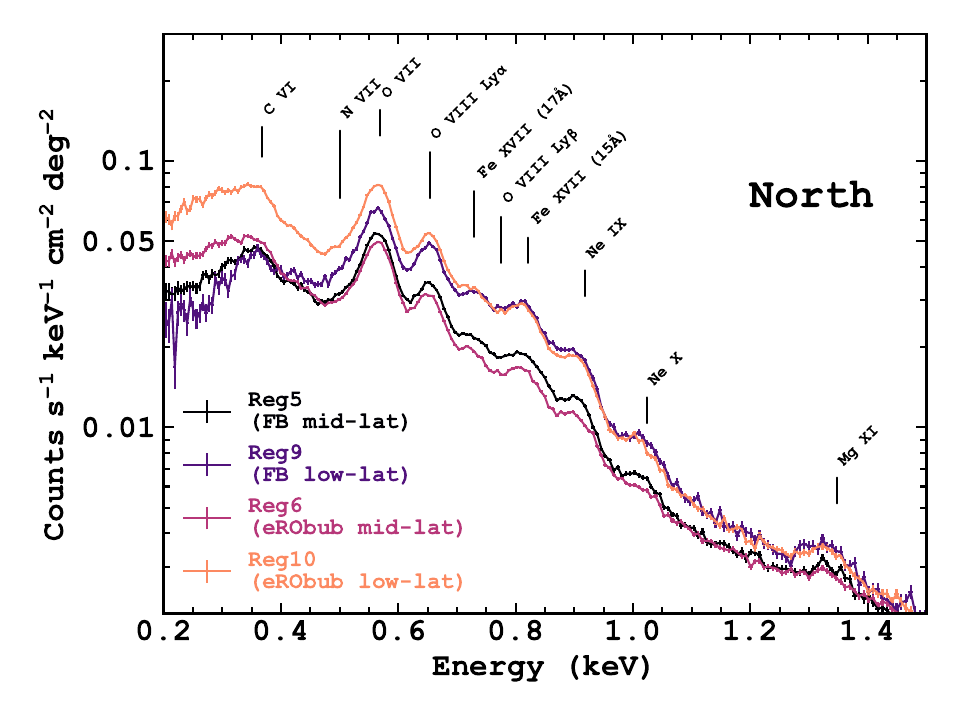}
    \caption{Comparison of northern eROSITA and Fermi bubble spectra, separated by two Galactic latitude ranges.}
    \label{fig:FB}
\end{figure}

We did not find evidence of the FB regions having an additional nonthermal component, compared to the rest of the bubbles. This is most obvious when comparing reg~24 (eRObub) and reg~23 (FB) in the south and reg~10 (eRObub) and reg~9 (FB) in the North (Fig.~\ref{fig:FB}). Their spectra are almost identical, featuring similarly enhanced \ion{Fe}{XVII} and \ion{Ne}{XI,X} lines.
The difference in the soft part between reg~9 and 10 is caused by a significantly higher $N_{\rm H}$. It is obvious that there is no need for an additional power-law component in the FB.

\subsubsection{Single CIE (1T) model and cool shell} \label{subsubsec:baseline}
We proceeded to fit the regions within the eRObub using an additional CIE component. We found that assuming all metals follow a single ratio to solar abundance is too rigid. After extensive testing, we converged on leaving O, Fe, and Ne abundances free as the standard template, while abundances of all other elements heavier than He were coupled to the O abundance. Similar to the background regions, allowing these elemental abundances to be free is driven by the high S/N spectra. We defer the justification of this modelling choice in Appendix.~\ref{app:erobub_abund}.
Additional flexibility in C and Mg abundances is required in some regions. These regions can be identified by filled $Z_{\rm C,eRObub}$ and $Z_{\rm Mg,eRObub}$ rows in Table~\ref{tab:erobub_params}.
The eRObub CIE component was subjected to the same absorption column density as the CGM, CXB and stellar components except for reg~13, 14, 15, 30 and 31, where a two-layer absorption model was adopted. The two-layer absorption model is where the (extragalactic) CXB is attenuated by the fixed, total Galactic $N_{\rm H}$, while the absorption for the CGM and stellar component is coupled and allowed to vary below the total Galactic value because they are physically within the Galaxy. This prescription has the same degrees of freedom as the one-layer absorption model because the CXB absorption is fixed, and more importantly, performs better than the one-layer model in low-latitude regions. A quantitative comparison between the two models in the relevant regions is shown in Appendix~\ref{app:CGM_abs}.

\begin{table*}[htbp]
    \small
    \caption{Parameters of regions within the eRObub. This table spans three pages.} \label{tab:erobub_params}
    \centering
    \renewcommand{\arraystretch}{1.2}
    {\small
    \begin{tabular}{c|ccccccc}
    \hline\hline
Reg & $l$ & $b$ & ${\rm EM_{LHB}}$ & ${\rm EM_{CGM}}$ & ${\rm CXB_{norm}}$ & $\log\left(\frac{N_{\rm H,CXB}}{{\rm cm^{-2}}}\right)$\tablefootmark{(a)}  & $\log\left(\frac{N_{\rm H,CGM}}{{\rm cm^{-2}}}\right)$\tablefootmark{(a)} \\
 & (\degr) & (\degr) & ($10^{-3}\,{\rm cm^{-6}\,pc}$) & ($10^{-2}\,{\rm cm^{-6}\,pc}$) & ($10^{-3}\,{\rm ph\,s^{-1}\,cm^{-2}\,deg^{-2}}$) &  &  \\\hline
$1$ & $335$ & $54$ & $4.1\pm0.1$ & $8.94^{+0.27}_{-0.22}$ & $3.63\pm0.03$ & $20.52\pm0.01$ & $\ldots$ \\ 
$2$ & $341$ & $69$ & $4.5\pm0.2$ & $11.21^{+0.43}_{-0.46}$ & $3.56\pm0.05$ & $20.45\pm0.01$ & $\ldots$ \\ 
$3$ & $291$ & $55$ & $5.7\pm0.3$ & $11.60^{+0.40}_{-0.39}$ & $3.67\pm0.02$ & $20.40\pm0.01$ & $\ldots$ \\ 
$5$ & $350$ & $37$ & $2.2\pm0.1$ & $4.17^{+0.22}_{-0.25}$ & $3.14\pm0.06$ & $20.76\pm0.01$ & $\ldots$ \\ 
$6$ & $324$ & $38$ & $3.0^{+0.1}_{-0.0}$ & $0.09^{+0.32}_{-0.04}$ & $3.54^{+0.05}_{-0.02}$ & $20.68\pm0.00$ & $\ldots$ \\ 
$7$ & $300$ & $35$ & $2.7^{+0.3}_{-0.2}$ & $2.02^{+1.10}_{-2.02}$ & $3.49^{+0.03}_{-0.04}$ & $20.58\pm0.02$ & $\ldots$ \\ 
$9$ & $352$ & $26$ & $2.5\pm0.2$ & $0.16\pm0.07$ & $2.75^{+0.13}_{-0.14}$ & $20.89\pm0.01$ & $\ldots$ \\ 
$10$ & $331$ & $25$ & $5.3\pm0.2$ & $0.38^{+0.02}_{-0.03}$ & $3.04^{+0.08}_{-0.09}$ & $20.66\pm0.01$ & $\ldots$ \\ 
$11$ & $314$ & $24$ & $6.8\pm0.4$ & $0.40^{+0.04}_{-0.03}$ & $3.12^{+0.11}_{-0.12}$ & $20.51\pm0.01$ & $\ldots$ \\ 
$13$ & $342$ & $18$ & $2.4\pm0.3$ & $0.12\pm0.09$ & $2.43\pm0.33$ & $21.15$ & $20.83\pm0.01$ \\ 
$14$ & $327$ & $18$ & $3.4^{+0.5}_{-0.3}$ & $0.35^{+0.05}_{-0.03}$ & $3.45^{+0.22}_{-0.50}$ & $21.02$ & $20.60\pm0.01$ \\ 
$15$ & $315$ & $18$ & $4.5\pm0.3$ & $0.24\pm0.03$ & $3.51^{+0.18}_{-0.19}$ & $21.01$ & $20.63\pm0.01$ \\ 
$17$ & $336$ & $-64$ & $6.8^{+0.2}_{-0.1}$ & $2.77^{+0.27}_{-0.74}$ & $3.57^{+0.04}_{-0.05}$ & $20.29\pm0.02$ & $\ldots$ \\ 
$19$ & $351$ & $-40$ & $4.7^{+0.2}_{-0.5}$ & $1.27^{+0.21}_{-0.13}$ & $3.32^{+0.10}_{-0.49}$ & $20.38^{+0.02}_{-0.08}$ & $\ldots$ \\ 
$20$ & $327$ & $-40$ & $4.5\pm0.2$ & $1.21\pm0.02$ & $3.04^{+0.08}_{-0.09}$ & $20.32\pm0.02$ & $\ldots$ \\ 
$21$ & $313$ & $-40$ & $4.1\pm0.2$ & $1.10^{+0.08}_{-0.07}$ & $3.40^{+0.03}_{-0.05}$ & $20.52\pm0.02$ & $\ldots$ \\ 
$23$ & $348$ & $-25$ & $4.0\pm0.3$ & $2.56^{+0.59}_{-0.28}$ & $3.24^{+0.10}_{-0.16}$ & $20.54\pm0.01$ & $\ldots$ \\ 
$24$ & $331$ & $-25$ & $4.2\pm0.2$ & $1.80^{+0.38}_{-0.39}$ & $3.06^{+0.09}_{-0.10}$ & $20.60\pm0.01$ & $\ldots$ \\ 
$25$ & $317$ & $-25$ & $2.4\pm0.2$ & $1.70^{+0.31}_{-0.35}$ & $3.63\pm0.05$ & $20.71^{+0.02}_{-0.03}$ & $\ldots$ \\ 
$27$ & $350$ & $-17$ & $4.4\pm0.5$ & $7.72^{+0.43}_{-0.71}$ & $3.08^{+0.16}_{-0.18}$ & $20.53\pm0.01$ & $\ldots$ \\ 
$28$ & $319$ & $-17$ & $6.1\pm0.4$ & $5.25\pm0.18$ & $3.56^{+0.08}_{-0.09}$ & $20.59\pm0.02$ & $\ldots$ \\ 
$30$ & $352$ & $-12$ & $2.7\pm0.3$ & $2.78^{+1.36}_{-1.69}$ & $3.10^{+0.24}_{-0.23}$ & $21.17$ & $20.77\pm0.01$ \\ 
$31$ & $334$ & $-12$ & $3.9^{+0.2}_{-0.3}$ & $0.33^{+1.09}_{-0.33}$ & $3.47^{+0.21}_{-0.19}$ & $21.25$ & $20.84\pm0.01$ \\ 
$33$ & $332$ & $-17$ & $3.2\pm0.3$ & $7.48^{+0.61}_{-0.84}$ & $2.98\pm0.14$ & $20.63\pm0.01$ & $\ldots$ \\ 
$34$ & $305$ & $24$ & $2.7^{+0.4}_{-0.6}$ & $0.34^{+0.02}_{-0.06}$ & $3.44^{+0.06}_{-0.05}$ & $20.56^{+0.01}_{-0.02}$ & $\ldots$ \\ \hline         
    \end{tabular}

\vspace*{0.2cm}\begin{tabular}{c|ccccccc}
\hline\hline
Reg & $kT_{\rm eRObub1}$ & ${\rm EM_{eRObub1}}$ & $kT_{\rm eRObub2}$\tablefootmark{($b$)} & ${\rm EM_{eRObub2}}$\tablefootmark{($b$)} & $Z_{\rm C, eRObub}$\tablefootmark{($c$)}& $Z_{\rm O, eRObub}$ & $Z_{\rm Ne, eRObub}$\tablefootmark{($c$)} \\
 & (eV) & ($10^{-3}\,{\rm cm^{-6}\,pc}$) & (eV) & ($10^{-3}\,{\rm cm^{-6}\,pc}$) & ($Z_{\rm C, \odot})$  & ($Z_{\rm O, \odot})$ & ($Z_{\rm Ne, \odot})$ \\\hline
$1$ & $224^{+6}_{-4}$ & $1.5\pm0.2$ & $614^{+17}_{-16}$ & $0.56\pm0.03$ & $6.83^{+0.85}_{-0.79}$ & $1.80\pm0.12$ & $1.78^{+0.18}_{-0.19}$ \\ 
$2$ & $259\pm8$ & $5.8^{+2.4}_{-2.2}$ & $589^{+18}_{-16}$ & $1.65^{+0.75}_{-0.68}$ & $\ldots$ & $0.95^{+0.63}_{-0.28}$ & $1.13^{+0.65}_{-0.30}$ \\ 
$3$ & $198^{+4}_{-3}$ & $2.6\pm0.2$ & $\ldots$ & $\ldots$ & $5.75^{+0.60}_{-0.55}$ & $1.49^{+0.08}_{-0.07}$ & $3.05\pm0.17$ \\ 
$5$ & $238\pm8$ & $38.0^{+4.6}_{-4.0}$ & $600^{+17}_{-16}$ & $6.82^{+0.79}_{-0.76}$ & $\ldots$ & $0.17^{+0.02}_{-0.01}$ & $0.27\pm0.02$ \\ 
$6$ & $187^{+2}_{-0}$ & $82.1^{+0.7}_{-4.0}$ & $622^{+32}_{-22}$ & $4.80^{+0.16}_{-0.20}$ & $0.42^{+0.02}_{-0.03}$ & $0.16\pm0.00$ & $0.27\pm0.01$ \\ 
$7$ & $172^{+14}_{-8}$ & $45.4^{+25.8}_{-14.8}$ & $\ldots$ & $\ldots$ & $0.38^{+0.11}_{-0.08}$ & $0.13^{+0.06}_{-0.04}$ & $0.27\pm0.05$ \\ 
$9$ & $198^{+5}_{-4}$ & $127.3^{+9.6}_{-9.7}$ & $575\pm13$ & $17.68^{+1.48}_{-1.41}$ & $\ldots$ & $0.14\pm0.01$ & $0.29\pm0.02$ \\ 
$10$ & $217\pm4$ & $107.2^{+6.0}_{-5.1}$ & $593^{+17}_{-15}$ & $13.12^{+0.99}_{-0.94}$ & $\ldots$ & $0.10\pm0.00$ & $0.20\pm0.01$ \\ 
$11$ & $204^{+5}_{-3}$ & $156.5^{+7.6}_{-9.1}$ & $653^{+31}_{-32}$ & $6.08^{+1.17}_{-1.12}$ & $\ldots$ & $0.07\pm0.00$ & $0.15\pm0.01$ \\ 
$13$ & $200^{+5}_{-4}$ & $334.9^{+25.6}_{-25.2}$ & $576^{+31}_{-34}$ & $23.78^{+3.46}_{-3.28}$ & $\ldots$ & $0.11\pm0.01$ & $0.22\pm0.02$ \\ 
$14$ & $204^{+14}_{-6}$ & $246.4^{+16.8}_{-27.5}$ & $556^{+316}_{-71}$ & $12.43^{+3.13}_{-6.41}$ & $\ldots$ & $0.04^{+0.00}_{-0.01}$ & $0.14^{+0.01}_{-0.03}$ \\ 
$15$ & $198^{+4}_{-3}$ & $238.6^{+12.0}_{-12.1}$ & $621^{+69}_{-68}$ & $6.20^{+1.90}_{-1.80}$ & $\ldots$ & $0.06\pm0.00$ & $0.16\pm0.01$ \\ 
$17$ & $206^{+8}_{-13}$ & $14.2^{+7.7}_{-2.7}$ & $652^{+27}_{-26}$ & $2.74^{+0.60}_{-0.33}$ & $\ldots$ & $0.12\pm0.01$ & $0.18\pm0.03$ \\ 
$19$ & $211^{+98}_{-12}$ & $29.1^{+7.4}_{-24.0}$ & $605^{+24}_{-77}$ & $7.98^{+10.96}_{-1.52}$ & $\ldots$ & $0.06\pm0.01$ & $0.05^{+0.03}_{-0.05}$ \\ 
$20$ & $360^{+7}_{-6}$ & $11.6\pm1.3$ & $605^{+25}_{-26}$ & $7.17^{+1.74}_{-1.45}$ & $\ldots$ & $0.18\pm0.01$ & $0.07\pm0.02$ \\ 
$21$ & $215^{+26}_{-14}$ & $17.6^{+4.0}_{-4.6}$ & $\ldots$ & $\ldots$ & $\ldots$ & $0.05\pm0.01$ & $\ldots$ \\ 
$23$ & $223^{+22}_{-6}$ & $61.5^{+7.0}_{-14.5}$ & $576^{+21}_{-22}$ & $8.06^{+1.22}_{-2.41}$ & $\ldots$ & $0.12^{+0.02}_{-0.01}$ & $0.17^{+0.02}_{-0.03}$ \\ 
$24$ & $213^{+6}_{-5}$ & $73.7^{+8.1}_{-7.9}$ & $584\pm14$ & $11.03^{+1.03}_{-0.99}$ & $\ldots$ & $0.14\pm0.01$ & $0.24\pm0.01$ \\ 
$25$ & $208^{+24}_{-13}$ & $0.9^{+0.4}_{-0.3}$ & $\ldots$ & $\ldots$ & $\ldots$ & $2.81^{+0.76}_{-0.53}$ & $\ldots$ \\ 
$27$ & $258^{+13}_{-15}$ & $55.6^{+13.4}_{-7.3}$ & $589^{+29}_{-33}$ & $9.65^{+2.21}_{-1.73}$ & $\ldots$ & $0.18^{+0.02}_{-0.03}$ & $0.24^{+0.03}_{-0.02}$ \\ 
$28$ & $297^{+22}_{-21}$ & $1.0^{+0.2}_{-0.1}$ & $\ldots$ & $\ldots$ & $\ldots$ & $2.31^{+0.51}_{-0.43}$ & $\ldots$ \\ 
$30$ & $200^{+5}_{-4}$ & $267.9^{+30.3}_{-26.5}$ & $601^{+23}_{-22}$ & $15.88^{+2.17}_{-2.05}$ & $\ldots$ & $0.10\pm0.01$ & $0.19\pm0.01$ \\ 
$31$ & $186^{+2}_{-1}$ & $405.8^{+16.0}_{-21.1}$ & $618\pm34$ & $9.19^{+1.83}_{-1.86}$ & $\ldots$ & $0.08^{+0.00}_{-0.01}$ & $0.17\pm0.01$ \\ 
$33$ & $212\pm7$ & $125.1^{+16.4}_{-12.6}$ & $590^{+23}_{-22}$ & $12.32^{+1.64}_{-1.54}$ & $\ldots$ & $0.07\pm0.00$ & $0.19\pm0.01$ \\ 
$34$ & $194^{+6}_{-8}$ & $86.9^{+12.3}_{-8.4}$ & $\ldots$ & $\ldots$ & $\ldots$ & $0.05^{+0.01}_{-0.00}$ & $0.15^{+0.03}_{-0.01}$ \\ 
\hline
\end{tabular}
}
\end{table*}

\begin{table*}[t]
  \ContinuedFloat
  \caption{Continued.} \label{tab:erobub_cont}
    \centering
\renewcommand{\arraystretch}{1.2}
{\small
\begin{tabular}{c|cccccc|ccc}
\hline\hline
Reg & $Z_{\rm Mg, eRObub}$\tablefootmark{($c$)} & $Z_{\rm Fe, eRObub}$\tablefootmark{($c,d$)} & $\chi^2/{\rm dof}$ & dof & Sky area & ${\rm Stars_{norm}}$ & 1T $\chi^2$/dof\tablefootmark{($g$)} & F-stat (1T)\tablefootmark{($h$)} & $p$ (1T)\tablefootmark{($h$)} \\
 & ($Z_{\rm Mg, \odot})$ & ($Z_{\rm Fe, \odot})$ &  &  & (${\rm deg^2}$) & ($10^3\,M_\odot\,{\rm cm^{-2}}$)  &&&\\\hline
$1$ & $\ldots$ & $1.00$ & $1.28$ & $174$ & $484.7$ & $17.3$ & $2.93$ & $114$ & $10^{-16}$\\ 
$2$ & $\ldots$ & $0.66^{+0.40}_{-0.18}$ & $1.63$ & $174$ & $369.9$ & $11.5$ & $3.60$ & $107$ & $10^{-16}$\\ 
$3$ & $\ldots$ & $1.00$ & $1.18$ & $176$ & $263.7$ & $7.8$ & $\ldots$ & $\ldots$ & \ldots\\ 
$5$ & $0.37\pm0.03$ & $0.16\pm0.01$ & $1.57$ & $173$ & $251.5$ & $14.4$ & $3.03$ & $82$ & $10^{-16}$\\ 
$6$ & $0.32^{+0.03}_{-0.02}$ & $0.14\pm0.00$ & $1.72$ & $172$ & $387.9$ & $20.0$ & $3.28$ & $79$ & $10^{-16}$\\ 
$7$ & $1.37^{+0.34}_{-0.38}$ & $0.06\pm0.02$ & $1.46$ & $174$ & $157.4$ & $7.6$ & $\ldots$ & $\ldots$ & \ldots\\ 
$9$ & $0.27\pm0.02$ & $0.14\pm0.01$ & $1.48$ & $173$ & $123.3$ & $12.1$ & $3.24$ & $106$ & $10^{-16}$\\ 
$10$ & $0.23\pm0.01$ & $0.12\pm0.00$ & $2.20$ & $173$ & $206.8$ & $18.9$ & $4.10$ & $77$ & $10^{-16}$\\ 
$11$ & $0.24\pm0.02$ & $0.06\pm0.00$ & $2.36$ & $173$ & $120.4$ & $9.4$ & $2.51$ & $6$ & $2\times10^{-3}$\\ 
$13$ & $0.22^{+0.03}_{-0.02}$ & $0.12\pm0.01$ & $1.23$ & $173$ & $25.8$ & $4.2$ & $1.70$ & $34$ & $3\times10^{-13}$\\ 
$14$ & $0.15^{+0.08}_{-0.02}$ & $0.06^{+0.01}_{-0.00}$ & $1.59$ & $173$ & $52.8$ & $7.3$ & $1.71$ & $7$ & $8\times10^{-4}$\\ 
$15$ & $0.28\pm0.03$ & $0.06\pm0.00$ & $1.65$ & $173$ & $53.8$ & $6.4$ & $1.73$ & $5$ & $6\times10^{-3}$\\ 
$17$ & $0.47\pm0.07$ & $0.09\pm0.01$ & $1.75$ & $173$ & $669.8$ & $22.1$ & $2.29$ & $28$ & $2\times10^{-11}$\\ 
$19$ & $\ldots$ & $0.07^{+0.01}_{-0.03}$ & $1.68$ & $174$ & $313.5$ & $18.1$ & $2.20$ & $28$ & $3\times10^{-11}$\\ 
$20$ & $\ldots$ & $0.11\pm0.01$ & $1.39$ & $174$ & $322.5$ & $16.8$ & $1.99$ & $39$ & $10^{-14}$\\ 
$21$ & $\ldots$ & $\ldots$ & $1.02$ & $178$ & $118.0$ & $6.1$ & $\ldots$ & $\ldots$ & \ldots\\ 
$23$ & $0.30^{+0.03}_{-0.05}$ & $0.14\pm0.01$ & $1.49$ & $173$ & $168.0$ & $17.1$ & $2.01$ & $32$ & $2\times10^{-12}$\\ 
$24$ & $0.29\pm0.02$ & $0.13\pm0.01$ & $1.48$ & $173$ & $177.9$ & $16.9$ & $2.61$ & $68$ & $10^{-16}$\\ 
$25$ & $\ldots$ & $1.00$ & $1.04$ & $178$ & $53.7$ & $4.5$ & $\ldots$ & $\ldots$ & \ldots\\ 
$27$ & $\ldots$ & $0.19\pm0.02$ & $1.29$ & $174$ & $78.8$ & $13.7$ & $1.71$ & $30$ & $8\times10^{-12}$\\ 
$28$ & $\ldots$ & $1.00$ & $1.15$ & $178$ & $26.5$ & $3.5$ & $\ldots$ & $\ldots$ & \ldots\\ 
$30$ & $0.24\pm0.02$ & $0.11\pm0.00$ & $1.50$ & $173$ & $75.7$ & $24.2$ & $2.07$ & $34$ & $3\times10^{-13}$\\ 
$31$ & $0.25\pm0.02$ & $0.07\pm0.00$ & $2.03$ & $173$ & $82.7$ & $21.1$ & $2.31$ & $13$ & $6\times10^{-6}$\\ 
$33$ & $0.23\pm0.02$ & $0.10^{+0.01}_{-0.00}$ & $1.44$ & $173$ & $99.8$ & $15.3$ & $1.90$ & $29$ & $10^{-11}$\\ 
$34$ & $\ldots$ & $0.03^{+0.01}_{-0.00}$ & $1.54$ & $176$ & $54.6$ & $3.9$ & $\ldots$ & $\ldots$ & \ldots\\ 
\hline
\end{tabular}
}
\normalsize
\tablefoot{The values adopted for $kT_{\rm LHB}$, $kT_{\rm CGM}$ and $Z_{\rm CGM}$ in each eRObub region can be found in Table~\ref{tab:bkg_param}, which listed the parameters of the background regions. These parameters are frozen during eRObub region fitting.\\
\tablefoottext{a}{Regions that have uncertainties associated with $N_{\rm H,CXB}$ have variable $N_{\rm H,CXB}$ and are fitted by a single-layer absorption model. In this model, all CXB, CGM, stellar and eRObub components are subjected to the same absorption column density. In contrast, regions with no associated uncertainties in $N_{\rm H,CXB}$ are fitted by a two-layer absorption model. With this treatment, the CXB is absorbed by a $N_{\rm H,CXB}$ that is fixed to the total Galactic value. Other components, including the CGM, stellar, and eRObub components, are located within the Galaxy and are thus absorbed by smaller $N_{\rm H}$ that are allowed to vary during the fit. The fit result of this $N_{\rm H}$ is reported in the next column ($N_{\rm H,CGM}$). Both absorption treatments contribute only one free parameter; as in the two-layer treatment, $N_{\rm H}$ remains frozen during the fit. We found the two-layer absorption model advantageous for reproducing spectra from regions closest to the Galactic plane.}\\
\tablefoottext{b}{If empty, the eRObub is modelled as a single temperature plasma.}\\
\tablefoottext{c}{Linked with $Z_{\rm O, eRObub}$ if left empty.}\\
\tablefoottext{d}{If $Z_{\rm Fe,eRObub}=1$, Fe is used as the reference element and is fixed at solar because the data provide inadequate constraints on the absolute abundance, and the abundances of other elements are relative to Fe.}\\
\tablefoottext{e}{eRObub abundance is fixed at solar.}\\
\tablefoottext{f}{Southern cool shell regions. The detection of the eRObub component is weak.}\\
\tablefoottext{g}{The $\chi^2/{\rm dof}$ of single temperature (1T) model. The dof of the 1T model is two less than the 2T model.}\\
\tablefoottext{h}{F-statistic and its corresponding probability $p$ when comparing the 1T and 2T models in the F-test.}
}
\end{table*}

\begin{table}[]
  \ContinuedFloat
  \caption{Continued.}
    \centering
\renewcommand{\arraystretch}{1.2}
{\small
\begin{tabular}{c|ccc}
\hline\hline
Reg & \texttt{neij}\tablefootmark{($i$)} $\chi^2$/dof & F-stat (\texttt{neij})\tablefootmark{($j$)} & $p$ (\texttt{neij})\tablefootmark{($j$)} \\
 &  &  &  \\\hline
$1$ & $1.40$ & $17$ & $7\times10^{-5}$\\ 
$2$ & $2.39$ & $82$ & $2\times10^{-16}$\\ 
$3$ & $\ldots$ & $\ldots$ & \ldots\\ 
$5$ & $2.48$ & $101$ & $10^{-16}$\\ 
$6$ & $3.19$ & $149$ & $10^{-16}$\\ 
$7$ & $\ldots$ & $\ldots$ & \ldots\\ 
$9$ & $2.13$ & $78$ & $10^{-15}$\\ 
$10$ & $3.83$ & $130$ & $10^{-16}$\\ 
$11$ & $2.61$ & $19$ & $3\times10^{-5}$\\ 
$13$ & $1.80$ & $82$ & $2\times10^{-16}$\\ 
$14$ & $1.69$ & $13$ & $5\times10^{-4}$\\ 
$15$ & $1.75$ & $11$ & $10^{-3}$\\ 
$17$ & $3.07$ & $133$ & $10^{-16}$\\ 
$19$ & $1.71$ & $4$ & $4\times10^{-2}$\\ 
$20$ & $1.65$ & $34$ & $2\times10^{-8}$\\ 
$21$ & $\ldots$ & $\ldots$ & \ldots\\ 
$23$ & $1.96$ & $56$ & $3\times10^{-12}$\\ 
$24$ & $1.96$ & $58$ & $2\times10^{-12}$\\ 
$25$ & $\ldots$ & $\ldots$ & \ldots\\ 
$27$ & $1.64$ & $49$ & $5\times10^{-11}$\\ 
$28$ & $\ldots$ & $\ldots$ & \ldots\\ 
$30$ & $2.03$ & $62$ & $3\times10^{-13}$\\ 
$31$ & $2.21$ & $17$ & $7\times10^{-5}$\\ 
$33$ & $1.95$ & $63$ & $2\times10^{-13}$\\ 
$34$ & $\ldots$ & $\ldots$ & \ldots\\ \hline
\end{tabular}
}
\tablefoot{
\tablefoottext{i}{The $\chi^2/{\rm dof}$ of the plane-parallel shock (\texttt{neij}) model. The dof of the \texttt{neij} model is one less than the 2T model.}\\
\tablefoottext{j}{F-statistic and its corresponding probability $p$ when comparing the neij and 2T models in the F-test.}
}
\end{table}

\begin{figure*}[htbp]
    \centering
    \includegraphics[width=0.49\textwidth]{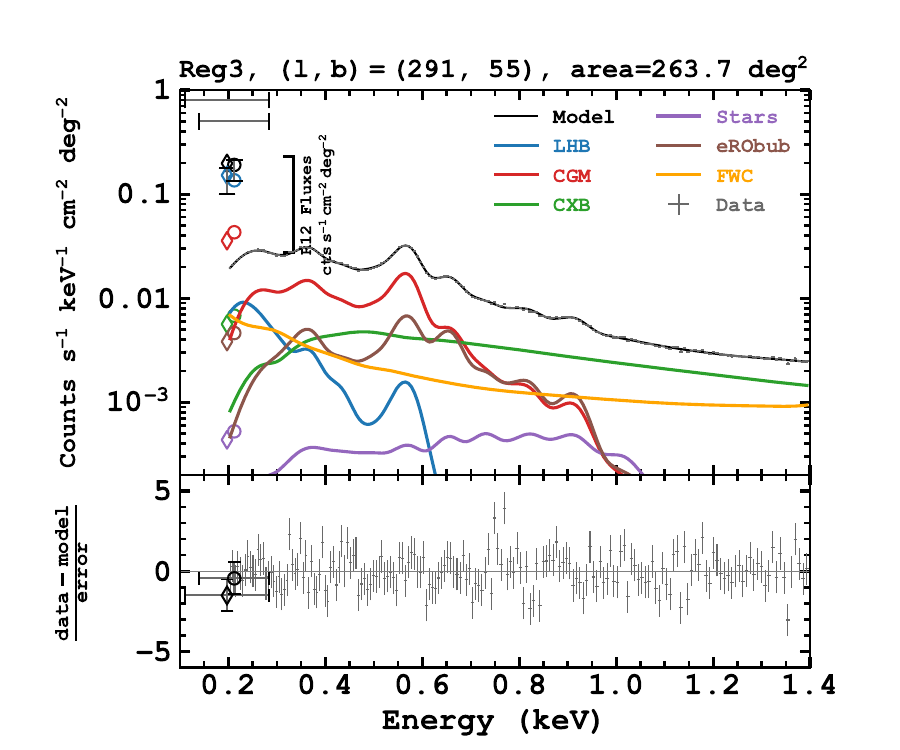}
    \includegraphics[width=0.49\textwidth]{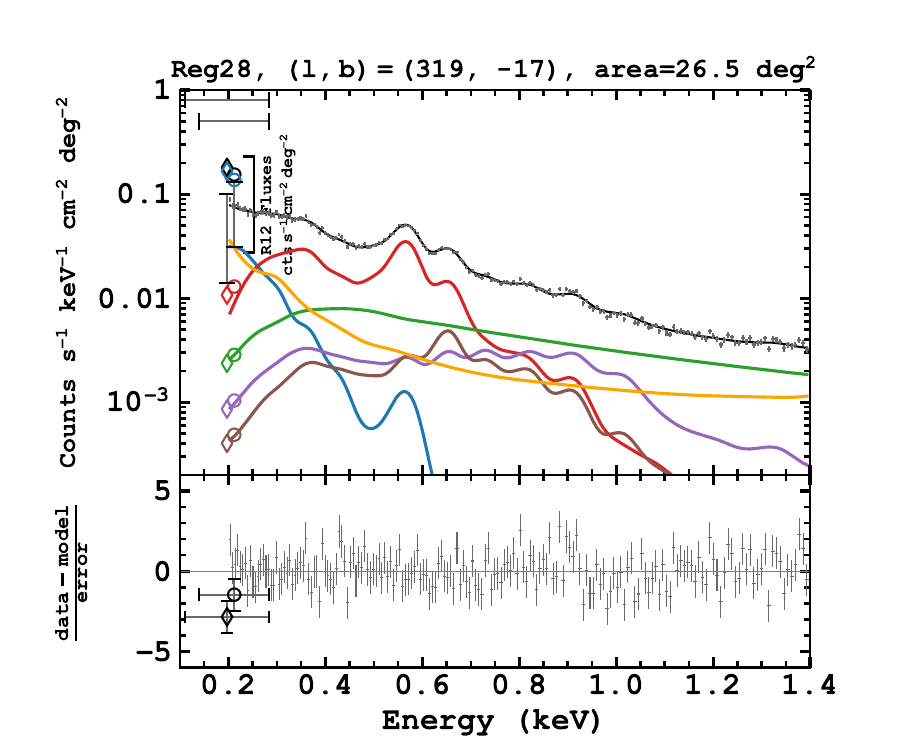}
    \caption{Example spectra from the cool shell region, showing reg~3 ({\it left}) in the north and reg~28 ({\it right}) in the south.}
    \label{fig:cool_shell_spec}
\end{figure*}

\begin{figure}[htbp]
    \centering
    \includegraphics[width=\linewidth]{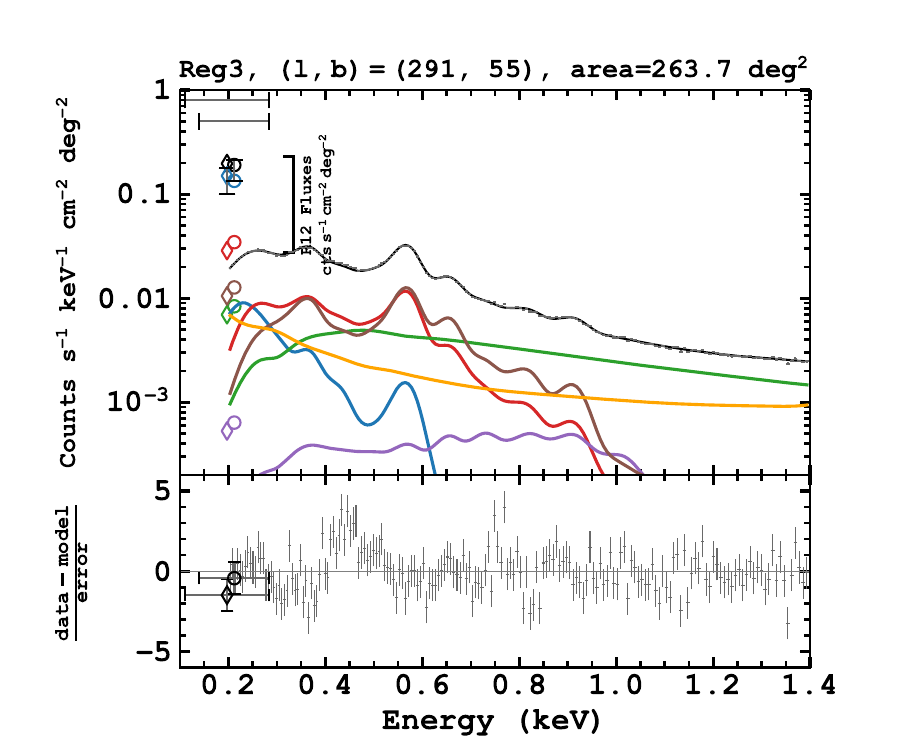}
    \caption{Spectrum, model and residual of the reg~3, similar to the left panel of Fig.~\ref{fig:cool_shell_spec} but the CGM EM is fixed to the background (reg~0).}
    \label{fig:cool_shell_fixCGMEM}
\end{figure}

We could get reasonable fits to some regions with the single CIE model, most notably for regions located in the cool shell (reg~3, 7, 15, 21, 25, 28, 32 and 34). The best-fit parameters and fit statistics of the cool shell are listed in Table~\ref{tab:erobub_params} (full parameter table of the eRObub interior regions in Table~\ref{tab:1T}). Figure~\ref{fig:cool_shell_spec} shows two example spectra of the cool shell. Region~3 is where the cool shell is the brightest and easily identifiable from the eRASS1 RGB map in Fig.~\ref{fig:bubble_rgb}.
Reg~3 is also particularly interesting because it encapsulates much of the cool shell that appears to have a cooler temperature than the surrounding CGM if taking the ratio of \ion{O}{VIII}/\ion{O}{VII} narrowband maps as a plasma temperature proxy \citep{line-ratio}. Indeed, using the same narrowband definition as \citet{line-ratio}, reg~3 has a lower \ion{O}{VIII}/\ion{O}{VII} ratio (0.583) than reg~0 (background; 0.588), hence corresponds to a lower temperature. There are several caveats to using the narrowband ratio as a temperature proxy, as the authors already cautioned, including the presence of other components that also emit in the oxygen lines and foreground absorption. In fact, our fits of reg~3 and reg~0 reveal that the cool shell does not have a lower temperature than the surrounding CGM.
As shown in the left panel of Fig.~\ref{fig:cool_shell_spec} and Table~\ref{tab:erobub_params}, the additional eRObub component (brown line; corresponding to the cool shell) has a higher temperature ($kT = 198^{+4}_{-3}\,$eV) than the CGM component ($kT=149^{+1}_{-2}\,$eV). A careful reader may notice that the EM of the CGM component has increased by $\approx65\%$ when compared to reg~0. While this is possible and is indeed the reason we allow the CGM EM to be free in all regions, the proximity of reg~0 and 3 suggests that the CGM component in this case likely has a non-negligible contribution from the cool shell. Therefore, we performed an additional fit, assuming the CGM EM in the cool shell equals the background (reg~0), and let the eRObub component fit the excess emission above the background. The fit is shown in Fig.~\ref{fig:cool_shell_fixCGMEM}. The temperature of the cool shell decreased to $179\pm1$\,eV, which is still higher than the CGM temperature. From this exercise, we conclude that the cool shell region is likely dominated by a plasma with a temperature between $\approx0.18-0.2\,$keV.

On the other hand, the southern cool shell regions (reg~21, 25, 28, 32) possess an eRObub component much fainter than the north; the surface brightness (flux per sky area) of the cool shell in $0.2$--$1.4$\,keV band in reg~3 is $>10$ times higher than the southern counterparts\footnote{It is difficult to see from the ${\rm EM_{eRObub}}$ column in Table~\ref{tab:erobub_params} because the fitted metal abundances and the corresponding reference element differ. We computed the surface brightness in \texttt{SPEX}. In reg~3, the surface brightness is 263 photons$\,{\rm s^{-1}\,m^{-2}\,deg^{-2}}$, while the highest surface brightness recorded in the southern region is 21\,photons$\,{\rm s^{-1}\,m^{-2}\,deg^{-2}}$.}. In the southern cool shell regions, this corresponds to a level comparable to the estimated unresolved stellar contribution. Given the uncertainty in predicting the stellar contribution in the SXRB from extrapolating from the volume-complete 10\,pc sample, as well as the unknown contribution from the ISM, currently, there is inadequate evidence that the cool shell is present in the southern eRObub.

\begin{figure*}[htbp]
    \centering
    \includegraphics[width=.49\textwidth]{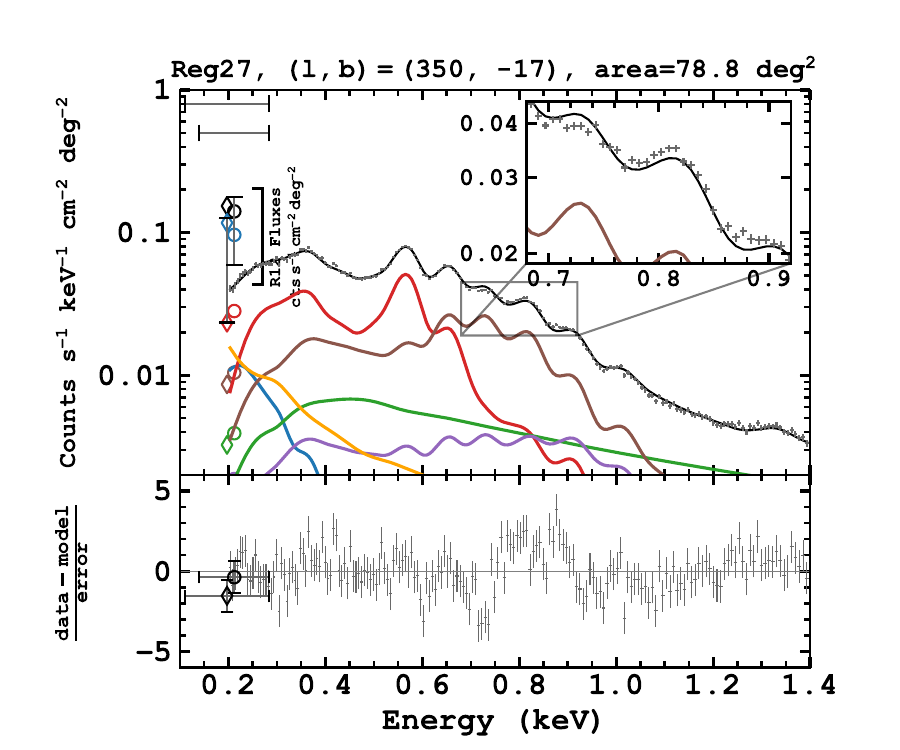}
    \includegraphics[width=.49\textwidth]{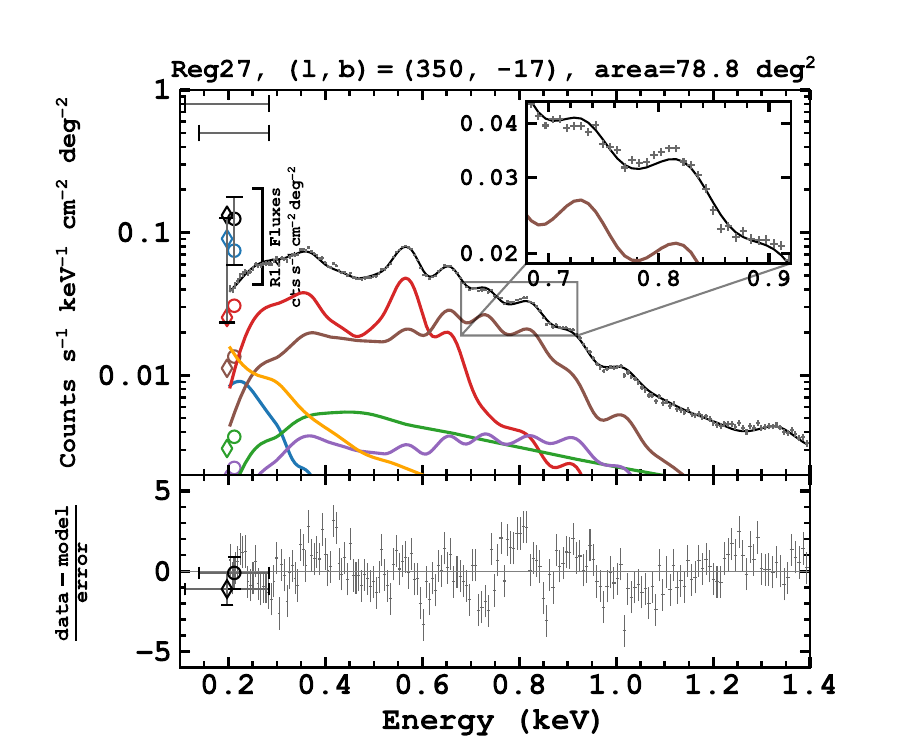}
    \includegraphics[width=.49\textwidth]{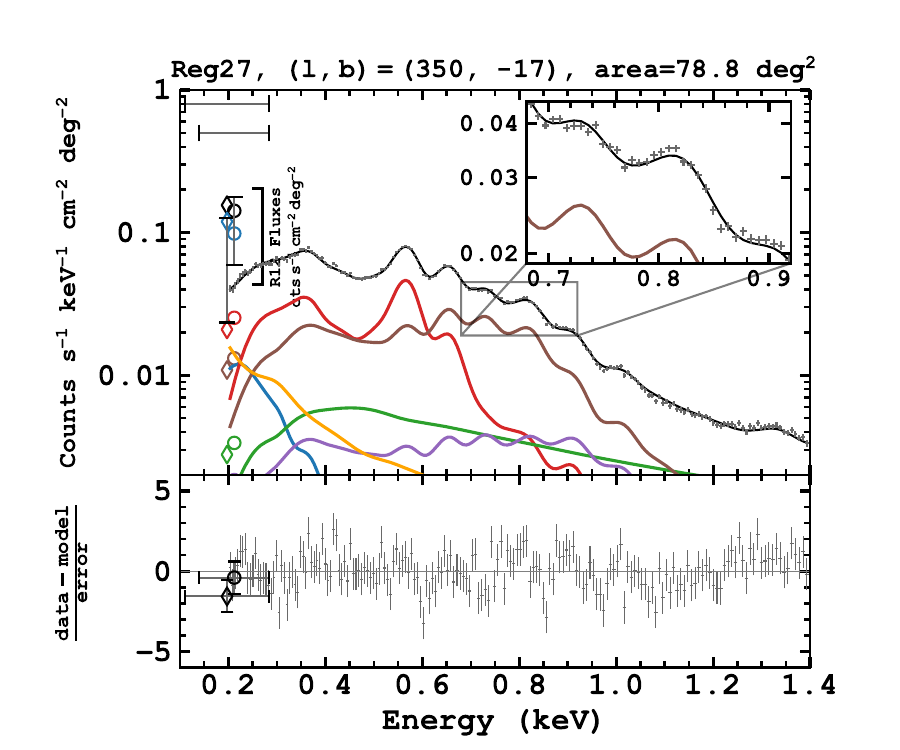}
\caption{Example comparison of 1T, plane-parallel shock \texttt{neij} and 2T models of the eRObub in reg~27 with insets showing the residuals near the \ion{Fe}{XVII} L shell transitions. The line colours follow the legends in Fig.~\ref{fig:bkg_spec} and \ref{fig:cool_shell_spec}. {\it Top:} 1T model ($\chi^2/{\rm dof}=1.70$; 176 dof). The dominant residual is around the 0.7--0.8\,keV range around the \ion{Fe}{XVII} L shell transitions (Table~\ref{tab:RS}). The data exhibit a higher \ion{Fe}{XVII} 15\,\AA/17\,\AA ~ratio than predicted by the 1T model. This is a general observation in most regions within the eRObub, excluding the cool shell. {\it Middle:} \texttt{neij} model ($\chi^2/{\rm dof}=1.64$; 175 dof). The residuals around the Fe L shell lines received only minor improvement. {\it Bottom:} 2T model ($\chi^2/{\rm dof}=1.28$; 174 dof). The model reproduces the data better, especially the \ion{Fe}{XVII} lines. In this particular region, the F-test yields a statistic of $29.7$ and a $p$-value of $\approx10^{-11}$, suggesting a highly significant improvement in using the 2T model to describe the eRObub over the 1T model. The improvement is similarly significant over the \texttt{neij} model, with a F-statistic of $49.3$ ($p$-value of $4\times10^{-11}$).}
    \label{fig:1T2T}
\end{figure*}

Besides the cool shell, our simplest assumption of a single temperature plasma in the eRObub does not fit the spectra well. As we inspect the fits further within the eRObub, there seems to be a persistent type of residual around the two main \ion{Fe}{XVII}\,L$\alpha$ peaks at $\sim$$17$\,{\AA} ($\sim$$0.7\,$keV) and $\sim$$15\,${\AA} ($\sim$$0.8\,$keV). We show an example of this residual in the top left panel of Fig.~\ref{fig:1T2T}. Elemental or ionic abundances do not affect the line ratio of these two lines, as these are transitions to the ground state from the same ion. We summarise the main transitions contributing to these two peaks in Table~\ref{tab:RS} under the entry of the \ion{Fe}{XVII} ion.

We investigated multiple causes for the anomalous \ion{Fe}{XVII}\,15\,\AA/17\,\AA~line ratio, including charge exchange, presence of an ionised absorber, and resonance scattering. We found none of these causes to be completely convincing. Therefore, we leave these investigations to Appendices~\ref{app:RS} and \ref{app:CX} for the interested readers. Instead, we found the most helpful way to resolve the anomalous line ratio was by invoking a second temperature component (Sect.~\ref{subsubsec:2T}). However, though unlikely, we cannot rule out the possibility of the anomalous line ratio being caused by the uncertainties in plasma codes, as we shall demonstrate in more detail in Sect.~\ref{subsec:plasma_code}.

\begin{table*}[htbp]
    \caption{List of resonance lines considered. Wavelengths are taken from the CHIANTI atomic database v10 \citep{Dere97,Delzanna21} and oscillator strengths $f$ are taken from the NIST database \citep[][and references therein]{NIST}. \ion{Fe}{XVIII} are not used in the spectral fitting (Appendix ~\ref{app:RS}), but potentially relevant \ion{Fe}{XVIII} transitions  are listed for completeness (if $f$ is available on the NIST database and $f>10^{-2}$).}
    \centering
    \renewcommand{\arraystretch}{1.2}
    {\small
    \begin{tabular}{lrrcccc}
    \hline\hline
      Ions   &  Lower level  & Upper level\tablefootmark{\textdagger} & Wavelength & Energy & $f$\\
             &               &             &  (\AA)     &  (eV) & \\\hline
    \ion{Ne}{IX} & $1s^2$\,$^1S_0$ & $1s2p$\,$^1P_1$ & 13.45 & 922 & $7.2\times10^{-1}$\\
    \ion{Fe}{XVII} & $2s^22p^6$\,$^1S_0$ & $2s^22p^53d$\,$^1P_1$ & 15.01 & 826 & 2.3 \\
    && $^3D_1$ & 15.26 & 812 & $6.3\times10^{-1}$ \\
    && $^3P_1$ & 15.45 & 802 & $9.7\times 10^{-3}$ \\
    && $2s^22p^53s$\,$^3P_1$ & 16.77 & 739 & $1.1\times10^{-1}$\\
    && $^1P_1$ & 17.05 & 727 & $1.2\times10^{-1}$ \\
    && $^3P_2$ & 17.10 & 725 &$4.4\times10^{-8}$ \\
    \ion{O}{VIII} (Ly$\beta$) & $1s$\,$^2S_{1/2}$ & $3p$\,$^2P_{3/2}$ & 16.01 & 775 & $5.3\times10^{-2}$ \\
    & & $^2P_{1/2}$ & 16.01 & 775 & $2.6\times10^{-2}$ \\
    \ion{O}{VIII} (Ly$\alpha$) &  & $2p$\,$^2P_{3/2}$ & 18.97 & 654 & $2.8\times10^{-1}$\\
    & & $^2P_{1/2}$ & 18.97 & 653 & $1.4\times10^{-1}$\\
    \ion{O}{VII} & $1s^2$\,$^1S_0$ & $1s2p$\,$^1P_1$ & 21.60 & 574 & $6.9\times10^{-1}$\\ \hline
    
    \ion{Fe}{XVIII} & $2s^22p^5$~$^2P_{3/2}$ & $2s^22p^4(^1D)\,3d$~$^2S_{1/2}$ &  14.26 & 870 & $2.4\times10^{-1}$\\
     && $3s$~$^2D_{5/2}$ & 15.62 & 794 & $6.0\times10^{-2}$ \\
     && $(^3P)\,3s$~$^2P_{1/2}$ & 15.63 & 793 & $2.6\times10^{-2}$ \\
     &  $^2P_{1/2}$ & $(^1D)\,3d$~$^2D_{3/2}$ & 14.35 & 864 & $9.3\times10^{-1}$ \\
     && $^2P_{3/2}$ & 14.42 & 860 & $2.0\times10^{-1}$ \\
     && $^2S_{1/2}$ &  14.47 & 857 & $8.5\times10^{-2}$\\
     && $(^1S)\,3s$~$^2S_{1/2}$ &  15.51 & 799 & $3.9\times10^{-2}$\\
     && $(^1D)\,^2D_{3/2}$ &  15.86 & 782 & $9.8\times10^{-2}$\\
     && $(^3P)\,^2P_{1/2}$ &  16.03 & 774 & $5.8\times10^{-2}$\\
    \hline
    \end{tabular}
    }
    \label{tab:RS}
    \tablefoot{
    \tablefoottext{\textdagger}{The parentheses specify the total spin and angular momentum of the four electrons in the $2p$ orbital.}
    }
\end{table*}

\subsubsection{Plane-parallel shock (\texttt{neij}) model} \label{subsubsec:neij}
Evidence of NEI has been reported in both the NPS and the south-eastern eRObub \citep[e.g.][]{Yamamoto22,Churazov26}. A particularly interesting and relevant model to replace the single temperature CIE model for the eRObub is a plane-parallel shock model, because it is parametrised by an ionisation parameter $u=n_et$, from which one can derive the gas density if one assumes a shock speed (hence a handle on $t$), or recover the shock speed if the gas density is known. In \texttt{SPEX}, a plane-parallel shock is one of the available modes in the \texttt{neij} model \citep{Kaastra93}. It is computed by summing the contributions of NEI plasmas resulting from a sudden temperature change from a logarithmic grid in $u$, which is expected to be more realistic in a propagating shock scenario, where gas layers immediately behind the shock would deviate more from CIE than those further behind. For the model parameters, we fixed the pre-shock temperature to the CGM temperature (fitted from the corresponding background region), and the post-shock temperature was left free and assumed to remain constant.

\begin{figure}[htbp]
    \centering
    \includegraphics[width=\linewidth]{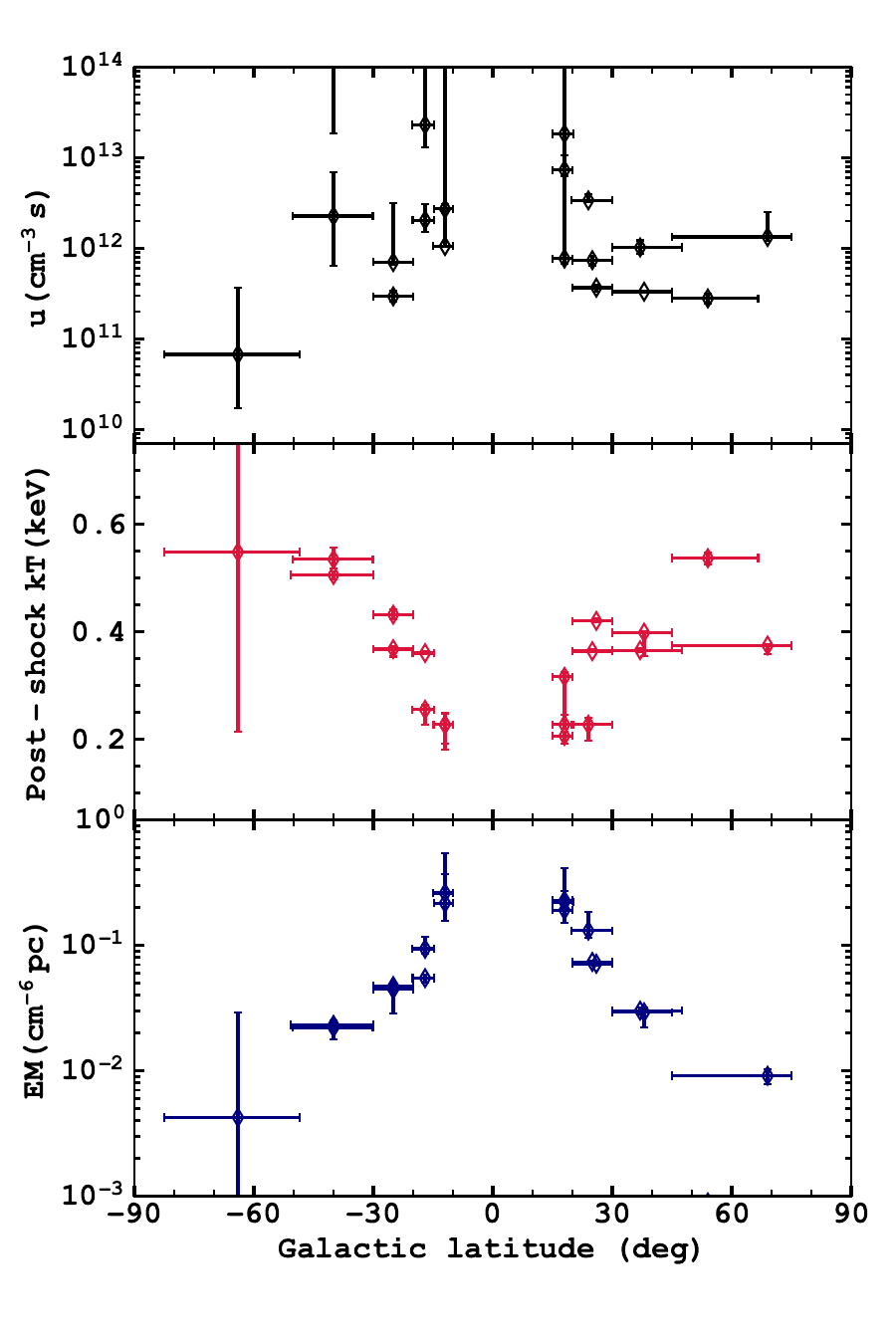}
    \caption{Latitudinal profiles of the ionisation parameter, post-shock temperature and EM with the \texttt{neij} model.}
    \label{fig:neij}
\end{figure}

Introducing the \texttt{neij} model undoubtedly improved the fits in all regions in the eRObub interior (statistical model comparison with our preferred 2T model (Sect.~\ref{subsubsec:2T}) in Table~\ref{tab:erobub_cont}; full parameter table in Table~\ref{tab:neij}). However, the fitted ionisation parameter shows substantial scatter, as shown in the top panel of Fig.~\ref{fig:neij}, complicating the interpretation of the \texttt{neij} model. At $u\gtrsim10^{12}\,{\rm cm^{-3}\,s}$, elements such as O, Ne and Fe dominating the spectrum are essentially at CIE \citep{Smith10}, which is the case for a large fraction of regions in the eRObub interior, especially at low latitudes. There is potential evidence of NEI in some regions with post-shock temperature reaching $\sim 0.6\,$keV, at high latitudes. It is worth mentioning that \citet{Churazov26} has recently reported that the south-eastern shell of the eRObub can be fitted reasonably well by a plane-parallel shock with $u=(2.8\pm0.2)\times10^{11}\,{\rm cm^{-3}\,s}$ and post-shock temperature of $0.64\pm0.02\,$keV. The shell on the south-western side is fainter than the SE, while the highest-latitude region (reg~17) shows a significant deviation from CIE; the constraint on the post-shock temperature in this region is weak. On the other hand, the other two regions at $b<-30\degr$ (reg~19 and 20) do not show clear evidence of NEI.

Upon closer inspection, despite performing better than the 1T model, it is clear that the \texttt{neij} model still cannot adequately explain the anomalous \ion{Fe}{XVII} ratio prevalent in the eRObub interior regions (Table~\ref{tab:residual_ratio} and one example spectrum shown in Fig.~\ref{fig:1T2T}). We show in the next Section that assuming the eRObub is dominated by gas of two characteristic temperatures at CIE can reproduce the data, especially the \ion{Fe}{XVII} ratio, much better than both the 1T and the \texttt{neij} model.

\begin{table*}[htbp]
    \caption{Data-to-model ratios of the 1T, \texttt{neij} and 2T models in the \ion{Fe}{XVII}\,15\,\AA~($E\in[0.77,0.87]$\,keV)and \ion{Fe}{XVII}\,17\,\AA~($E\in[0.70,0.77]\,$keV) bands, and the resulting deviation in the \ion{Fe}{XVII}\,15\,\AA/17\AA~line ratio. The regions shown are the eRObub interior regions. The 2T model reproduces the closest \ion{Fe}{XVII} 15\AA/17\AA\,ratio to the data among the three models.}
    \centering
    \renewcommand{\arraystretch}{1.2}
    {\small
    \begin{tabular}{c|ccc|ccc|ccc}
    \hline\hline

Region & \multicolumn{3}{c|}{1T} & \multicolumn{3}{c|}{\texttt{neij}} & \multicolumn{3}{c}{2T} \\
& Ratio 15\AA & Ratio 17\AA &  Offset (\%)& Ratio 15\AA  & Ratio 17\AA & Offset (\%) & Ratio 15\AA & Ratio 17\AA & Offset (\%) \\\hline
$1$ &$1.023$ &$0.982$ &$-4.2$ &$0.998$ &$1.004$ &$0.6$ &$0.998$ &$1.004$ &$0.6$ \\
$2$ &$1.027$ &$0.978$ &$-5.0$ &$1.013$ &$0.989$ &$-2.4$ &$0.998$ &$1.003$ &$0.4$ \\
$5$ &$1.037$ &$0.972$ &$-6.8$ &$1.024$ &$0.983$ &$-4.2$ &$1.007$ &$0.997$ &$-1.1$ \\
$6$ &$1.017$ &$0.982$ &$-3.5$ &$1.019$ &$0.982$ &$-3.8$ &$1.001$ &$0.999$ &$-0.3$ \\
$9$ &$1.040$ &$0.959$ &$-8.4$ &$1.024$ &$0.974$ &$-5.2$ &$1.003$ &$0.994$ &$-0.9$ \\
$10$ &$1.039$ &$0.966$ &$-7.6$ &$1.028$ &$0.976$ &$-5.3$ &$1.007$ &$0.991$ &$-1.6$ \\
$11$ &$1.025$ &$0.977$ &$-4.9$ &$1.028$ &$0.975$ &$-5.5$ &$1.016$ &$0.985$ &$-3.1$ \\
$13$ &$1.024$ &$0.946$ &$-8.2$ &$1.052$ &$0.955$ &$-10.2$ &$1.012$ &$0.986$ &$-2.7$ \\
$14$ &$1.003$ &$0.957$ &$-4.8$ &$1.022$ &$0.978$ &$-4.5$ &$1.011$ &$0.989$ &$-2.2$ \\
$15$ &$1.000$ &$0.959$ &$-4.2$ &$1.021$ &$0.977$ &$-4.5$ &$1.012$ &$0.986$ &$-2.7$ \\
$17$ &$1.014$ &$0.993$ &$-2.1$ &$1.004$ &$0.996$ &$-0.8$ &$0.999$ &$1.005$ &$0.6$ \\
$19$ &$1.026$ &$0.983$ &$-4.3$ &$1.003$ &$1.002$ &$-0.1$ &$1.003$ &$1.003$ &$-0.0$ \\
$20$ &$1.025$ &$0.983$ &$-4.3$ &$0.993$ &$1.010$ &$1.7$ &$1.003$ &$1.002$ &$-0.0$ \\
$23$ &$1.028$ &$0.970$ &$-6.0$ &$1.019$ &$0.978$ &$-4.2$ &$1.004$ &$0.991$ &$-1.3$ \\
$24$ &$1.032$ &$0.976$ &$-5.8$ &$1.014$ &$0.990$ &$-2.4$ &$1.000$ &$1.003$ &$0.3$ \\
$27$ &$1.036$ &$0.976$ &$-6.2$ &$1.021$ &$0.989$ &$-3.2$ &$1.008$ &$0.999$ &$-1.0$ \\
$30$ &$0.989$ &$0.933$ &$-6.0$ &$1.034$ &$0.966$ &$-7.0$ &$1.009$ &$0.989$ &$-2.0$ \\
$31$ &$0.975$ &$0.931$ &$-4.7$ &$1.038$ &$0.962$ &$-7.8$ &$1.013$ &$0.984$ &$-3.0$ \\
$33$ &$1.024$ &$0.980$ &$-4.5$ &$1.027$ &$0.978$ &$-5.0$ &$1.001$ &$1.000$ &$-0.1$ \\
\hline
    \end{tabular}
    }
    \label{tab:residual_ratio}
\end{table*}

\subsubsection{Two-temperature CIE (2T) model} \label{subsubsec:2T}
Invoking an additional CIE component with freely varying temperature and EM generally improved the fits within the eRObub. The second CIE component has abundance and absorption, which are always linked with the first to limit the number of free parameters. We listed the best-fit parameters with the two-temperature (2T) model as well as the statistical improvement of the 2T model in terms of the F-test in Table~\ref{tab:erobub_params}. Figure~\ref{fig:1T2T} compares 1T and 2T fits to the eRObub spectrum in reg~27. The latter perform significantly better in reproducing the \ion{Fe}{XVII}~15\,\AA/17\,\AA~ratio as well as a better fit to the \ion{Ne}{IX,X} lines. The improvement brought by the extra temperature component is significant ($>99\%$ confidence level) across all eRObub interior regions.

\begin{figure}[htbp]
    \centering
    \includegraphics[width=0.45\textwidth]{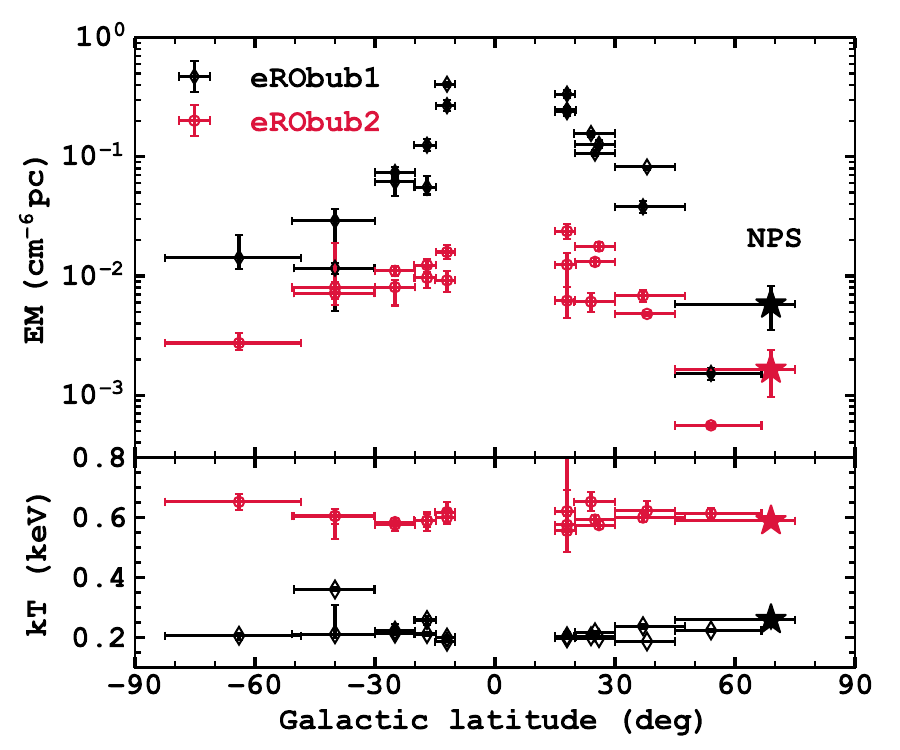}
    \caption{Latitudinal profiles of $kT_{\rm eRObub}$ and ${\rm EM_{eRObub}}$ within the \mbox{eRObub} using the large extraction regions. The cool shell and background regions are not shown in the figure.}
    \label{fig:kT_EM}
\end{figure}

A general trend emerged from this exercise: both temperature components have consistent temperatures throughout the eRObub, as shown in Fig.~\ref{fig:kT_EM}. No significant temperature gradient is observed after discarding regions in the southern cool shell because of the inconclusive detection of eRObub. The low (eRObub1) and high (eRObub2) temperature components have median temperatures of $kT_{\rm low}=0.21^{+0.03}_{-0.01}\,$keV and $kT_{\rm high}=0.60\pm0.02\,$keV (uncertainties reflecting the 16$^{\rm th}$ and 84$^{\rm th}$ percentiles in the population, not accounting for the statistical uncertainties of each fit), respectively. 
The low-temperature component dominates the flux, with EM $\sim5^{+6}_{-2}$ times larger than that of the hot component on average. The EM of both temperature components increases towards the Galactic plane, with the low-temperature component increasing slightly more sharply than the high-temperature component. The EM modulates approximately one order of magnitude from high to low latitudes.

\begin{figure}[htbp]
    \centering
    \includegraphics[width=0.45\textwidth]{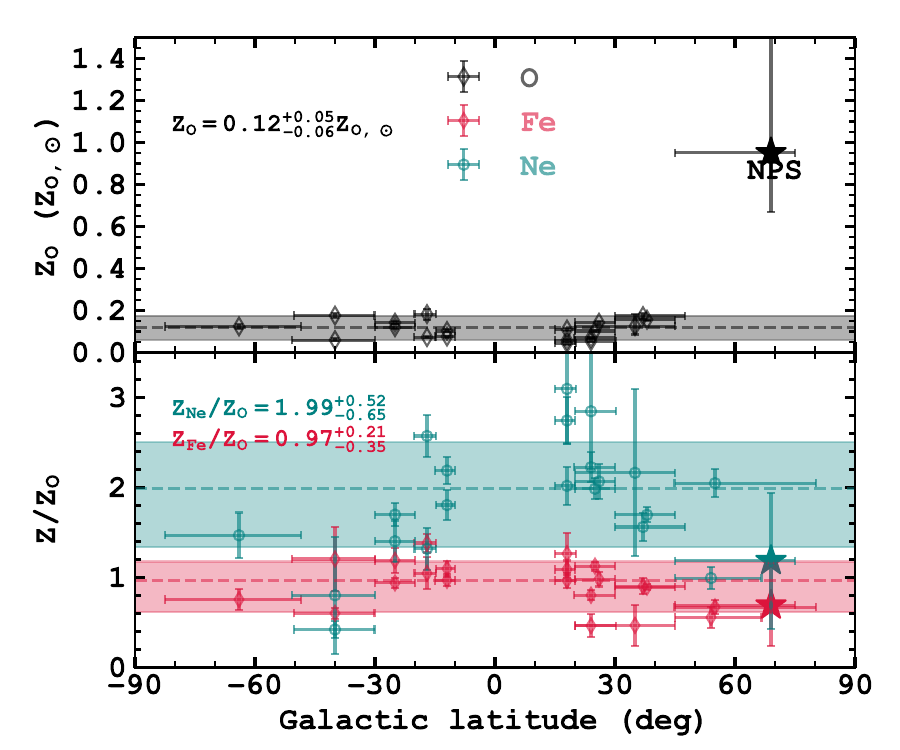}
    \caption{Latitudinal profile of O, Fe and Ne abundances. Ne and Fe abundances are shown with respect to O in the lower panel. The unit of the $y$-axis in the lower panel means the ratio $Z_{x}/Z_{\rm O}$ to the ratio in our assumed abundance table from \citet{Lodders2003} (proto-Solar; which is $\approx0.15$ in number fraction), but not the absolute number fraction of Fe or Ne to O. We removed reg~1 in the top panel if the absolute abundance could not be constrained.}
    \label{fig:abund}
\end{figure}

The data prefer subsolar abundance, with O abundance (linked with all other elements except those left free in the fits; Table~\ref{tab:erobub_params})
generally lying within 0.05--0.2 solar. As shown by Fig.~\ref{fig:abund}, the O abundance does not exhibit a clear trend with latitude. Fe abundance is marginally consistent with the O abundance. There are some indications of supersolar abundance for Ne, especially near the Galactic plane.
It is important to note that uncertainties arising from the choice of abundance tables can differ by a few to 20\% in the Ne/O ratio \citep[e.g.][]{anders89,Asplund09,Lodders09}. Therefore, we think there is still inadequate evidence to suggest that the eRObub has a non-solar Ne/O or Fe/O ratio. Even if the Ne/O overabundance is indeed physical, this appears not to be specific to the eRObub, as the Ne/O ratio in the CGM of the background region is similarly high, with the median about $1.5^{+0.5}_{-0.1}$.

The NPS (reg~2) is a special case in terms of abundance, where O, Fe and Ne abundances are enhanced compared to other eRObub regions (Fig.~\ref{fig:NPS}). A higher abundance is consistent with the idea from \citet{Zhang24} and \citet{Churazov24}, who pointed out that the NPS could originate in active star-forming regions, which naturally produce metal-enriched hot gas that rises to the halo due to buoyancy. To test whether the data indeed prefer a high abundance, we refitted the NPS region, this time with O and all other metals fixed at 0.2$\,Z_{\odot}$\footnote{Trials with setting the abundance even lower resulted in even worse fits, especially below $0.5$\,keV.}, except for Ne and Fe to accommodate for the anomalous \ion{Fe}{XVII} ratio. The new $\chi^2$/dof is 2.54 (175 dof), which is worse than the $\chi^2$/dof of the model with O abundance free (1.63; 174 dof). Statistically, the free-O abundance model is strongly preferred as shown by the F-statistic of 99 and the $p$-value in the order of $10^{-19}$. We show the model comparison in Appendix~\ref{App:NPS}. The comparison reveals degradation in the fit quality around 0.35\,keV and between 1.0--1.4\,keV; these could be remedied by individually enhancing the C, N and Mg abundances, indeed corroborating an enhanced elemental abundance.
It is worth noting that an abundance of $\sim0.5$--$1$ solar, dependent on the element, is not dissimilar to previous observations and analyses of the east part of the NPS \citep{Willingale03a,Miller08,Gu16}.

\subsubsection{Plasma codes} \label{subsec:plasma_code}
\begin{figure}[htbp]
    \centering
    \includegraphics[width=0.49\textwidth]{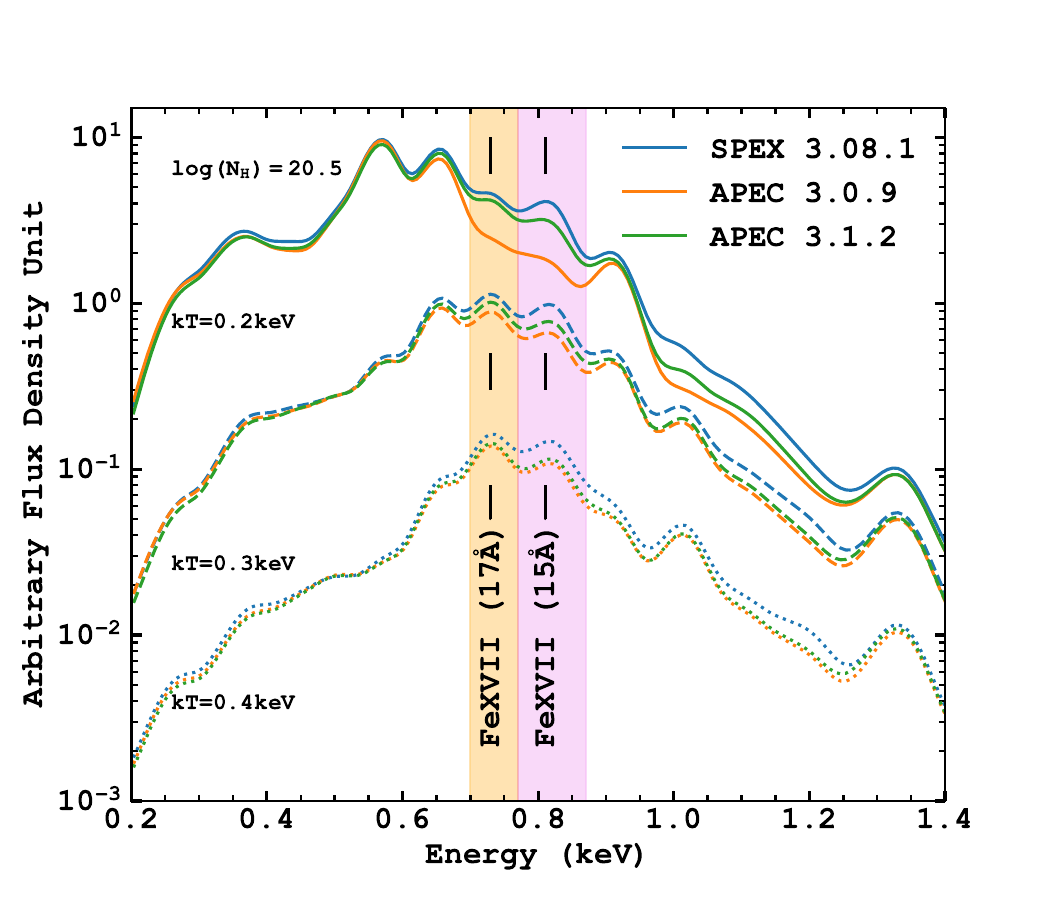}
    \caption{Comparison of the spectral shape of CIE models created by different plasma codes, subjected to a typical Galactic absorption column density of $\log{(N_{\rm H}/{\rm cm^{-2}})}=20.5$, and folded with the eROSITA response and effective area. The two main \ion{Fe}{XVII} L shell peaks are labelled. The prediction of their line ratio differs significantly. \texttt{SPEX} produces the closest line ratio shown by the data. The definitions of the \ion{Fe}{XVII}\,15\,\AA~(violet) and 17\AA~(orange) narrow bands are also shown.}
    \label{fig:CIE_compare}
\end{figure}

While we presented evidence for two-temperature plasma in the interior of eRObub, it is instructive to compare the uncertainties in the model spectra predicted by plasma codes with the level of our residuals.
We tested the two most commonly used plasma codes, \texttt{AtomDB/apec} \citep{apec,AtomDB} and \texttt{SPEX} \citep{SPEX, SPEX3.08}, in their latest releases v3.1.2 and v3.08.1, respectively. In addition, we also included an older \texttt{AtomDB/apec} release, v3.0.9, which ships with the commonly used \texttt{Xspec} release 12.14. The comparison is shown in Fig.~\ref{fig:CIE_compare}. Abundances were set to solar using the  \citet{Lodders09} values, which are common in both codes. The comparison was done using the spectral fitting programme \texttt{SPEX}, and the \texttt{AtomDB/apec} models were imported as user-defined models computed using \texttt{PyAtomDB} \citep{Pyatomdb}. We folded the model spectra with the eROSITA response and effective area and introduced a typical Galactic absorption of $\log{N_{\rm H}}=20.5$. Three discrete temperatures (0.2, 0.3 and 0.4\,keV) around the expected eRObub temperature were tested, showing noticeable differences between the three implementations within $0.7$--$0.9$\,keV.

\begin{table}[htbp]
    \caption{Narrowband ratio $E_2/E_1$ taken from Fig.~\ref{fig:CIE_compare}, where $E_1\in[0.70, 0.77]$\,keV (\ion{Fe}{XVII}\,17\,\AA~peak) and $E_2\in[0.77, 0.87]$\,keV (\ion{Fe}{XVII}\,15\,\AA~peak). The ratio was computed after dividing the flux by the width of the narrowband.}
    \centering
    \renewcommand{\arraystretch}{1.2}
    \begin{tabular}{cccc}
    \hline\hline
      $kT$ (keV)   &  \texttt{SPEX 3.08.1} & \texttt{apec 3.1.2} & \texttt{apec 3.0.9}\\\hline
       0.2  & 0.784 & 0.694 & 0.686 \\
       0.3 & 0.804 & 0.732 & 0.714 \\
       0.4 & 0.859 & 0.777 & 0.757 \\\hline
    \end{tabular}
    \label{tab:line_ratio}
\end{table}

The \texttt{SPEX} model predicts higher emissions than the other two \texttt{apec} versions, most notably between 0.7--1.3\,keV. The difference is the most pronounced at the \ion{Fe}{XVII}\,15\AA~peak, where \texttt{SPEX} consistently predicts a higher peak regardless of temperature. It is worth noting that \texttt{SPEX} produces a \ion{Fe}{XVII} 15\,\AA/17\,\AA~ratio that is closer to the data. This is the reason why we chose to use the \texttt{SPEX cie} model in our spectral analysis. The residual at these energies would be worse if \texttt{apec} is chosen. The line ratios of the two Fe peaks predicted by the three models are summarised in Table~\ref{tab:line_ratio}. It shows that about a $\sim$$10\%$ difference in the \ion{Fe}{XVII} 15\,\AA/17\,\AA~ratio can be expected from using different plasma codes, regardless of the temperature of the plasma within the relevant range.

It is instructive to compare the variation in \ion{Fe}{XVII} $15\,\text{\AA}$/$17\,\text{\AA}$ ratio introduced by the choice of plasma codes with the amplitudes of residuals we observe from fitting with the 1T, \texttt{neij} and 2T models in the eRObub interior regions (excluding the cool shell regions, which were well fitted by the 1T model). The data-to-model offset was computed by first taking the data-to-model ratios within the two bands (15\,\AA~band: $E\in [0.77, 0.87]\,$keV, 17\,\AA~band: $E\in[0.70, 0.77]\,$keV, see Fig.~\ref{fig:CIE_compare}), then dividing the data-to-model ratios of the two bands to quantify how much the data line ratio differ from the model line ratio. A positive line ratio deviation means that the data \ion{Fe}{XVII}\,15\,\AA/17\,\AA~ratio is higher than the model prediction, which is indeed the case for all in the 1T model as shown in Table~\ref{tab:residual_ratio}. However, quantitatively, the mostly positive deviation is in all cases less than $\sim$$10\%$, which is less than the variations introduced by the choice of plasma codes. Of course, this does not change the fact that we were already using the \texttt{SPEX} model, which produces the highest \ion{Fe}{XVII}\,15\,\AA/17\,\AA~ratio in the three plasma codes tested, and still failed to explain the enhanced \ion{Fe}{XVII}\,15\,\AA/17\,\AA~ratio in the data. The discrepancy is similar for the \texttt{neij} model. The 2T model could reproduce the data \ion{Fe}{XVII} ratio to a $\lesssim3\%$ level.
While it is possible that the 2T model or the alternative explanations explored in Appendices~\ref{app:RS} and \ref{app:CX} for the enhanced \ion{Fe}{XVII}\,15\,\AA/17\,\AA~ratio could become irrelevant with updated plasma codes with a raised Fe line ratio, recent laboratory measurements in fact suggest a lower $3d$/$3s$ intensity ratio (see transitions in Table~\ref{tab:RS}), indicating an even lower \ion{Fe}{XVII}\,15\,\AA/17\,\AA~compared to theoretical predictions \citep{Shah19}.

\subsection{Global fits to constant S/N regions} \label{subsec:global_fit}
Following the exploration of high-S/N spectra, we aim to produce smoother temperature and emission measure maps of the eRObub by defining smaller and spatially continuous regions, which might reveal structures that could otherwise be overlooked in the large regions.

\begin{figure}[htbp]
    \centering
    \includegraphics[width=0.49\textwidth]{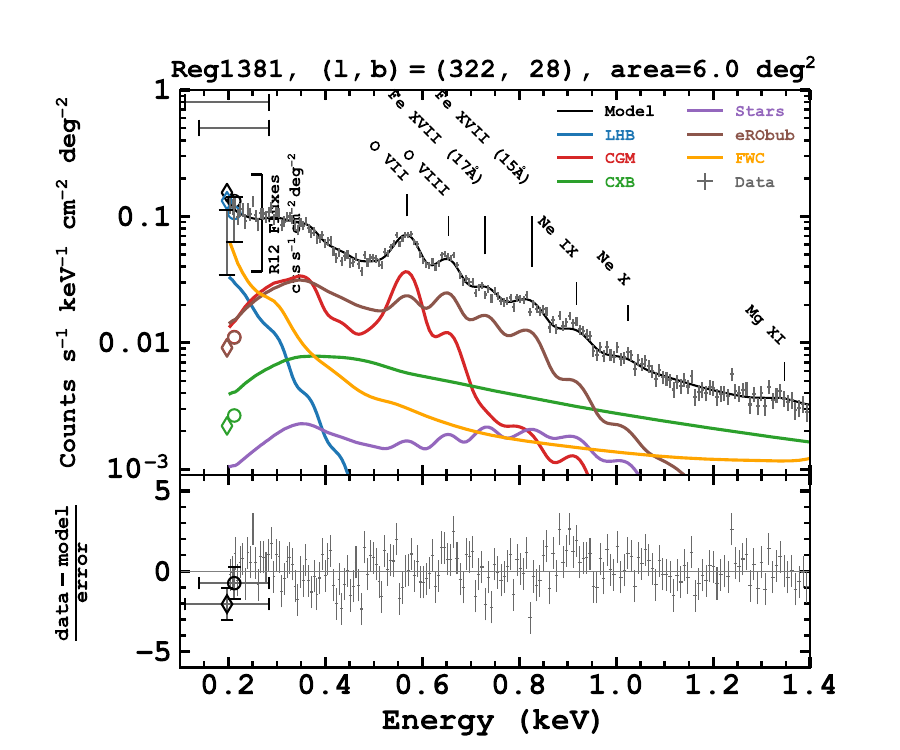}
    \caption{Example spectrum of one constant S/N region. The most prominent emission lines and contributing components of the spectrum are labelled. The two horizontal error bars at the top left corner indicate the bandwidths of the ROSAT R1 and R2 bands. There is a small hint of enhanced Ne abundance here, which is much clearer in the spectra of the large regions.}
    \label{fig:erobub_spec}
\end{figure}

We began with the same regions that were defined by \citet{Yeung2024} in the western Galactic hemisphere in eRASS1, based on a constant S/N criterion (S/N$\sim80$ in the 0.2--0.6\,keV band; each region subtends $\sim5$--$10$\,deg$^2$ in sky area). From these regions, we identified those that overlap with the large regions in the northern cool shell and the eRObub as defined in Fig.~\ref{fig:reg_num} (Sect.~\ref{subsubsec:reg}) and fitted them with the 1T model. Figure~\ref{fig:erobub_spec} shows one example of these regions with the best-fit model and prominent emission lines labelled.

Some simplifications were necessary to avoid model degeneracies in fitting the smaller regions due to lower S/N: 
\begin{itemize}
    \item The LHB and CGM temperatures and abundances were fixed to the best-fits in the large region (defined in Fig.~\ref{fig:reg_num}) that enclosed them (hence from Table~\ref{tab:bkg_param}).
    \item The eRObub component was modelled by a single CIE model, as we found the S/N in these spectra was insufficient to either identify NEI signature or constrain two-temperature plasmas.
    \item The eRObub abundance is fixed to $Z=0.1\,Z_\odot$ for the same reason. This is motivated by the low abundances obtained from the large regions (Table~\ref{tab:erobub_params}).
\end{itemize}

\begin{figure*}
    \centering
    \includegraphics[width=0.49\textwidth]{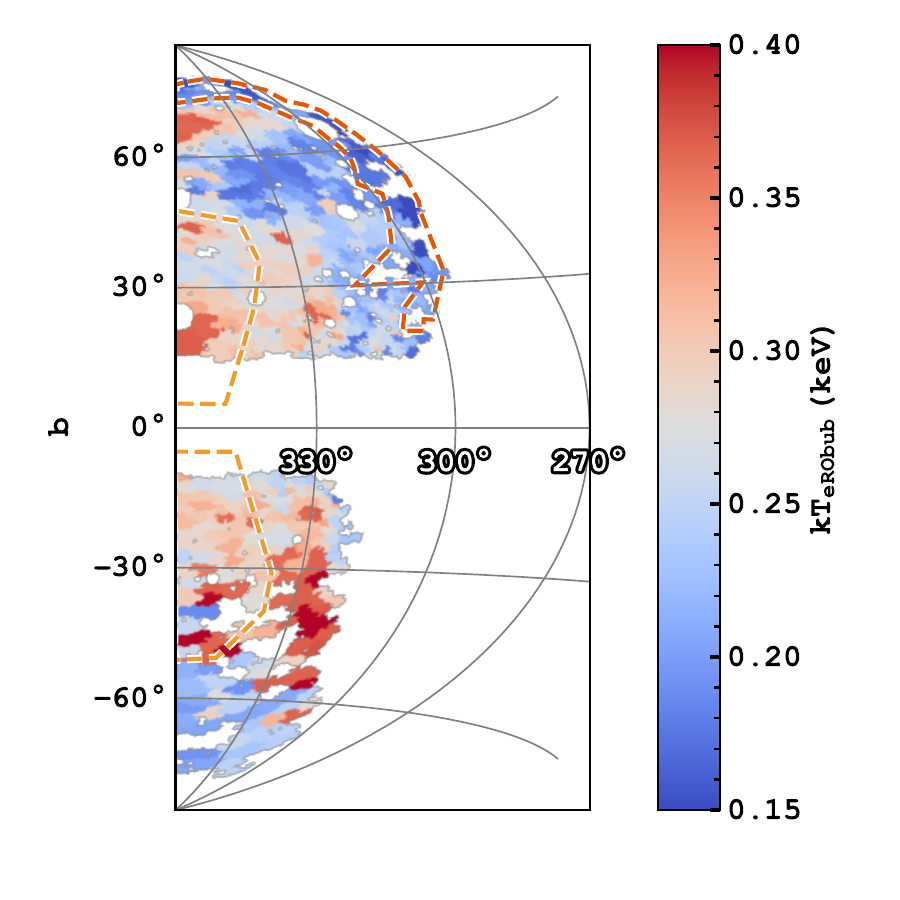}
    \includegraphics[width=0.49\textwidth]{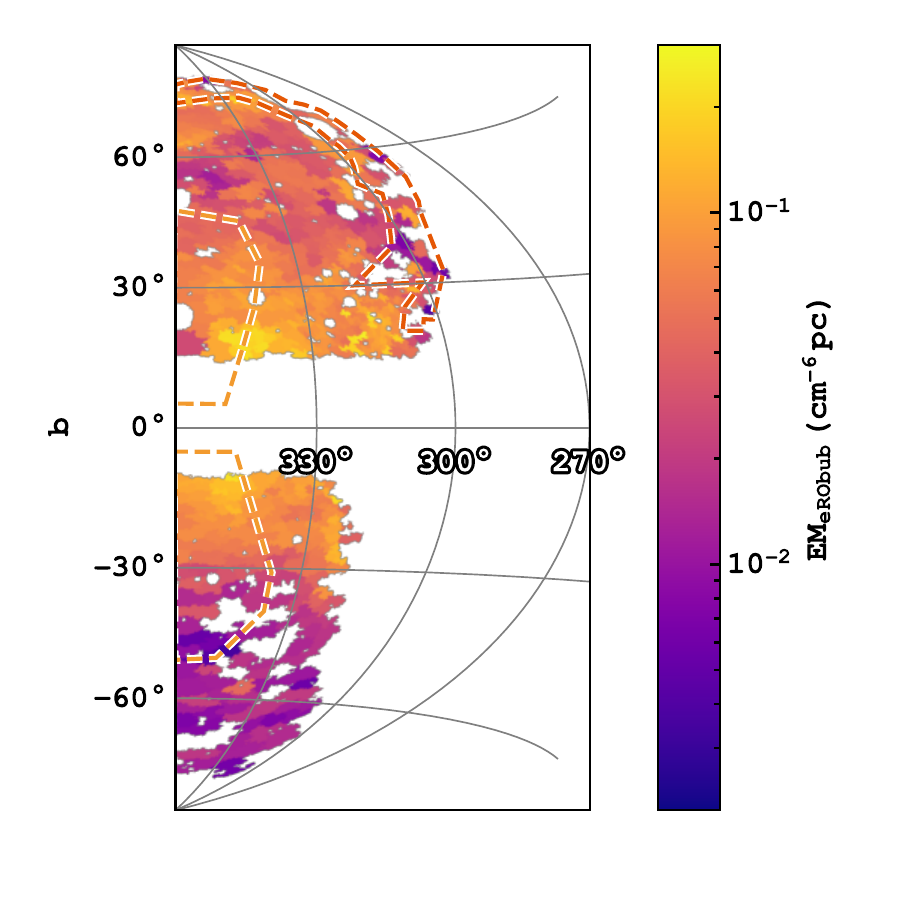}
    \caption{Temperature ({\it left}) and emission measure profile ({\it right}) of the eROSITA bubbles fitted with the 1T model at fixed $Z=0.1\,Z_\odot$. Crimson and orange dashed lines indicate the region of the cool shell and the FB as defined in \citet{Ackermann14}, respectively. The uncertainty maps of the two parameters and the $\chi^2/{\rm dof}$ map are shown in Fig.~\ref{fig:contbin_err}.}
    \label{fig:erobub_kT_EM}
\end{figure*}

\begin{figure*}[htbp]
    \centering
    \includegraphics[width=0.49\textwidth]{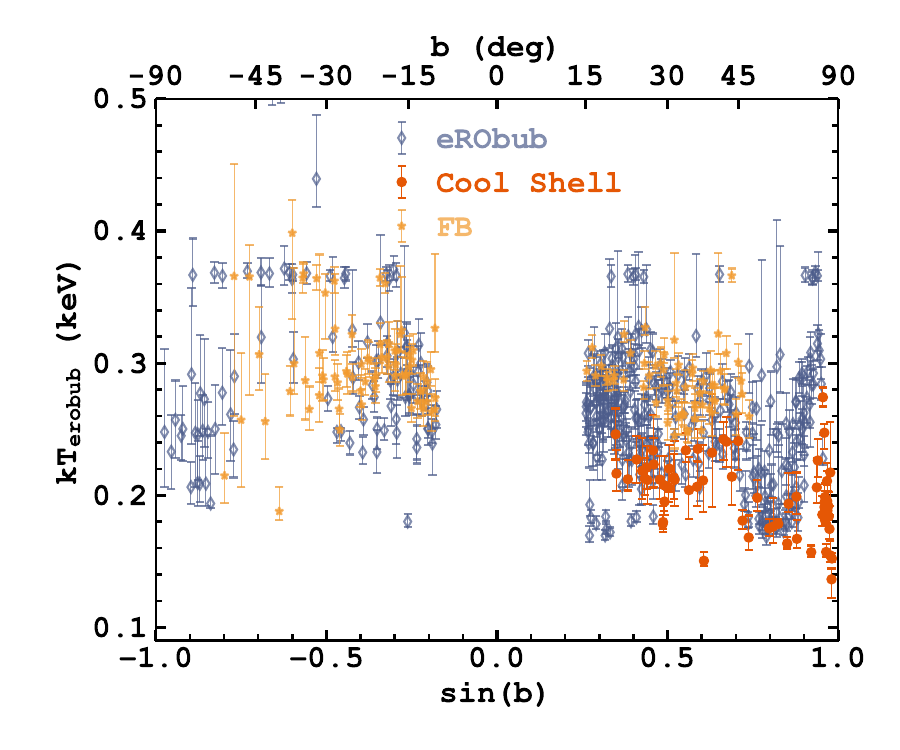}
    \includegraphics[width=0.49\textwidth]{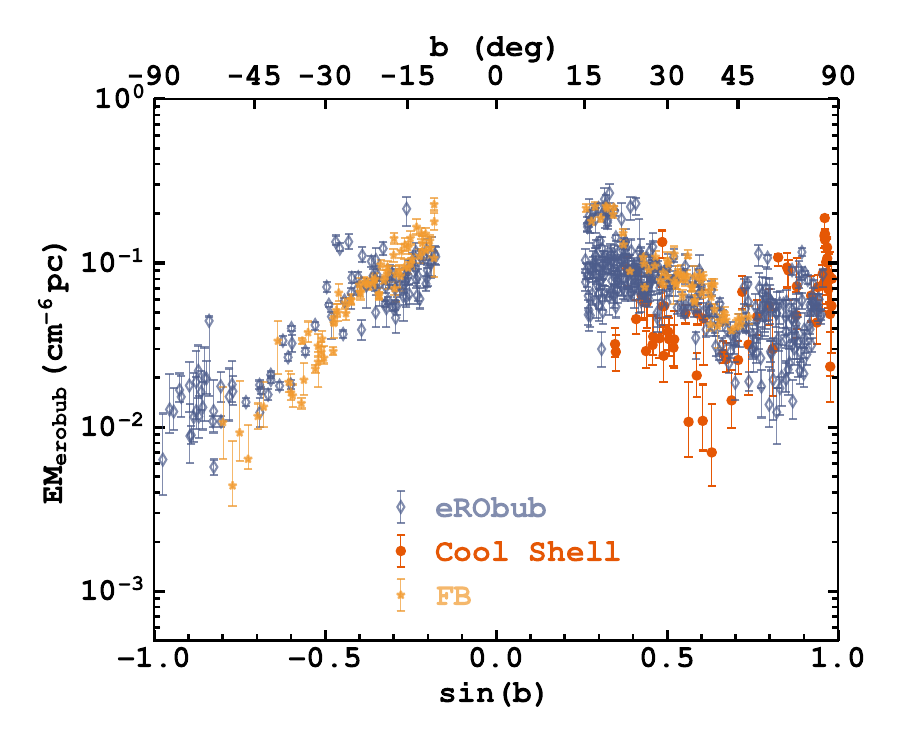}
    \caption{Latitudinal profiles of $kT_{\rm eRObub}$ ({\it left}) and ${\rm EM_{eRObub}}$ ({\it right}) within the eROSITA bubbles (eRObub) using the constant S/N regions. The spectral bins are divided into three groups: eRObub spectra outside (blue) and inside (yellow) of the Fermi bubbles, and within the cool shell (orange). The boundaries of the Fermi bubbles and the cool shell are shown respectively by the dashed polygons in their respective colour in Fig.~\ref{fig:erobub_kT_EM}. Spectral bins that have large uncertainties ($\sigma_{kT}>0.1\,$keV or $\sigma_{\rm EM}/{\rm EM} > 0.5$) or poor $\chi^2/{\rm dof}$ ($>1.5$) are not shown.
    }
    \label{fig:lat_profile}
\end{figure*}

We present the temperature and EM maps obtained from this exercise in Fig.~\ref{fig:erobub_kT_EM}. Figure~\ref{fig:contbin_err} provides the corresponding uncertainty maps and the $\chi^2/{\rm dof}$ map to help assess the significance of the temperature and EM maps. The majority of regions have temperatures between $0.2$--$0.3$\,keV, with only a weak positive gradient towards the Galactic plane. This is shown more clearly in the left panel of Fig.~\ref{fig:lat_profile}.
The EM profile shows an order-of-magnitude increase from high to low latitudes, with the NPS producing a prominent enhancement at $60\degr\lesssim b \lesssim 80\degr$. As shown also in the right panel of Fig.~\ref{fig:lat_profile}, the northern FB regions generally exhibit a higher EM than those outside, but their decay slope with latitudes is entirely consistent with each other.
The slight difference in EM is not seen in the southern bubble. The cool shell is cooler by comparison but has about half the EM at low latitudes as those in the interior, and progresses to similar or even higher levels at higher latitudes.
These findings generally echo the results in Sects.~\ref{kT_jump} and \ref{subsubsec:2T}.

One interesting feature not highlighted in previous Sections is a shell-like structure in the temperature map in the SW part of the southern bubble, extending from $b\sim-30\degr$ to $-45\degr$. This shell-like structure is not as obvious in terms of EM, but is more visible in the RGB map in Fig.~\ref{fig:bubble_rgb}. The enhanced temperature ($\approx 0.36$\,keV) is also reflected in the large region reg~20, which is an anomaly in the 2T model. It is very likely that this is the western counterpart of the shock front identified in the east by \citet{Churazov26}. 

\section{Understanding the bubble morphology} \label{Morph}
In this section, we describe our efforts to reproduce the X-ray shape and brightness profile of the eRObub to infer their possible three-dimensional geometry. To achieve this, we construct a physically motivated but analytically describable geometrical model of the emission. Similar work was recently performed by \citet{Liu24}, with the main difference that their work focused exclusively on the bubble borders derived from radio and X-ray data, but not on the emission filling the bubbles. 

\subsection{Morphological model}
\subsubsection{Geometry}

\begin{figure}[htbp]
    \centering
    \includegraphics[width=0.8\linewidth]{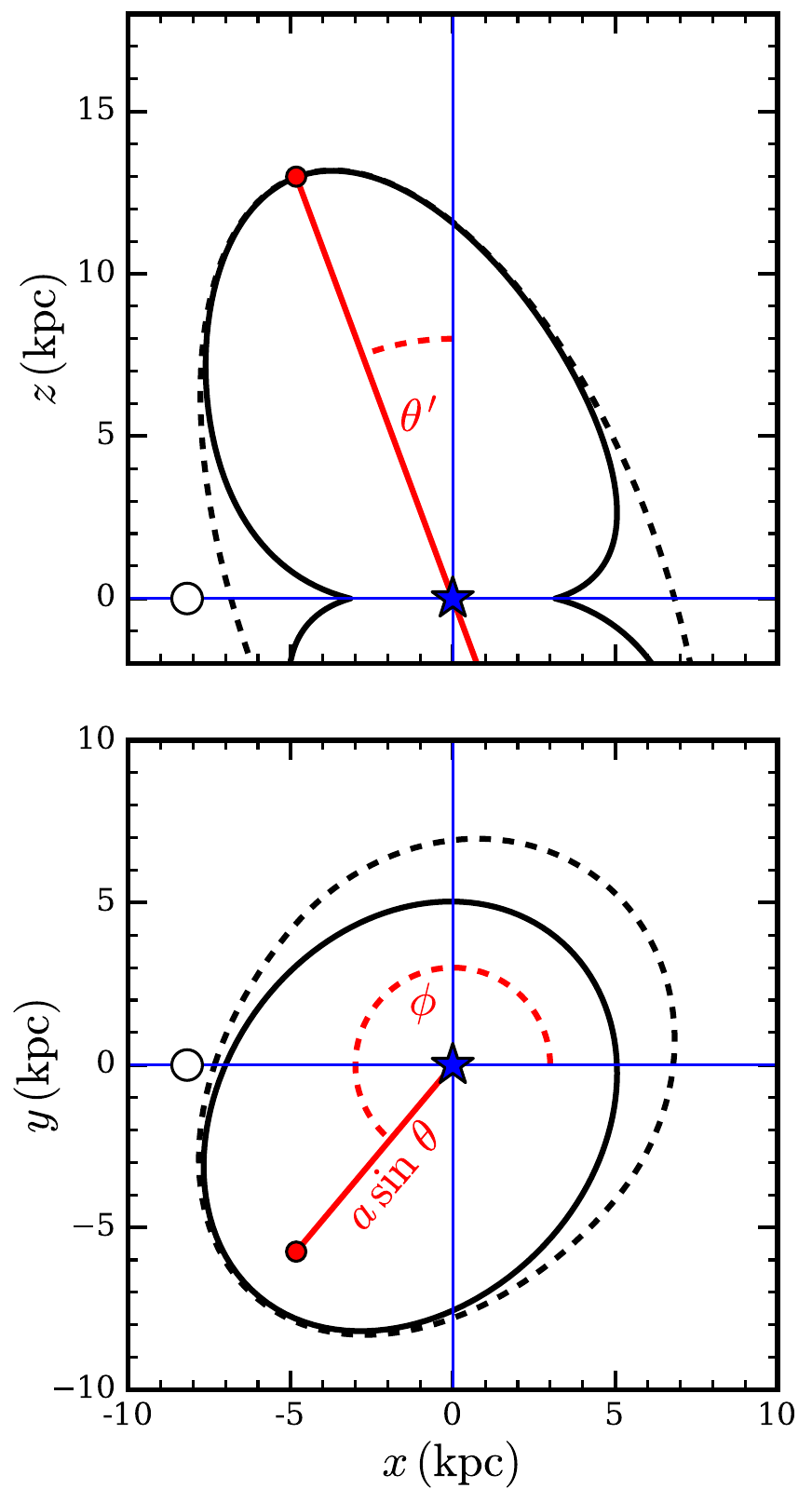}
    \caption{Sketch of the geometry of our empirical model of the eROSITA bubbles, with the parameters given in Table \ref{tab:bubble_params}. The top panel shows the side view (from the direction of the positive $y$-axis) of the northern bubble, with the ellipsoidal shell indicated with the black dashed line, and the constricted ellipsoid with the solid line. The Sun and the GC positions are indicated with a white circle and a blue star, respectively. The red line shows the projection of the ellipsoid major axis, and $\theta^{\prime}$ is the projected tilt angle with \mbox{$\tan \theta^{\prime} = \tan \theta \, \cos \phi$.} The lower panel shows the corresponding view of the northern bubble silhouette (in the range $z>0$) from the top, indicating the role of the polar and azimuthal tilt angles $\theta$ and $\phi$.}
    \label{BubbleGeometry}
\end{figure}

While the general shape of the bubbles can be approximately reproduced with two thick spherical shells symmetric about the Galactic plane \citep{Predehl20}, such models have shortcomings from a physical point of view: first, using two spherical shells implies two separate centres for the two bubbles, above and below the Galactic plane, respectively. This configuration is likely unrealistic, given that the origin of both bubbles is most likely related and located in the Galactic plane. Second, instead of a constant-density shell with finite thickness, one likely expects a continuous density increase up to the outer radius of the bubbles, if they are truly caused by a shock expanding into the CGM. The situation is similar to what is observed for X-ray bright shells of supernova remnants (SNRs) \citep[e.g.,][]{Heiles64,Cox82}. 

The general idea of our model is that of a shell caused by a bipolar point explosion in the GC, which takes the shape of a prolate ellipsoid with semi-major and semi-minor axes $a$ and $b$, where the larger extent is approximately perpendicular to the Galactic plane.
As apparent in the X-ray image (see Fig.~\ref{fig:masking_allsky} in Appendix.~\ref{sec:ero_recons}), the bubbles appear to be slightly tilted with respect to the normal on the Galactic plane, hence we allow for a rotation of the ellipsoid, meaning its semi-major axis direction is given by a polar and an azimuthal angle ($\theta$ and $\phi$), viewed from the GC. This yields the following expression for the coordinates of an ellipsoidal shell:
\begin{eqnarray}
    1&=& \left( \frac{x^{\prime}}{b} \right)^2 + \left(\frac{y^{\prime}}{b}\right)^2 + \left(\frac{z^{\prime}}{a}\right)^2 \,\text{, where} \label{eq1}\\
    \begin{pmatrix}
        x^{\prime} \\ y^{\prime} \\ z^{\prime} \label{eq2}
    \end{pmatrix}
    &=&
    \begin{pmatrix}
       \cos\theta\cos\phi  & \cos\theta\sin\phi  & -\sin\theta \\ 
       -\sin\phi & \cos\phi & 0 \\ 
       \sin\theta \cos\phi & \sin\theta \sin\phi& \cos\theta
    \end{pmatrix}
    \begin{pmatrix}
        x \\ y \\ z
    \end{pmatrix}.
    \label{eq:tilt}
\end{eqnarray}
Here we have used a right-handed Galactocentric reference system, with the origin at the GC and the Sun located at $(x,y,z)_{\odot} = (-8.18, 0, 0)\,\mathrm{kpc}$ \citep[assuming the Sun lies on the Galactic disk and the geometric distance to Sgr~A$^*$ taken from][]{galcen_d}, and the $y$-axis increasing toward the east at the GC. 
The azimuthal angle $\phi$ of the ellipsoid is defined such that $\phi = 0\degr$ ($\theta>0$) corresponds to the northern (southern) bubble tilted away from (toward) the Sun, and $\phi=90^{\circ}$ implies a tilt toward the east (west).

Realistically, the assumed shock wave forming the bubble could likely not expand freely, but its expansion was impeded by the surrounding medium, most significantly in the dense Galactic plane. Therefore, we imposed an equatorial narrowing of the bubble, dependent on the height above the Galactic plane. This `constricted bubble' was parametrized as a $z$-dependent reduction of the bubble radius $R^{\prime}$, with respect to the radius $R = \sqrt{x^2 + y^2 + z^2}$
of the unmodified ellipsoid, with a scale height $z_{n}$ and a `waist radius' in the Galactic plane $c_{n}$:
\begin{equation}
    R^\prime\left( z \right) = \left[1 - \frac{b-c_{n}}{b}\,\exp \left(-\frac{\vert z\vert}{z_{n}}\right)\right] R(z).
\end{equation}
A morphology of two shells centred on the GC naturally emerges, as is visualised in Fig.~\ref{BubbleGeometry}. 
As seen in the eRASS1 all-sky image (Fig.~\ref{fig:bubble_rgb}), the northern eROSITA bubble is tilted toward the west. It also appears to show an increased extent in that direction, as is traced in particular by radio emission \citep{Liu24}. In contrast, the southern bubble shows little tilt and a smaller horizontal extent. 
Therefore, separate geometries (i.e., size and orientation) likely have to be used for the northern and southern hemispheres to reproduce the observed bubble shape satisfactorily. 

The second ingredient for our model is the assumption of a realistic density profile of the X-ray-bright material within the bubble. Here, we can borrow from simple analytical solutions developed for modelling point explosions (such as SNRs) in the Sedov-Taylor phase \citep{Sedov, Taylor}. This assumption appears sensible as no strong optical line emission, indicative of radiative cooling, has been observed from the bubbles, thus far. However, a key difference is that for SNRs, an infinitely strong shock can usually be assumed (when ${\mathcal M\gtrsim5}$), as the shock velocity strongly exceeds the ambient sound speed ($\approx15\sqrt{{T/10^4\,{\rm K}}}\,{\rm km\,s^{-1}}$), which is far from the truth for the eRObub. Instead, the shock is likely only weakly supersonic \citep[e.g., $\mathcal{M} \approx 1.5$;][]{Predehl20}, due to the large sound speed in the hot Galactic halo. 
The Rankine-Hugoniot jump conditions yield the following density increase $\chi_s$ behind a nonrelativistic shock:
\begin{equation} \label{eq:compression}
    \chi_{s} \coloneqq \frac{n_2}{n_1} = \frac{4\mathcal{M}^2}{\,\mathcal{M}^2+3},
\end{equation} 
which is $\chi_{s} \sim 1.7$ if $\mathcal{M}=1.5$, meaning the gas is compressed much less than in an SNR where $\chi_{s} = 4$. 
The analytic solutions by \citet{Cox82} for the density profile of a Sedov-Taylor blast wave allow for the inclusion of $\chi_{s}$ (or equivalently $\mathcal{M}$) as a free parameter, and show that lower Mach number shocks exhibit a much shallower density profile. Ideally, this implies that the apparent thickness of the X-ray bright shell can be used as a tracer of the shock Mach number, motivating our usage of their solutions to model the location-dependent gas density inside the bubbles. For its evaluations, the radius of the narrowed ellipsoidal shell $R^{\prime}$ measured from the GC was used as the forward shock radius to evaluate the radial density profile $\chi(r/R^\prime)$ in each direction \citep[Eq. 16 in][]{Cox82}. 
We note that, even if the scenario of a point explosion at the GC does not correspond to the true origin of the bubbles, the main point of our assumption is to `populate' the geometric shape of the bubbles with a plausible density profile, whose exact shape (i.e. shell thickness) can be regulated by a free parameter, the Mach number.   

A further ingredient for our model is that the density of the CGM, with which the shock wave interacts, is not uniform but likely strongly decreasing with Galactocentric radius, meaning that emission at low Galactic latitudes is expected to be strongly brightened. To include this effect, we used the density model combining a spherical halo and a disk component derived by \citet{Locatelli24}, to estimate the `ambient' density of the CGM: 
\begin{eqnarray}
\label{eq:CGM}
    n_1(x,y,z) &=& C \, \left(x^2+y^2+z^2+r_0^2 \right)^{-3/4} \\   \nonumber
    &+& n_0\,\exp\left( -\frac{\sqrt{x^2+y^2}}{\rho_{h}}-\frac{\vert z\vert}{z_{h}} \right)   
\end{eqnarray}
where $C=4.6\times10^{-2}\,\mathrm{cm^{-3}\,kpc^{3/2}}$, $n_0 = 3.2\times10^{-2}\,\mathrm{cm^{-3}}$, the disk scale height $z_{h}=1.1\,\rm kpc$ and radius $\rho_{h}=6.2\,\rm kpc$, and the halo core size chosen to be $r_0=3\,\rm kpc$. 
The blast wave profile $\chi(r)$ along a given direction from the GC can then be multiplied by this value to obtain the post-shock density $n_2 = \chi n_1$. This naturally produces a higher density of emitting material in regions close to the Galactic disk, reproducing the larger brightness observed at low latitudes.
An analogous approach was used for the post-shock temperature profile $T_2/T_1(r)$  \citep[following Eqs. 17$-$20 of][]{Cox82}, assuming a uniform unshocked CGM temperature of $kT_1 = 0.163\,\rm keV$, consistent with our findings from spectroscopy (see Table \ref{tab:bkg_param}).

After initial tests of our model, we found a considerable improvement compared to the all-sky image when allowing for a direction-dependent modulation in the density profile of the CGM with which the blast wave interacts. We parametrised this using a quadrupolar relative modulation in density, parametrised as 
\begin{equation}
    n_2 = \left[1 + (f_{\rm quad} - 1) \cos^2 \psi \right]\,\chi\, n_1.
\end{equation}
Here, $n_1$, $n_2$, and $\chi$ are as defined above, and $\psi$ describes the great-circle distance between the direction of a particular position $(x, y, z)$ seen from the GC, and the direction of the quadrupole $(\theta_{\rm quad}, \phi_{\rm quad})$. This direction and the relative quadrupole amplitude $f_{\rm quad}$ were included in our fit as free parameters, and including this component in the model allowed us to reproduce much better `one-sided' morphological features, such as the NPS.

\subsubsection{Evaluation and fitting}
In order to reproduce the observed X-ray image, we used an image grid corresponding to that of the half-sky maps by \citet{Zheng2024_broad}, rebinned by a factor of two, yielding a pixel size of $6\arcmin$. Along a given line of sight $(l,b)$, the Galactocentric coordinates of a point at a distance $s$ from the observer are given by:
\begin{equation}
   \begin{pmatrix}
        x \\ y \\ z
    \end{pmatrix}
    =
    \begin{pmatrix}
        -8.18\,\mathrm{kpc} \\ 0 \\ 0
    \end{pmatrix}
    +s
    \begin{pmatrix}
        \cos b \cos l \\ \cos b \sin l \\ \:\sin b
    \end{pmatrix}.
\end{equation}
For the line of sight corresponding to each pixel, we numerically determined the points of intersection $s_{\rm min}$ and $s_{\rm max}$ with our geometric model of a constricted ellipsoidal shell. 
The final expression for the predicted X-ray brightness $\Sigma$ of the bubbles in our model, is then given by:
\begin{equation}
    \mathrm{\Sigma} = \int_{s_{\rm min}}^{s_{\rm max}} n_2(s)^2 \Lambda(T_2(s)) \,\mathrm{d}s, \label{eq:EM}
\end{equation}
where $\Lambda(T)$ describes the temperature dependence of the flux, expressed as eROSITA count rate ($0.6-1.0\,\rm keV$) per area per unit emission measure  \citep[e.g., $\Lambda = 284.3 \,\rm ct\,s^{-1}\,deg^{-2}\,cm^6\,pc^{-1}$ for a $0.3\,\rm keV$ plasma with abundances of $0.2\,Z_{\odot}$ for \texttt{apec v3.1.2};][]{apec,AtomDB}.

In order to quantitatively compare our model to the all-sky data, a few further steps were necessary. 
First, we related the observed count rate to this model emission measure as \mbox{$R_{i} = A_{i}\,\mathrm{EM}_{i}$}.  
The factor $A_{i}$ is a free parameter, acting as a model normalisation, and can be understood as the squared ratio of true density to model density, although it may also be sensitive to temperature and abundance variations. 
Second, to avoid overpredicting the X-ray emission at low Galactic latitudes, we estimated the absorbing column density $N_{\rm H}$ from the HI4PI map of \ion{H}{i} emission \citep{HI4PI}. Using the T\"ubingen-Boulder absorption model \citep{Wilms00} acting on a $0.3\,\rm keV$ plasma represented by an \texttt{apec} model \citep{apec}, we estimated the transparency of this absorbing layer (i.e., the fraction of unabsorbed photons) in the $0.6-1.0 \,\rm keV$ band. By multiplying this location-dependent fraction $f_{A}$ with our model count rate from above, we obtained a proxy for the detected X-ray emission from the eRObub, which can be compared to the observed half-sky map.

We found that background emission from the Galactic plane, cosmic X-ray background, and instrumental background needed to be explicitly included in our model for the fit to be able to trace the actual bubble morphology. Practically, this implied a purely empirical background count rate parametrised as  
\begin{equation}
    R_{B}(l, b) = B_{U} + B_{A}\,f_{A}(l, b)\,\exp\,\left( -\,\frac{l^2}{2\rho_{l}^2}-\frac{b^2}{2\rho_{b}^2}\right),
\end{equation}
where $B_{U}$ reflects a spatially constant and unabsorbed background, and $B_{A}$ traces the absorbed emission from the Galactic plane, with a its angular extent $\rho_{l}$ within the plane, and $\rho_{b}$ perpendicular to it, as free parameters.\footnote{Note that $l$ is defined here to be in the range $[-180^{\circ}, 180^{\circ}]$.} The LHB is unimportant in this energy band and thus can be neglected in this empirical description. The main aim of this parametrisation is to capture the unresolved stellar contribution using the disk component, and the CXB using the uniform component. We note that the CGM rate is already accounted for using Eq.~\ref{eq:CGM}.
The overall model for the count rate can then be written as 
\begin{equation}
    R(l, b) = R_{B}(l, b) + \begin{cases} 
    A_{N}\,\mathrm{EM}(l,b|a_{N}, b_{N}, \theta_{N}, \phi_{N}, \mathcal{M}_{N}) \,\text{for}\, b\geq 0, \\
    A_{S}\,\mathrm{EM}(l,b|a_{S}, b_{S}, \theta_{S}, \phi_{S}, \mathcal{M}_{S}) \,\,\,\text{for}\, b< 0,\end{cases}
\end{equation}
where subscripts $N$ and $S$ stand for north and south, and the morphological parameters follow Eqs.~\ref{eq1} to \ref{eq:compression}.
Finally, we compared our model prediction to the all-sky count rate measurements with the following Gaussian likelihood: 
\begin{equation}
    \mathcal{L}= -\frac{1}{2} \sum_{l, b} \left\{ \log\,\left[ \sigma^2(l,b) + s^2 R^2(l,b) \right]
    + \frac{\left[ F(l,b)-R(l,b)\right]^2}{\sigma^2(l,b) + s^2 R^2(l,b)}\right\},     
\end{equation}
where $F$ and $\sigma$ are the measured count rates and their statistical errors, $R$ represents the total model predictions, meaning the combination of bubble and background components, and the sum runs over all pixels in the modelling region (see Appendix.~\ref{sec:fit_details}). Finally, $s$ is a free parameter introduced to trace systematic scatter around the model emission. This systematic scatter is necessary since our simplistic model clearly cannot capture the complexity necessary to reproduce the observed `clumpy' morphology of the bubbles. Therefore, local over- and underpredictions of our model are expected. 
Given its complicated and high-dimensional character, a statistical sample from this likelihood was drawn using the Markov chain Monte Carlo sampler \texttt{emcee} \citep{ForemanMackey13} to constrain the parameters.

\begin{figure}[htbp]
    \centering
    \includegraphics[width=\linewidth]{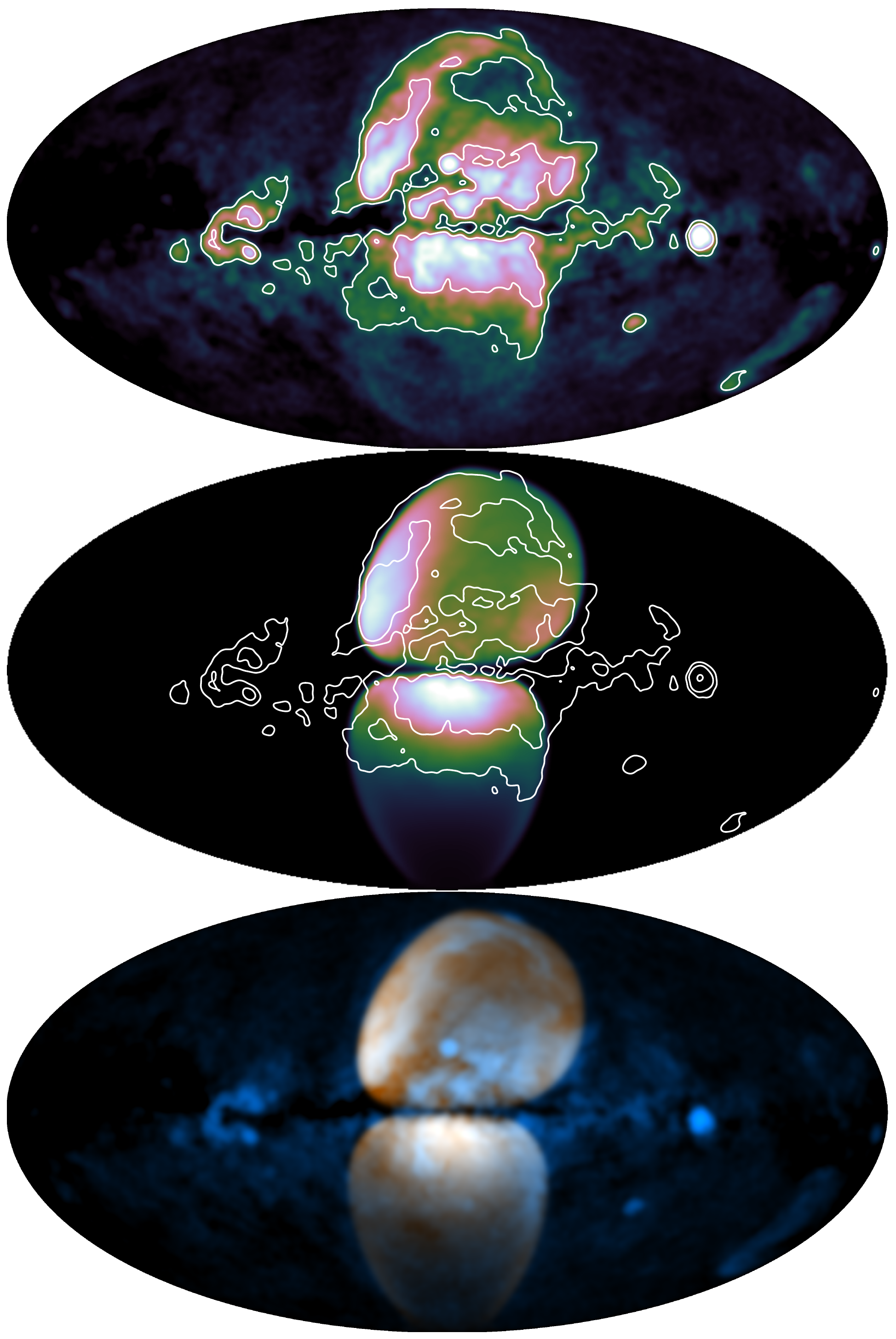}
    \caption{
    Comparison of observed and modelled emission from the \mbox{eRObub}. The top panel shows the observed all-sky diffuse X-ray emission in the $0.6$--$1.0\,\rm keV$ band, smoothed with a $1^{\circ}$ Gaussian kernel. The middle panel shows the predicted X-ray emission from our geometric model of the bubbles (without background components). The overlaid contours trace the observed morphology of the emission. 
    The bottom panel shows an overlay of the two images, with orange reflecting our model, and cyan the observed emission.}
    \label{fig:bubble_overlay}
\end{figure}

\begin{figure*}[t!]
    \centering
    \includegraphics[width=12cm]{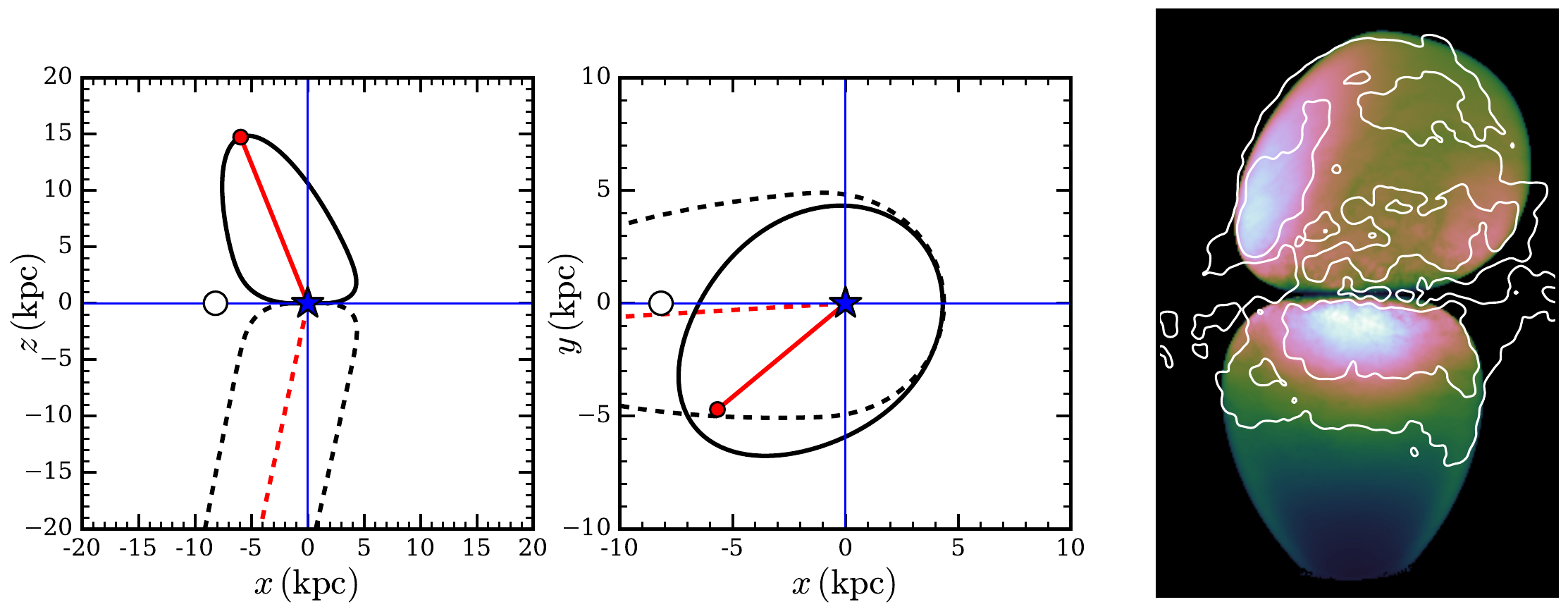}
    \includegraphics[width=12cm]{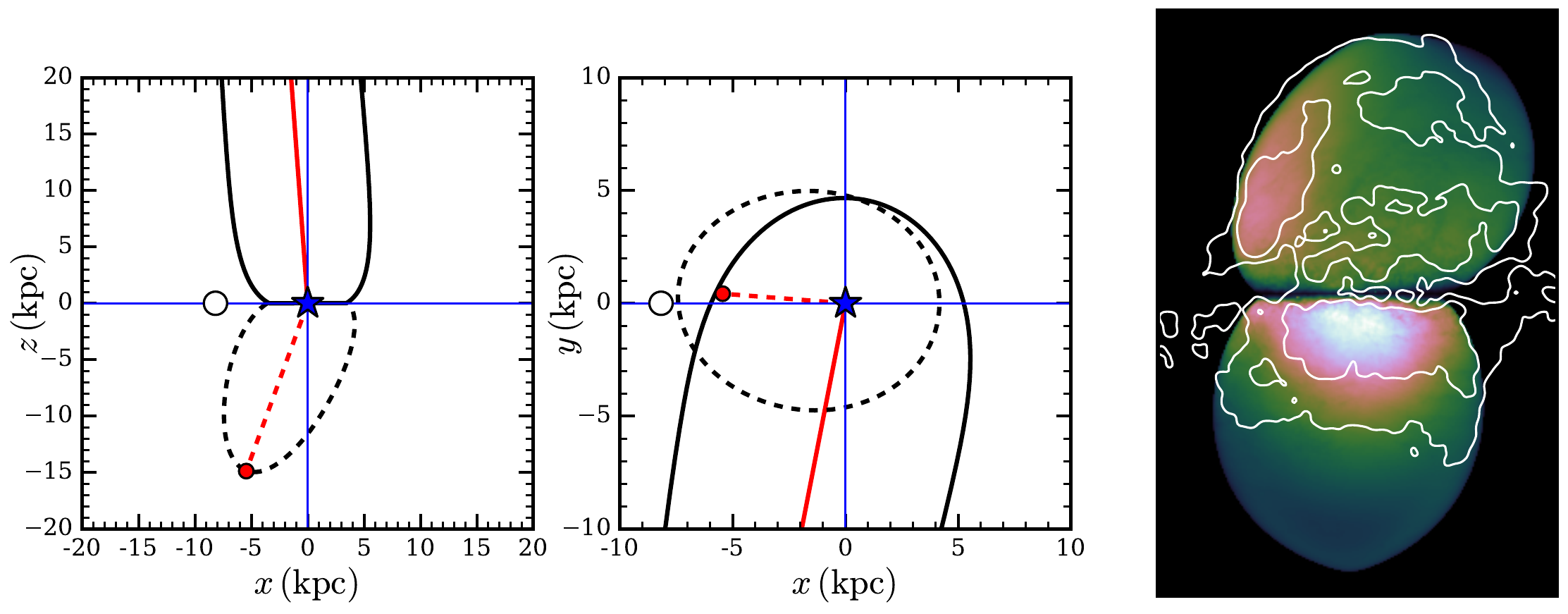}
    \includegraphics[width=12cm]{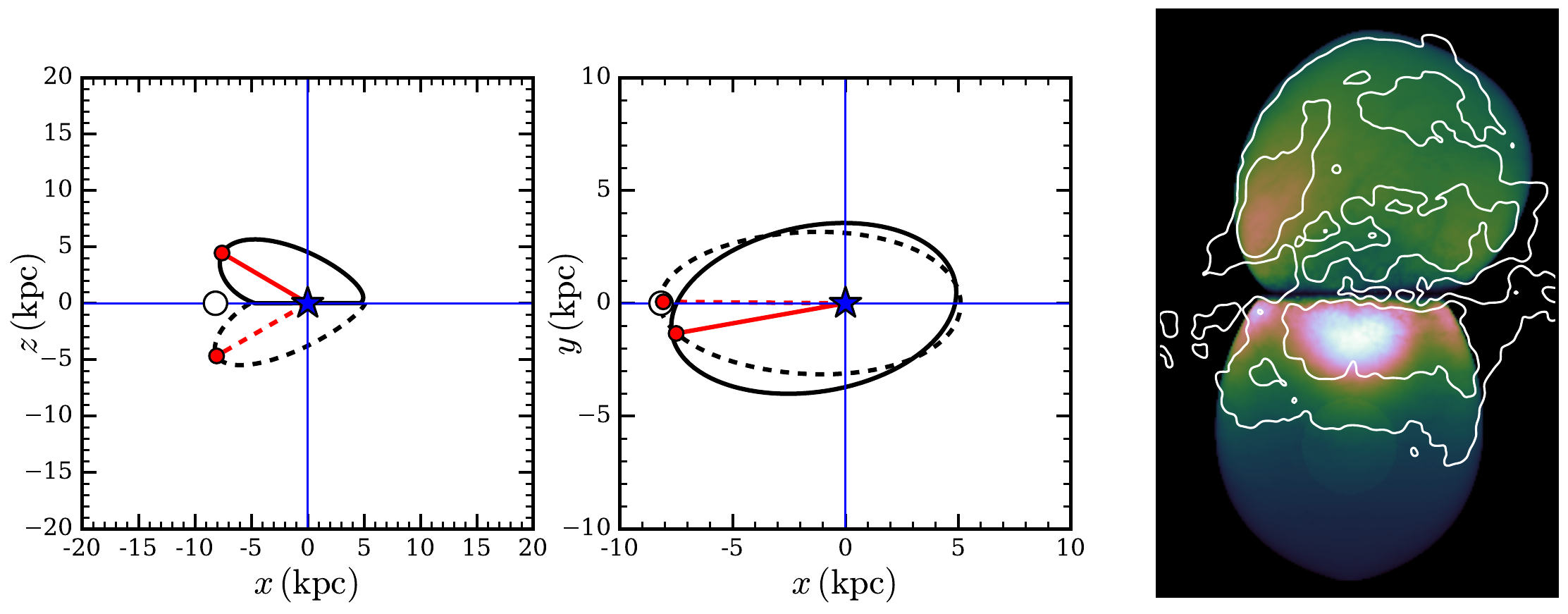}
    \caption{Visualisation of the impact of projection effects on the bubble morphology. The left and centre panels show the cross-section of the northern (solid line) and southern (dashed line) bubble shell from the side (along the $y$-axis), and from the top ($z$-axis). The right panels show the resulting very similar morphologies. 
    The top row corresponds to our best-fit model (Fig.~\ref{fig:bubble_overlay}), whereas the other two rows show geometrically different models resulting in very similar bubble shapes.}
    \label{fig:degeneracies}
\end{figure*}

\begin{table}[htbp]
    \renewcommand{\arraystretch}{1.2}
    \centering
    \caption{Parameters of the bubble model displayed in Figs.~\ref{fig:bubble_overlay} and \ref{fig:postpred}.} 
    \begin{tabular}{cc|cc}
        \hline\hline
        Parameter & Value & Parameter & Value \\ \hline 
         $\log\,A_{N}$ & $-2.47_{-0.07}^{+0.06}$ & $z_{n}$ & $0.84_{-0.12}^{+0.16}$ \\
         $\log\,a_{N}$ & $1.22_{-0.03}^{+0.04}$ & $c_{n}$ & $1.4_{-0.6}^{+0.5}$ \\
         $\log\,b_{N}$ & $0.703_{-0.008}^{+0.010}$ & $\log\,B_{U}$ & $0.593_{-0.003}^{+0.003}$ \\
         $\theta_{N}$ & $27.5_{-1.4}^{+1.5}$ & $\log\,B_{A}$ & $0.957_{-0.029}^{+0.028}$ \\
         $\phi_{N}$ & $219_{-3}^{+4}$ & $\rho_{b}$ & $16.8_{-0.4}^{+0.4}$ \\
         $\log\,\mathcal{M}_{N}$ & $0.268_{-0.010}^{+0.011}$ & $\rho_{l}$ & $36.5_{-1.3}^{+1.4}$ \\
         $\log\,A_{S}$ & $-2.23_{-0.06}^{+0.05}$ & $\theta_{\rm quad}$ & $38.5_{-1.2}^{+1.3}$ \\
         $\log\,a_{S}$ & $1.92_{-0.10}^{+0.06}$ & $\phi_{\rm quad}$ & $355.5_{-1.3}^{+1.3}$ \\
         $\log\,b_{S}$ & $0.696_{-0.010}^{+0.013}$ & $\log\,f_{\rm quad}$ & $0.78_{-0.03}^{+0.04}$ \\
         $\theta_{S}$ & $12.2_{-2.0}^{+1.7}$ & $s$ & $0.2475_{-0.0021}^{+0.0021}$ \\
         $\phi_{S}$ & $3.1_{-2.7}^{+2.9}$ \\
         $\log\,\mathcal{M}_{S}$ & $0.46_{-0.04}^{+0.04}$ \\
       \hline
    \end{tabular}
    \tablefoot{The size and orientation of the bubble semi-axes for the northern and southern bubbles are described by $a$, $b$, $\theta$, and $\phi$, whereas the equatorial constriction is parametrised by the scale height $z_{n}$ and waist radius $c_{n}$ (see description in the text). Units are kpc for spatial scales ($a_{i}, b_{i}, z_{n}, c_{n}$), degree for all angles ($\theta_{i}, \phi_{i}, \rho_{i}$), and $\rm ct\,s^{-1}\,deg^{-2}$ for the background levels $B_{i}$, with the remaining parameters being unitless.
    }
    \label{tab:bubble_params}
\end{table}

\subsection{Comparison to observed morphology}
The best-fit model from our analysis is compared to the smoothed all-sky data in Fig.~\ref{fig:bubble_overlay}, with a more quantitative comparison to the fitted data provided in Fig.~\ref{fig:postpred}. The corresponding physical parameters are given in Table \ref{tab:bubble_params}, and Fig.~\ref{fig:corners} illustrates degeneracies between the different parameters describing the bubble geometry. 
Several interesting insights can be gained from inspecting the morphological aspects driving our best-fit model, even though we caution against overinterpreting the quantitative values of our fit parameters, as our model is clearly not sufficiently complex to reproduce the emission from the eRObub fully. 

Some of the geometrical parameters describing our best-fit model immediately stand out, such as the (likely unrealistic) contrast between a maximum three-dimensional extent of $a_{N} = 16.6\pm1.6\,\rm kpc$ in the north (indicated by the red line in Fig.~\ref{BubbleGeometry}), and $a_{S} > 70\,\rm kpc$ in the south. Our fit prefers such different sizes because of the relatively strong brightness difference at high latitudes, with the northern bubble being significantly brighter than the southern one, which can be reproduced by an origin of the high-latitude emission in the south at larger distances, meaning in a lower-density environment. In addition, the southern bubble edge appears to reach somewhat larger Galactic latitudes than the north, reaching down to $b=-90^{\circ}$, requiring either a significant tilt in our direction or a considerable vertical extent. 
The horizontal extent of the two bubbles is constrained comparatively reliably and to quite similar values at semi-minor axes of $b_{N} = 5.05^{+0.12}_{-0.09}\,\rm{kpc}$ and $b_{S} = 4.97^{+0.15}_{-0.11}\,\rm{kpc}$. These values are somewhat smaller than the spherical shell radius of $7\,\rm kpc$ originally estimated by \citet{Predehl20}. The more recent estimates by \citet{Liu24} give horizontal radii around $9\,\rm kpc$ for the north, in combination with a strong tilt away from us, and around $6\,\rm kpc$ for the south.
Further, we note that the fitted `waist' radius of the bubbles $c_{n} = 1.4^{+0.5}_{-0.6}\,\rm kpc$ indicates a significant extent of the emitting gas within the Galactic plane. However, this depends on the geometric model used, since extrapolation into a highly absorbed region is required. Nevertheless, this radius closely matches the well-known `Expanding 3\,kpc arm' traced by 21\,cm and CO observations \citep[$v_r\approx55\,{\rm km\,s^{-1}}$; e.g.][]{vanWoerden57,Oort58,Dame01,Dame08}, potentially arguing for a common origin for the eRObub and 3\,kpc arm \citep[e.g.][]{Sanders74,Sofue23}, although such a radial velocity could also be driven by elliptical orbits in the Galactic bar \citep[e.g.][]{Kumar25}.

Regarding the measured bubble orientation angles $\theta$ and $\phi$, it appears convincing that the northern bubble is tilted westward (negative $y$-direction) quite significantly. The tilt is evident in its morphology in the all-sky X-ray image (Fig.~\ref{fig:bubble_overlay}) and might also be traced by counterparts in polarised radio emission \citep{Liu24}. The quite significant tilt in our direction implied by $(\theta_{ N},\phi_{ N}) \sim (28^{\circ}, 220^{\circ})$ is connected to a rather compact bubble geometry needing to reach $b\sim +80^{\circ}$, but dependent on our assumptions for the radial emission profile. 
On the other hand, the southern bubble appears tilted more weakly, especially in the east-west direction, at $\theta_{ S} \lesssim14^{\circ}$, $\phi_{ S} \sim 0^{\circ}$. Hence, it is implied that the two bubbles exhibit different physical sizes and their semi-major axes are not co-aligned. This finding may, however, not be surprising, even in the case of shells driven by outbursts in the GC: the simulations by \citet{Pillepich21} show that Milky-Way-like galaxies in cosmological simulations exhibit quite chaotic bubble-like structures that need not be limited to a single bubble pair, and are frequently very asymmetric.
Our fit implies that both bubbles are tilted towards the Sun. While, if physical, this could be pure coincidence, one would also expect to find this behaviour if the true bubble geometry is conical. In this case, the nearby portion of the base of the X-ray emitting gas would dominate the emission, and give rise to shell-like structures in projection only \citep{Bland03}.    

To further illustrate the above points, we demonstrate the model degeneracies introduced by projection effects in Fig.~\ref{fig:degeneracies} \citep[see also in Ext.~Fig.~9 of][]{Zhang24}. This figure displays the three-dimensional geometry of our best-fit model, in comparison with two vastly different bubble geometries, with vertical extents ranging between 7 and 100\,kpc, and tilt angles up to $\theta = 60^{\circ}$. Despite their striking intrinsic differences, all models produce apparent bubbles of very similar sizes and shapes to the observed ones, with the main difference lying in the emission profile within the bubble. 
The main reason for this inability to constrain the true vertical extent and inclination of the bubbles is our location in the Galactic disk, potentially quite close to the edge of the bubbles. Our location prohibits us from obtaining a proper `side view' as we could for any other edge-on galaxy, and we are dominated by the `base' of the bubbles close to the Galactic plane. Crucially, the emission attributed to the `top' of the bubbles is expected to be extremely faint due to the low densities, so the observed X-ray emission depends little on its location, or even the existence of a closed shell \citep{Bland03}. Hence, what we observe as a shell in projection depends strongly on the horizontal bubble extent, but very little on its true vertical size. 
To summarise, the vertical extent and bubble orientation towards/away from us exhibit strong degeneracies, while the horizontal extent, inside and above the Galactic plane, and east/west orientation, drive the observed emission, and can be treated as more reliable. 

Apart from the geometrical shape, our model also qualitatively reproduces a few of the key features in the emission profiles of the bubbles (See Fig.~\ref{fig:bubble_overlay}). For instance, our model very well reproduces the very X-ray bright region in the latitude range $-20^{\circ}\leq b\leq-5^{\circ}$. Similarly, the northeastern edge of the bubbles is strongly brightened, so it qualitatively reproduces the NPS. Both of these features are reproducible due to the quadrupolar density enhancement in the direction $(\theta_{\rm quad}, \phi_{\rm quad}) \approx (40^{\circ}, 355^{\circ})$ allowed by our model. Hence, the prominent X-ray morphology of the NPS alone is not necessarily an argument for a physical separation of the NPS from the bubbles, as density inhomogeneities in the CGM could provide a viable alternative scenario. 
Similarly, we note a subtle difference in the amplitude scaling factors of the two bubbles, with best-fit values of $A_S/A_N = 1.74 \pm 0.37$, implying that the southern bubble emits more flux per {\it model} emission measure (Eq.~\ref{eq:EM}) than the northern bubble, likely driven by the respective brightness contrasts between low and high latitudes. While highly tentative due to the simplicity of the model, one could attribute this to intrinsically higher CGM densities in the south, by around $30\%$, for instance due to material released during past tidal interactions between the Milky Way and the Magellanic Clouds \citep[e.g.,][]{Besla10, Lucchini21,Carr25,Oprea26}.

Finally, our fit produces a significantly higher typical Mach number $\mathcal{M}$ for the shock in the south than in the north. While the exact values were derived under the assumption of a point explosion into a homogeneous medium, their relative difference can be attributed to the respective radial emission profiles. A higher Mach number produces a sharper density decline behind the shock, meaning a thinner emitting shell in X-rays. Hence, the thicker appearance of the shell in the north (in particular at the NPS) likely drives the lower Mach number there, whereas the opposite is true for the southern bubble, which generally appears more limb-brightened. Nonetheless, a few edges in the observed data are much sharper than reproducible by our model, for example, the western rim of the southern bubble (see also Figs.~\ref{fig:postpred} and \ref{fig:erobub_kT_EM}), which may be attributed to inhomogeneities in CGM density or shock velocity modifying local emission behaviour.

\section{Discussion}\label{Discussion}

\subsection{Temperature structure}
From the spectral analysis, a relatively clear picture has emerged. The eRObub is generally characterised by two temperature components, one at $\sim0.2\,$keV and another at $\sim0.6\,$keV, which are uniform across hemispheres. The elemental abundance of the eRObub gas is also consistent with the expectation of shock-heated halo gas between $Z=0.1$--$0.3\,Z_{\odot}$, not immediately obvious of metal enrichment from the star formation activities towards the inner part of the Galaxy. The exceptions to this general picture are the presence of the NPS, the cool shell surrounding the northern bubble and a hot shell ($kT\sim0.35\,$keV) between $-30\degr\lesssim b\lesssim-45\degr$ in the south. The northern eRObub is surrounded by a shell (reg~3, 7, and 34), which appears to have a slightly cooler temperature than the low-temperature component in the eRObub interior. The hot component is not present in the cool shell. In terms of its surface brightness, the NPS is an anomaly in the northern eRObub that does not follow the latitudinal decay trend exhibited by other eRObub regions. Our spectral analysis suggests that a higher elemental abundance primarily raises the NPS's brightness without an obvious jump in EM.

 \subsubsection{Intrinsic multi-temperature gas}
A two-temperature or even a multi-temperature structure is not intrinsically surprising, as repeated shock heating, continuous cooling or turbulent mixing of plasmas could naturally produce a multi-temperature structure \citep[e.g.][]{Lancaster21,Wang21}. A spread in plasma temperature has been commonly found in the outflow or halo of nearby edge-on disc galaxies \citep[e.g.][]{Strickland04,Ranalli08}. We did test a lognormal temperature distribution to model the eRObub in the early stage of the spectral analysis. There were two reasons not to pursue this further. The first was that the lognormal distribution could not produce a \ion{Fe}{XVII}\,15\,\AA/17\,\AA~ratio as high as the data, without worsening the fit elsewhere. The second reason was that it is computationally expensive to fit a lognormal temperature distribution, as it is, in essence, a sum of dozens of individual CIE components, which are calculated on-the-fly in \texttt{SPEX}.

Our temperature measurement with the 1T model is broadly consistent with previous findings by \citet{Kataoka13,Kataoka2015}, who measured a relatively uniform temperature of $\sim0.3$\,keV throughout the FB, despite different modelling assumptions, most notably the inclusion of the unresolved stellar contribution. In addition, we were able to show that this is also the property of gas outside of the FB but inside of the eRObub, even though a two-temperature fit would benefit all regions within the eRObub as revealed by high-S/N spectra.
The hotter component in our model is usually a few times lower in EM and only distinguishable from the large regions with sufficient S/N. When we were analysing the smaller, lower-S/N regions, the hot component was not readily apparent. The constancy of the temperature of both components within the bubbles is somewhat puzzling, which may reflect the fact that the X-ray emissions primarily trace the shell region of the eRObub due to the sharply increasing density there \citep[e.g.][]{Weaver77, Cox82}.
eROSITA has large effective area between $0.5$--$2$\,keV and should be sensitive to gas at $kT\sim1$--$2\,$keV. The absence of such a phase appears on face value to argue against an AGN-burst scenario where the gas is expected to be heated to a few $10^7\,$K \citep[e.g.][but see the multiple caveats listed within]{Yang22}. Our measurement also disfavours a scenario where the FB and eRObub were formed by two AGN jet episodes \citep{RZhang25}, as a temperature jump would be expected at the boundary of the FB.

\subsubsection{Efficient post-shock radiative cooling?}
A natural explanation for the cool shell could be that it consists of dense material, which has significantly cooled due to the emitted radiation. Such processes are observed in stellar wind bubbles and old SNRs, where radiative cooling leads to efficient energy loss and the formation of thin dense shells \citep{Raymond79}. However, it seems unlikely that radiative cooling at the low ambient densities of the Milky Way CGM is sufficiently efficient to significantly lower the temperature of $\sim0.3\,\rm keV$ gas within a few Myr \citep[cooling time $\gtrsim$ Hubble time for $n\sim 10^{-4}\,{\rm cm^{-3}}$ and $Z=0.3\,Z_\odot$;][]{Sutherland93}.  
An alternative mechanism for producing a cool shell could be cooling due to adiabatic expansion. If the ambient density of the CGM decreases relatively rapidly (i.e., over a small spatial scale compared to the bubble size), as could be conceivable for a shock moving away from the Galactic plane into the halo, the plasma temperature would be expected to decrease, ideally as $T \propto n^{2/3}$ \citep{Yamaguchi18}. 

As pointed out by \citet{line-ratio}, the apparent cool shell could also be explained by a projection effect, with the cold portion of the shell at high latitudes being located rather nearby, interacting with higher density material. In this scenario, the hotter gas inside the bubbles is located at larger distances from the Galactic plane and projected at lower apparent radii to the spherical geometry of the bubbles.  

\subsubsection{Non-equilibrium ionisation in the cool shell?}\label{subsubsec:NEI}
An alternative explanation for the apparent cool shell could be that the hot plasma is not in CIE. A simple estimate of the relevant ionisation timescale, quantifying the number of collisions experienced by an average ion, is given by the product of an approximate age of the bubbles and the typical halo density $n_{e}t \sim 5\times 10^{-4}\,\mathrm{cm^{-3}}\times 2\times 10^{7}\,\mathrm{yr} \approx 3\times10^{11}\mathrm{cm^{-3}\,s}$ (see below), indicating a slightly underionised plasma. Tentative evidence for NEI along a sight line towards the NPS has, in fact, been presented by \citet{Yamamoto22}. 
An underionised plasma is characterised by a preferential occupation of lower ionisation states, compared to the CIE expectation for a given plasma temperature, which would mimic a lower temperature when fitted with a CIE model, and leave clear signatures in an oxygen line ratio map \citep{line-ratio}. Hence, the observed cool shell may in fact be a shell of underionised plasma behind the forward shock, which has experienced shock heating sufficiently recently to not have fully equilibrated. 
Even though this scenario bears a strong resemblance to processes observed in young SNRs \citep[e.g.][]{Borkowski01}, a fundamental difference is that the observed shock strength in the eRObub is much lower due to the interaction with tenuous hot CGM, implying less drastic differences between pre- and post-shock temperatures. Hence, the effect of NEI on the observed spectrum is likely different from that observed in SNRs.

To test this scenario, we replaced the CIE eRObub model with the plane-parallel shock model (\texttt{neij)} in \texttt{SPEX} for the cool shell spectrum (reg~3) as in Sect.~\ref{subsubsec:neij}.
We found that it is possible to achieve a good fit with $u=1.20^{+0.18}_{-0.14}\times10^{11}\,{\rm cm^{-3}\,s}$ and a post-shock temperature of $0.41\pm0.01$\,keV, in addition to the solution where $u\gtrsim10^{12}\,{\rm cm^{-3}\,s}$ and $kT\approx0.2\,$keV, which returns to the CIE case. This new eRObub component at eROSITA spectral resolution is indistinguishable from the CIE  case due to small changes in C, O and Ne abundances (left panel of Fig.~\ref{fig:cool_shell_spec}), and therefore the fitted spectrum is not shown again. Statistically, the new $\chi^2/{\rm dof}$ is 1.17 (175 dof), which means the improvement is insignificant compared to the 1T CIE model ($\chi^2/{\rm dof}=1.18$ with 176 dof, corresponding to a F-statistic of 2.02 with a $p$-value of $0.16>0.01$). Despite the similarity in the final spectral shape, the EM in the \texttt{neij} model is $\sim4.5$ times less than the 1T CIE model, corresponding to a density about a factor of two lower, assuming the same geometry. The lack of a clear limb-brightening morphology also casts doubt on this proposition, but it is nevertheless possible for a weak shock as found in our geometrical analysis, where the northern bubble prefers a lower Mach number than the south (Table~\ref{tab:bubble_params}).

\subsection{The cool shell as a foreground structure?}\label{subsubsec:foreground}
Another possibility for the cool shell is that it is not a physical part of the eRObub but a foreground structure to the eRObub. The most popular candidate would be the Loop~I superbubble, visible in cold gas, whose relevance has been jeopardised after discovering the (southern) eRObub \citep{Predehl20}. Naturally, the Loop~I superbubble would express itself in the cool shell and as a general foreground to the eRObub. Then the cool shell is the region where the Loop~I superbubble does not overlap with the background eRObub in projection. The EM profile in Fig.~\ref{fig:erobub_kT_EM} of the eRObub EM could become more symmetric about the Galactic plane if the Loop~I contribution is subtracted out. However, admittedly, such an asymmetry is less evident from the large region, likely because abundance effects have introduced a scatter in the EM.

\citet{Egger95} identified and linked the observation of a local ring-like structure in \ion{H}{I} \citep{Dickey90} with simulation results from \citet{Yoshioka90}, where \citet{Yoshioka90} showed that the formation of a dense ring or wall between two spherical shock waves is possible if both have reached the radiative stage. This observation led \citet{Egger95} to conclude the existence of the Loop~I superbubble, and its current interaction with the LHB.

\begin{figure*}[htbp]
    \centering
    \includegraphics[width=0.7\textwidth]{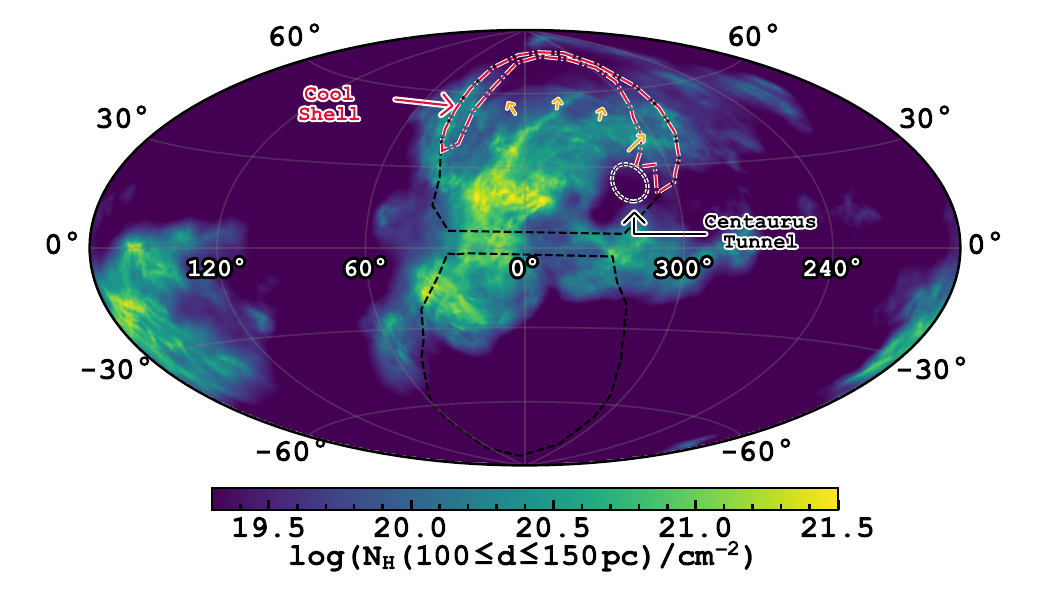}
    \caption{Column density map integrated between $100$--$150$\,pc inferred from Gaia stellar extinction \citep{Edenhofer24}. The dust edge is highlighted by orange arrows, which may mark the end of the LHB-Loop~I supberbubble interface. The cool shell could be a part of the Loop~I superbubble that is not overlapping with the background outflow. We also indicate the location of the Centaurus tunnel, which appears to be a channel between the LHB and the Loop~I superbubble that lies within the extent of this interface \citep{Yeung2024}. The overall silhouette of the eRObub is shown by the black dashed lines.}
    \label{fig:dust}
\end{figure*}

While the main proposal of \citet{Egger95} is to invoke the Loop~I superbubble as the origin of the NPS, which remains a topic of great debate, we draw attention to a tentative observation which provides further evidence to the existence of a dust wall and hence the existence of the Loop~I superbubble, revealed by a state-of-the-art 3-dimensional dust map inferred from Gaia stellar extinction \citep{Edenhofer24}. Figure~\ref{fig:dust} shows the column density between a radial distance of 100 to 150\,pc from us, where we converted the $A_V\,{\rm pc^{-1}}$ information in \citet{Edenhofer24} to the unit of $N_{\rm H}\,{\rm pc^{-1}}$ using $N_{\rm H} = 2.21\times10^{21}\,A_V$ by \citet{Guver09}. We point out a prominent curved edge highlighted by orange arrows in the figure, located at a distance of 100--150\,pc. The edge appears to mark the end of a dense wall, which further supports the scenario proposed by \citet{Egger95} if they are real features but not artefacts in the inversion. In addition, \citet{Yeung2024} found that the LHB has a higher EM towards the direction of Centaurus, potentially caused by a tunnel of hot gas linking the Loop~I superbubble with the LHB. This tunnel can be seen in the dust map as a hole with little $N_{\rm H}$, located just within the northern dust edge as labelled in Fig.~\ref{fig:dust}. One possibility for the cool shell (red outline in Fig.~\ref{fig:dust}) is that the Loop~I superbubble is more extended in projection than the extent of the dust edge/wall and the Galactic outflow in the background, hence producing a shell-like structure in front of the eRObub. A foreground component also would help to reconcile the somewhat higher EM in the northern bubble as seen in Fig.~\ref{fig:erobub_kT_EM} and \ref{fig:lat_profile}, when a uniform abundance is assumed everywhere.

\begin{figure}[htbp]
    \centering
    \includegraphics[width=0.49\textwidth]{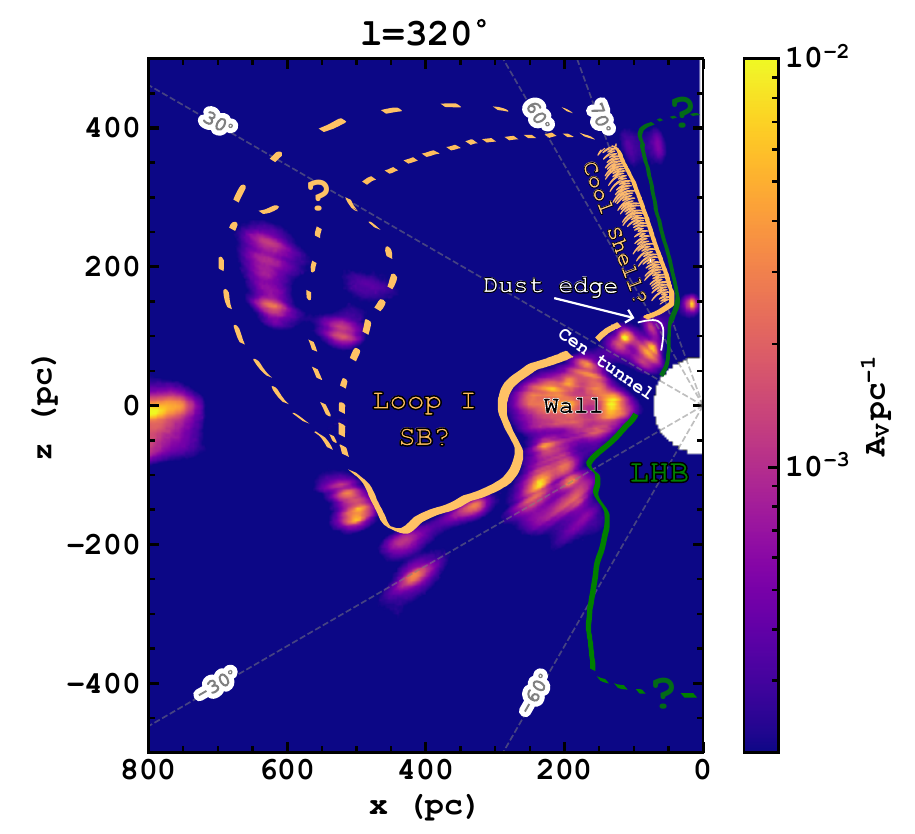}
    \caption{Illustration of the possible scenario of the cool shell as a part of the Loop~I superbubble. The extinction map is taken from \citet{Edenhofer24}. The approximate outlines of the LHB and the Loop~I superbubble are drawn. The LHB extent is the most uncertain towards the Galactic poles. The boundary of the Loop~I superbubble on the far side and towards the Galactic north is also uncertain. If the cool shell is indeed foreground, its strong enhancement against the background means the boundary of the Loop~I superbubble is well-defined at this latitude ($b\approx70\degr$) and is probably still expanding.}
    \label{fig:LoopI}
\end{figure}

A sketch of the possible scenario is illustrated in Fig.~\ref{fig:LoopI}, which shows a slice of the 3D dust extinction map on the plane of $l=320\degr$ \citep{Edenhofer24}. It shows a wall of dense materials that is commonly referred to as the wall of the LHB. At about $b\sim30\degr$ is the location of the Centaurus tunnel, which is a gap in the LHB wall that is potentially connected to the Loop~I superbubble. One can see that the wall is inclined towards the LHB at higher latitudes, with the end marked by the dust edge in white at $b\sim60\degr$. At longitude $l=320\degr$, the cool shell is located at $b\sim70\degr$. As the cool shell is a clear enhancement in the X-ray, the boundary of the Loop~I superbubble should be sharp there. In the extreme case that the Loop~I superbubble contributes solely to the cool shell, we can estimate its density and pressure from the fitted temperature and EM. If we assume the line-of-sight distance through the cool shell is $\sim 200\,{\rm pc}$ ($400\,{\rm pc}$)
and the EM from the cool shell region is $\approx 4\times10^{-2}\,{\rm cm^{-6}\,pc}$ (reg~3 in Table~\ref{tab:erobub_params}), the density of the cool shell would be $n_e\approx 1.4\times 10^{-2}\,{\rm cm^{-3}}$ ($10^{-2}\,{\rm cm^{-3}}$). The corresponding thermal pressure is $P/k\sim 3\times10^4\,{\rm cm^{-3}\,K}$ ($2\times10^4\,{\rm cm^{-3}\,K}$)
This estimate is in the same order as the LHB and the Orion-Eridanus superbubble also in the solar neighbourhood ($n_e\sim10^{-2}\,{\rm cm^{-3}}$ and $P/k\sim5\times10^{4}\,{\rm cm^{-3}\,K}$; \citealt{Guo95,Joubaud19}).
However, the fitted abundance of the cool shell is more consistent with other eRObub regions than the NPS, which is lower than one would expect for a superbubble originating from a cascade of supernova explosions in the Sco-Cen OB association \citep{Egger98}, casting doubt on this scenario.

Shadowing experiments of clouds beyond the Loop~I superbubble, yet in front of the eRObub, are perfect targets to decompose the emissions from the two, given there are identifiable differences in their spectral shapes. Two such prime targets are molecular clouds CG~12 and Dobashi~6193. Both are sufficiently distant at about 650--700\,pc \citep{Reipurth21,Yeung24b}, which is likely beyond the Loop~I superbubble. They shadow the eRObub at about $b\sim20\degr$, at the western base of the eRObub. One could quantitatively measure or rule out the X-ray emission from the Loop~I superbubble with X-ray observations towards these targets.

\subsection{Shock speed and energetics}
We showed that our eRObub spectra are best reproduced by a two-temperature model with temperatures of $\approx0.21\,$keV and $\approx0.60\,$keV, respectively.  While a 2T model fits the spectra best, it is unclear how to properly interpret the two characteristic temperatures as a single propagating shock. However, it is clear that the eRObub temperature is unlikely to exceed $0.6$\,keV, as shown by both the 2T and plane-parallel shock models. Therefore, we estimate the corresponding shock speeds from the temperature component dominating the EM in the 2T model, $0.21\,$keV, as well as the temperature of the hotter component at $0.6\,$keV, in parentheses below.

A post-shock temperature of $0.21\,$keV ($0.6$\,keV) is a factor $1.3$ ($3.5$) higher than the temperature of the CGM of $\simeq0.16 \,\rm keV$ measured in the background regions, with which the eRObub interact. Following the Rankine-Hugoniot jump conditions implies a weak shock with a Mach number $\mathcal{M} \sim 1.3$ ($\sim3$), driven into the CGM. 
In Sect.~\ref{Morph}, we showed that a Mach number of $1.3$ would be in a similar range as implied by the bubble morphology in the north ($\mathcal{M}_{N} = 1.85\pm0.05$), whereas the emission appears more limb-brightened in the south, pointing toward a higher Mach number ($\mathcal{M}_{S} = 2.9\pm0.3$). 
For the corresponding CGM sound speed $c_{s} \approx 200\,\rm km/s$, $\mathcal{M} \sim 1.3$--$1.9$ ($\mathcal{M}\sim3$) would imply a shock velocity $v \approx 260$--$390\,\rm km\,s^{-1}$ ($600\,\rm km\,s^{-1}$), inflating the eRObub. Furthermore, this implies a relative increase in pressure by a factor of $3.0$--$4.3$ ($11$) at the shock front.

The eRObub predominantly emit in the 0.2--1.5\,keV band. To estimate the energetics of the eRObub within this band, we summed the fluxes from both the low and high temperature components of our large regions (Table~\ref{tab:flux}). They totalled $\sim5\times10^{-8}\,\rm erg\,s^{-1}\,cm^{-2}$. The flux from the low-temperature component of the northern eRObub is $1.4$ times that of the southern one. In contrast, the hemispheric difference is small for the hot component. We remind the reader that the region coverage of the eRObub is not complete at low latitudes (Sect.~\ref{subsubsec:reg}), so this is likely an under-estimation of the fluxes from the full eRObub. For comparison, our geometrical model reported in Table~\ref{tab:bubble_params} produces integrated (absorption-corrected) count rates of $2.8 \times 10^4\, \rm ct\,s^{-1}$ and $2.1 \times 10^4\, \rm ct\,s^{-1}$ for the northern and southern bubbles, respectively.
This suggests a slightly larger total flux of $\sim 6.4\times10^{-8}\,{\rm erg\,s^{-1}\,cm^{-2}}$ in the 0.2--1.5\,keV band, as expected due to the regions excluded in the flux estimate from spectral fitting.

\begin{table}[htbp]
    \caption{Breakdown of the total flux from the eRObub, by hemispheres and temperature components (Sect.~\ref{subsubsec:2T})}    \label{tab:flux}
    \centering
    {\small
    \renewcommand{\arraystretch}{1.2}
    \begin{tabular}{ccc}
    \hline\hline
      \multicolumn{3}{c}{Total flux (0.2--1.5\,keV)}\\
      \multicolumn{3}{c}{(${10^{-9}\,\rm erg\,s^{-1}\,cm^{-2}}$)}\\\hline
       Component & cool (0.21\,keV)& hot (0.60\,keV)     \\
       North  &  22.3 & 4.8\\
       South  &  15.9 & 4.9 \\\hline
    \end{tabular}
    }
    \tablefoot{
    Fluxes within the $-10\degr<b<15\degr$ were not taken into account as they were not covered by the spectral extraction regions.
    }
    
\end{table}
To estimate the luminosity of the bubbles, we computed an emission-weighted `effective distance' for the two bubbles by evaluating our geometric model: 
\begin{equation}
    d_{\rm eff}^2 = \frac{\int{n(s)^2 \Lambda(T(s))^2 s^2\,\mathrm{d}s}}{\int{n(s)^2\Lambda(T(s))^2\, \mathrm{d}s}},
\end{equation}
which corresponds to an emission-weighted mean squared distance to the bubbles, along the line of sight. 
When averaged over the two respective hemispheres, the geometrical model has effective distances of $d_{\rm eff}=6.3$\,kpc and $d_{\rm eff}=10.2$\,kpc for the northern and southern bubbles. Assuming the fluxes determined in our spectral fits, this yields luminosities of $1.3\times10^{38}\,{\rm erg\,s^{-1}}$ and $2.6\times10^{38}\,{\rm erg\,s^{-1}}$, respectively. For comparison, the fluxes of our geometrical model imply $1.7\times10^{38}\,{\rm erg\,s^{-1}}$ and $3.5\times10^{38}\,{\rm erg\,s^{-1}}$. Our luminosity estimates are slightly lower than the estimation by \citet{Predehl20}, mostly driven by a lower effective distance to the northern bubble in our model, but not a difference in flux or surface brightness. 

Given our weak constraints on emission from the bubble cap in the south, and the model degeneracies (Fig.~\ref{fig:degeneracies}), we assume the maximum extent of 15--18\,kpc of the northern bubble to be more representative of the typical eRObub size.
Combined with our estimates of the current shock speed of $260$--$390\,{\rm km\,s^{-1}}$, we recover an age of the eRObub of $\sim15$--$27$\,Myr, similarly to \citet{Predehl20}. Taking the hot component would reduce this age estimate to $\sim10\,$Myr.
This computation is highly uncertain, as both the true bubble geometry and shock wave trajectory need to be assumed, and it only indicates a rough estimate of the characteristic age of the bubbles.
Due to the strongly asymmetric morphology of the eRObub in an inhomogeneous ambient medium, we refrained from using any spherically symmetric model (e.g. Sedov-Taylor) to constrain the energy which inflated the bubbles. Instead, we rescaled our best-fit model of the X-ray emitting density profiles for the two bubbles using their normalisations (i.e., $\propto A_{i}^{1/2}$), and integrated over the bubble volume to obtain a census of the entire hot gas mass. We obtained average densities of $\Bar{n}_{N} = 8.8\times10^{-4}\,\rm cm^{-3}$ and $\Bar{n}_{S} = 2.5\times10^{-4}\,\rm cm^{-3}$ for the two hemispheres (with the difference mainly due to the larger volume attributed to the southern bubble), and an integrated hot gas mass of $5.1\times 10^7\,M_{\odot}$.
Assuming our average spectrally measured temperatures of $kT = 0.21\,\rm keV$ and $0.6$\,keV and a simplified assumption that their ratio in EM reflects their difference only in average density (hence $\frac{M_{\rm cold}}{M_{\rm hot}} = \frac{n_{\rm cold}}{n_{\rm hot}}=\sqrt{\frac{\rm EM_{cold}}{\rm EM_{hot}}}\sim \sqrt{5}$), this yields a total thermal energy of $\sim5.5\times10^{55}\,\rm erg$, yielding an excess of $1.5\times10^{55}\,\rm erg$ over the pre-shock CGM. In contrast, assuming the concrete temperature structure fit in Sect.~\ref{Morph}, we obtain an excess of $1.66\times10^{56}\,\rm erg$, recovering an injected energy in a similar range as \citet{Predehl20}, despite vastly different assumed geometries.

\subsection{Origin of the eRObub}
Our results, including energetics, temperature, and abundances of the X-ray emitting gas, could constrain formation models of the eRObub and FB, invoking past star formation or AGN activity.  
Most prominently, our analysis measures a sub-solar composition of the emitting material of $\sim 0.1$--$0.3\,Z_{\odot}$, which is not dissimilar to the expected CGM abundance. There is tentative evidence for an elevated Ne/O abundance ratio of about $\approx {\rm 2\,Ne_{\odot}/O_{\odot}}$ which unfortunately has large systematic uncertainties dependent on the adopted abundance table, which is similar to the finding in selected \textit{Suzaku} sight lines by \citet{Gupta23}.

Generally, Ne overabundance in the eRObub would be unexpected in an AGN outburst model. However, it would also not be trivial to ascribe such behaviour to a fuelling of the bubbles by star formation \citep[cf.][]{Gupta23}. 
Generally, the stellar winds and core-collapse supernovae associated with star formation bursts are expected to produce large quantities of light $\alpha$-elements, including O, Ne, and Mg, and smaller amounts of Fe-peak elements.
However, relative enhancements of Ne to O in stellar nucleosynthesis compared to the solar composition are highly model-dependent \citep[e.g., see][]{Nomoto13,Sukhbold16}. An important aspect to consider is that the spectra of the background regions have shown that the possible $\alpha$-enhancement is not limited to the eRObub, but is likely present also in the CGM (Sect.~\ref{subsubsec:bkg_reg}).
Therefore, we believe further theoretical modelling is necessary before quantitatively interpreting the relative abundances in the X-ray bright plasma in the eRObub in favour of a particular physical origin. 

As shown above, their X-ray morphology and temperature imply an energy around $10^{56} \, \rm erg$ necessary for inflating the bubbles, which requires a mechanism with an average power of $P \gtrsim 1.5\times10^{41} \, \rm erg\,s^{-1}$ to produce the bubbles within an age $\lesssim 2\times10^7 \, \rm yr$. Such an energy output is certainly feasible for an AGN outburst \citep[e.g.,][]{Yang22}, even over much shorter time spans, as the Eddington luminosity of the Milky Way black hole is $L_{\rm Edd} \approx 5\times10^{44}\,\rm erg \,s^{-1}$. 
Achieving the required power from a star formation burst requires energy input corresponding to around one supernova per century, or a total of $10^5$ supernovae during the lifetime of the bubbles. While these numbers are in principle feasible for a nuclear starburst \citep{Sarkar15, Nguyen22}, such a large supernova rate within a few Myr would imply significantly elevated gamma-ray emission from the GC at the radioactive decay energy of $\rm ^{26}Al$, which is in contrast with the observed distribution \citep{Diehl06, Su10}. Furthermore, such a high supernova rate appears to be contradicted by the relatively low current star formation rate of $0.1\, M_{\odot} \, \rm yr^{-1}$ measured for the GC \citep{Ponti15,Barnes17}, which implies a supernova rate around $1 \, \rm kyr^{-1}$ when assuming a Kroupa initial mass function \citep{Kroupa01} and a minimum exploding mass of $8\,M_{\odot}$.
A possible alternative scenario invoking star-formation fuelling is that of a conical Galactic wind originating in star-forming regions, which produces a bubble-like morphology only in projection \citep{Zhang24, Churazov24}. This scenario would both reduce the total energy of the hot gas and increase its likely lifetime, reducing the required power output sufficiently to allow fuelling by the observed star formation activity in the Galaxy. 

Generally, a shortcoming of an AGN outburst model is that a short, strong energy injection predicts much higher present-day temperatures (or equivalently oxygen line ratios) than observed in X-ray data \citep[see also][]{Mondal22, Sarkar23}. 
In particular, the multi-keV plasma behind the forward shock predicted by \citet{Yang22} does not appear to be observed in eROSITA spectra of any portion of the bubbles. 
This problem would likely be avoided by a model invoking long-term, low-power fuelling of the bubbles, such as through AGN wind or star formation \citep{Mou14,Crocker15,Sarkar15, Sarkar17, Nguyen22, Zhang24}, as the involved temperatures would be expected to be much lower.   

A further interesting puzzle piece is given by the orientation of the bubbles. While the FB appear to be relatively orthogonal to the Galactic plane \citep{Ackermann14}, the eRObub appear to be somewhat tilted, as the northern and southern orientation angles appear to be offset from each other in the east-west direction (see Sect.~\ref{Morph}). A hypothetical AGN jet can, in principle, have an arbitrary orientation with respect to the Galactic plane, and thus a tilted bubble morphology appears unsurprising. Recent measurements indeed suggest the accretion disk of Sgr~A$^*$ is significantly inclined \citep[$i\sim160\degr$;][]{EHT22,Gravity22}, in agreement with such a scenario. However, one may expect a nuclear starburst to produce a perpendicular structure, particularly in the presence of collimation through dense surrounding gas in the Galactic disk. However, as shown by \citet{Nguyen22}, a non-uniform distribution of stellar mass within the nuclear region may be capable of producing similarly tilted morphologies as observed. Furthermore, the apparent `tilt' of both bubbles towards the Sun may also be interpreted as a signature of a conical star-formation-driven outflow \citep{Zhang24, Churazov24} producing a bubble morphology in projection.

A final important aspect to consider is the presence of leptonic cosmic rays, in the form of the FB, on the inside of the eRObub. Such a morphology appears to be well reproduced in simulations of an energetic short-term jet activity from Sgr~A$^*$, in which cosmic rays are injected at the base of the jet, and are advected with the hydrodynamic flow \citep{Yang22, Mondal22}. Thermal X-ray emission from CGM gas shocked and heated by the jet forms the eRObub, which is separated by a contact discontinuity from the decelerated outflow in which the leptonic cosmic rays produce radio and GeV $\gamma$-ray emission via synchrotron and inverse Compton processes. 
In contrast, in the model of starburst powering, it is not intuitively explained why the most energetic cosmic rays would be located behind the reverse shock (i.e. in the FB), rather than the forward shock, corresponding to the eRObub. A multiwavelength morphology similar to the observed one would require cosmic rays to be accelerated in the nuclear region, rather than at the shock front expanding into the CGM. 

While an origin close to the GC is required if one assumes that the eRObub and the FB are physically connected, it is worth reminding that the two structures need not to share a common origin. Although a significant portion of the eRObub emission is likely from the inner Galaxy, the active star-forming ring at a Galactocentric distance of 3--5\,kpc has been shown to be powerful enough to easily drive galactic winds to a vertical distance of 10\,kpc \citep{Zhang24}. Magnetic ridges observed in radio polarised intensity maps also appear to trace the silhouette of the eRObub closely. The fact that these magnetic ridges seemingly originate from the locations of high star formation rates in the star-forming ring seems to lend credibility to this model. It is also unclear whether the edge of the eRObub is a forward shock front or just a contact discontinuity. The fact that we found the NEI signature to be unimportant in interpreting the eRObub spectra may point to the latter, as proposed by \citet{Churazov24} who suggested the NPS (or more generally for different parts of the eRObub) could be raised to a high vertical distance due to the buoyancy of hot gas stemming from active star-forming regions. Our measurement of an enhanced elemental abundance of the NPS compared to the other parts of the eRObub agrees with this scenario for the NPS. However, an alternative explanation might be required for the rest of the eRObub.

\section{Summary} \label{Summary}
This work presents a detailed analysis of the physical properties of the gigantic X-ray emitting eROSITA bubbles in the western Galactic hemisphere, based on the first {\it SRG}/eROSITA All-Sky Survey dataset.  
This analysis includes detailed modelling of spatially resolved X-ray spectra, complemented by comparisons of the observed all-sky emission map with geometrical models.

Our spectral analysis used customised, larger regions to boost S/N and isolate interesting features such as the FB and NPS, while accounting for latitudinal variations in the background. Even without fitting, it was immediately apparent from the enhanced \ion{Fe}{XVII}, \ion{Ne}{IX,X} and \ion{Mg}{XI} lines that the temperature is significantly enhanced above the CGM temperature. However, the FB regions do not exhibit a noticeable difference in X-ray spectral signature, such as evidence of nonthermal emission, compared to the eRObub regions outside. 

The high S/N from the large regions allowed us to determine the abundances of O, Fe, and Ne. Despite the greater number of degrees of freedom, a single temperature model could not fit most regions satisfactorily, except for the cool shell, which was characterised by temperatures of $0.18$--$0.20$\,keV. The greatest difficulty in matching the observed spectra arose from the \ion{Fe}{XVII} L shell lines between $\sim$0.70--0.85\,keV that consistently showed an enhanced 3$d$/3$s$ (or 15\,\AA/17\,\AA) line ratio. 

We showed that, while predictions differ slightly across popular plasma codes \texttt{SPEX} and \texttt{AtomDB/apec}, none can reproduce the observed ratio using a single temperature component, indicating the need for a second temperature component to resolve the discrepancy. We found alternative models or explanations for a boosted \ion{Fe}{XVII}\,15\,\AA/17\,\AA~ratio, involving plane-parallel shock, resonant scattering, charge exchange, or an ionised absorber, are less convincing than a two-temperature model. The low- and high-temperature components exhibit remarkable constancy across latitudes, at $0.21^{+0.03}_{-0.01}\,$keV and $0.60\pm0.02\,$keV respectively, with the cooler component on average exhibiting $\sim5^{+6}_{-2}$ times higher EM. The presence of two temperatures could indicate a broader distribution of gas temperatures in the outflow, a phenomenon commonly observed in X-ray observations of nearby edge-on galaxies.

We produced smoother eRObub temperature and EM maps using smaller, constant S/N regions with sky area $\sim$$5$--$10$\,deg$^2$ throughout the eRObub. A necessary trade-off is that the eRObub emission was modelled as a CIE plasma. Under this scheme, the eRObub temperature is consistent with $\sim$$0.3\,$keV gas with EM increasing towards the Galactic plane, consistent with previous work \citet{Kataoka13,Kataoka2015}. The maps reaffirm the existence of the cool shell surrounding the northern bubble, which is almost as bright as the eRObub interior.

The cool shell could form from rapid radiative cooling following the passage of a forward shock that forms the eRObub, as seen in some old SNRs. This scenario is challenged by the low ambient density in the outer halo, casting doubt on the efficiency of radiative cooling. In principle, NEI effects could also be important, especially in low-density environments after shock heating, implying underionisation in the cool shell, which may mimic a lower temperature plasma in CIE. We demonstrated that NEI remains a viable option by showing that a plane-parallel shock model with $n_et\approx10^{11}\,\rm cm^{-3}\,s$ heated up to $kT\approx0.4\,$keV by the shock performs as well as the CIE assumption.

The cool shell does not require a two-temperature model to fit and has a temperature (in CIE) similar to that of the low-temperature component within the eRObub, raising the possibility that the low-temperature component is at least partially contributed by a foreground structure that does not entirely overlap with the Galactic outflow in the background. The natural candidate is the Loop~I superbubble. We identified a prominent dust edge within 100--150\,pc from us in the recent three-dimensional dust map \citep{Edenhofer24}, marking the end of a wall of dust at high latitude ($b\approx60\degr$), that could be compressed by the interaction of the Loop~I superbubble and the LHB -- echoing the original idea by \citet{Egger95}.
Future shadowing observations of molecular clouds beyond the distance of the Loop~I superbubble can quantify or eliminate the foreground contribution of the eRObub.

The eRObub spectra favour a subsolar elemental abundance similar to the CGM ($\sim0.1$--$0.3\,Z_{\odot}$), consistent with a structure that extends to large vertical distance. However, our data provide tentative evidence of light $\alpha$-enhancement (especially Ne) compared to Fe, which would be suggestive of a star-formation driven scenario involving metal enrichment through core-collapse supernovae. However, the evidence of metal enrichment seems strongest towards the NPS, in which our fitting showed a statistically significant increase in abundances compared to the rest of the eRObub. This could indicate a separate origin for the NPS from the eRObub, for instance, stemming from active star-forming regions on the Galactic disc.

In our morphological analysis, we constructed an empirical geometrical model of the eRObub, consisting of two half-ellipsoids with flexible orientation, filled with three-dimensional hot-gas densities following the profile of a blast wave originating from the GC, and propagating in an idealised CGM halo. We demonstrated that such a model can reasonably reproduce the eRASS1 count rate map \citep{Predehl20}.
By fitting our geometrical model in projection to the observed emission map, we found that the eRASS1 data provide a good handle on the horizontal radius of the eRObub ($\sim 5\,$kpc). However, limited by our vantage point within the Galactic plane and the low density in the outer CGM, the eRASS1 is almost insensitive to the vertical extent of the eRObub. Our model requires a significant tilt of the northern bubble towards the west and the Sun ($(\theta,\phi)\approx(30\degr,220\degr)$) in order to reproduce the protruded contour of the eRObub in the west. In contrast, the southern bubble appears to exhibit a much weaker tilt towards the Sun, indicating a significant north-south geometrical asymmetry. The fact that our model prefers an orientation of both bubbles in our direction might be seen as evidence for an outflow that in reality is conical and originating at $\gtrsim 1\,\rm kpc$ from the GC, and need not exhibit a closed shell-like morphology.  

Under the assumption of our blast-wave density profile, the shell thickness is a tracer of the shock Mach number, which would indicate a faster shock in the southern bubble, whereas the outflow appears only weakly supersonic in the north ($\mathcal{M} \sim 1.6$--$1.9$). The higher Mach number in the south is echoed by spectra showing hotter gas near the south-western edge.
Furthermore, if the observed brightness variations are taken to trace the ambient CGM density, significant deviations from a simple azimuthally symmetric density profile are required, for instance, to reproduce the bright NPS.

\begin{acknowledgements}
MCHY thanks Jelle de Plaa for his constant \texttt{SPEX} support throughout this project.
MCHY and MJF acknowledge support from the Deutsche Forschungsgemeinschaft (DFG) through the grant FR 1691/2-1.
M.G.F.M. acknowledges support from the DFG through the grant MA 11073/1-1. 
M.S. acknowledges support from the DFG through the grants SA 2131/13-2, SA 2131/14-2, and SA 2131/15-2.
This work is based on data from eROSITA, the soft X-ray instrument aboard SRG, a joint Russian-German science mission supported by the Russian Space Agency (Roskosmos), in the interests of the Russian Academy of Sciences represented by its Space Research Institute (IKI), and the Deutsches Zentrum für Luft- und Raumfahrt (DLR). The SRG spacecraft was built by Lavochkin Association (NPOL) and its subcontractors, and is operated by NPOL with support from the Max Planck Institute for Extraterrestrial Physics (MPE).
The development and construction of the eROSITA X-ray instrument was led by MPE, with contributions from the Dr. Karl Remeis Observatory Bamberg \& ECAP (FAU Erlangen-Nuernberg), the University of Hamburg Observatory, the Leibniz Institute for Astrophysics Potsdam (AIP), and the Institute for Astronomy and Astrophysics of the University of Tübingen, with the support of DLR and the Max Planck Society. The Argelander Institute for Astronomy of the University of Bonn and the Ludwig Maximilians Universität Munich also participated in the science preparation for eROSITA.
The eROSITA data shown here were processed using the eSASS/NRTA software system developed by the German eROSITA consortium. 

MCHY thanks the developers of the following software and packages that made this work possible:  \texttt{numpy} \citep{numpy}, \texttt{matplotlib} \citep{matplotlib}, \texttt{astropy} \citep{astropy:2013, astropy:2018, astropy:2022}, \texttt{PyXspec} \citep{xspec}, \texttt{HEAsoft} \citep{ftools2}, \texttt{FTOOLS} \citep{ftools1}, \texttt{contbin} \citep{contbin}, \texttt{emcee} \citep{ForemanMackey13}, \texttt{scipy} \citep{scipy}.
\end{acknowledgements}

\bibliographystyle{aau}
\bibliography{References}

\begin{appendix}
\section{Reconstruction of a smoothed eRASS1 all-sky map} \label{sec:ero_recons}

\begin{figure}[htbp]
    \centering
    \includegraphics[width=8.5cm]{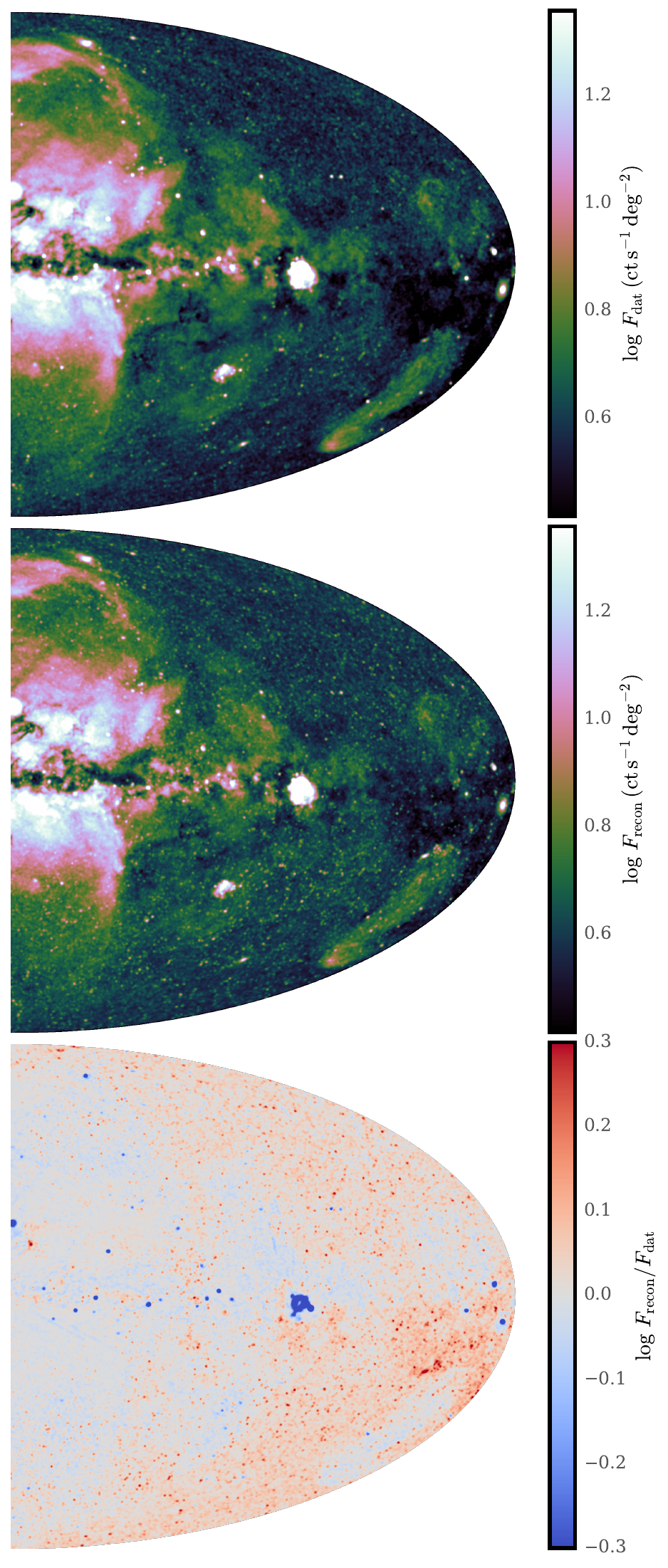}
    \caption{Calibration of reconstructed all-sky map. The top panel shows the publicly available eRASS1 0.6--1.0$\,\rm keV$ all-sky image \citep{Zheng2024_broad, Merloni24}, and the middle panel shows our best attempt at recalibrating the all-sky eRASS1 image \citep{Predehl20} on a quantitative scale. The bottom panel shows the ratio of the two, with areas appearing in blue/red having under-/overestimated reconstructed fluxes. The average deviation is $\Delta \log F_{\rm recon}/F_{\rm dat} \approx 7.7\times 10^{-2}$, corresponding to relative deviations around $20\%$. 
    All maps were smoothed with a Gaussian kernel of $15\arcmin$ for display purposes.}
    \label{fig:reconstruction}
\end{figure}

\begin{figure}[htbp]
    \centering
    \includegraphics[width=0.49\textwidth]{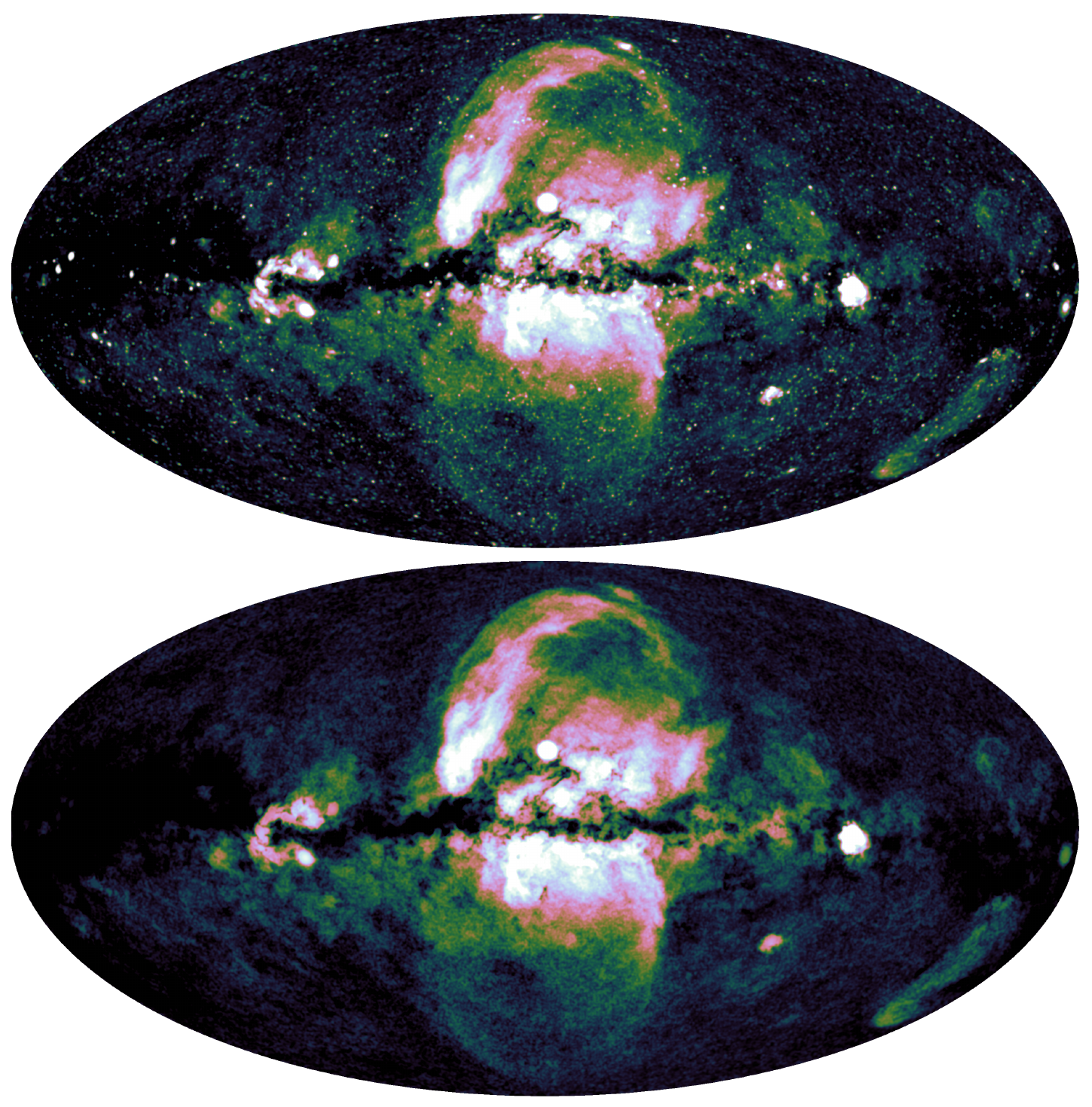}
    \caption{Masking of point and point-like source in the eRASS1 all-sky map. The top and bottom panels show the eRASS1 all-sky image \citep{Zheng2024_broad, Merloni24} in the 0.6--1.0\,keV band, before and after our masking of point sources, respectively. Both maps were smoothed with a Gaussian kernel of $15\arcmin$ for display purposes.
    }
    \label{fig:masking_allsky}
\end{figure}

A three-colour all-sky map based on eRASS1 was published in image form in the eROSITA bubble discovery paper \citep{Predehl20}\footnote{See also \url{https://www.mpe.mpg.de/7461761/news20200619}}, but was not released in a directly machine-readable format. Since the eROSITA all-sky maps strongly outperform the imaging quality of all previous X-ray all-sky surveys, it is worthwhile to reconstruct the eRASS1 map on a quantitative scale.
In order to achieve this, we mapped the individual images in the RGB channels ($0.3$--$0.6\,\rm keV$, $0.6$--$1.0\,\rm keV$, $1.0$--$2.3\,\rm keV$) of \citet{Predehl20} from their Hammer-Aitoff projection onto the coordinate system of the published eRASS1 half-sky broad-band maps \citep{Zheng2024_broad}\footnote{These half-sky maps are publicly available at \url{https://erosita.mpe.mpg.de/dr1/AllSkySurveyData_dr1/HalfSkyMaps_dr1/}}.
These half-sky maps were then smoothed to mimic the processing performed on the available image \citep{Predehl20}, and used to recalibrate the all-sky emission from arbitrary image units back into physical count rates. Since the eRObub appear most prominently in the intermediate energy band $0.6$--$1.0\,\mathrm{keV}$ \citep{Predehl20}, this energy band was the primary focus of our reconstruction.
To achieve this, we sorted the all-sky image into narrow bins of increasing image value, and for each bin, we computed the corresponding median count rate in the overlapping area in the available half-sky image in the same band. The resulting nonlinear relation between image values and corresponding physical units was then used to place the all-sky map on a physical scale of count rate per unit area.

This process of reconstruction is illustrated in Fig.~\ref{fig:reconstruction}. While the reconstructed map is visually almost indistinguishable from the half-sky map, quantitative offsets are present. On the one hand, these are caused by the saturation of the all-sky image around the brightest sources. However, a relative excess can also be seen in the reconstruction of faint regions, which can likely be attributed to the removal of the emission from detected point sources in \citet{Zheng2024_broad}, lowering the apparent flux in the areas without much diffuse emission. These shortcomings have little impact on our results, since we are interested only in large-scale structures ($\gtrsim1^{\circ}$) with significant diffuse emission. Furthermore, we refrain from interpreting any quantitative flux scales in our analysis, since we are primarily interested in the geometry of the bubbles and relative flux differences.
The large spatial scales of our analysis target also imply that little improvement would be expected from including additional eRASSs, which would only lower statistical fluctuations, while systematics are our main source of uncertainty.

The resulting recalibrated map (top panel in Fig.~\ref{fig:masking_allsky}) contains numerous bright extended objects with much smaller angular sizes than the target of our analysis, the eRObub. These objects, which include Galactic SNRs, galaxy clusters, and misclassified emission around bright point sources \citep{Merloni24}, are treated as contaminants for our morphological analysis. In order to remove these, we followed a pragmatic approach similar to \citet{Predehl20}, in which we successively convolved the half-sky map with Gaussian filters of scales increasing from $15\arcmin$ to $1\fdg5$. After each step, the input image was compared to the convolved one, and pixels were masked which exceeded the smoothed image by a fixed threshold. The `holes' left by the masked pixels were filled by the local value of the respective smoothed image. In this way, the majority of X-ray emitting structures with scales $\ll 1^{\circ}$ were excluded from our analysis.  
In Fig.~\ref{fig:masking_allsky}, we show our resulting cleaned map with bright point and point-like sources removed (see Fig.~\ref{fig:bubble_rgb} for an RGB representation). As can be seen, our approach is very effective at removing compact sources with typical extents $\ll 1^{\circ}$ from the map. However, intrinsically highly extended features, such as the Vela SNR, the Large Magellanic Cloud, the Virgo galaxy cluster, or bright straylight around Sco X-1, persisted in the map and had to be masked manually before fitting.

\section{North-south comparison of eRObub spectra} \label{app:NS_com}
\begin{figure}[htbp]
    \centering
    \includegraphics[width=.49\textwidth]{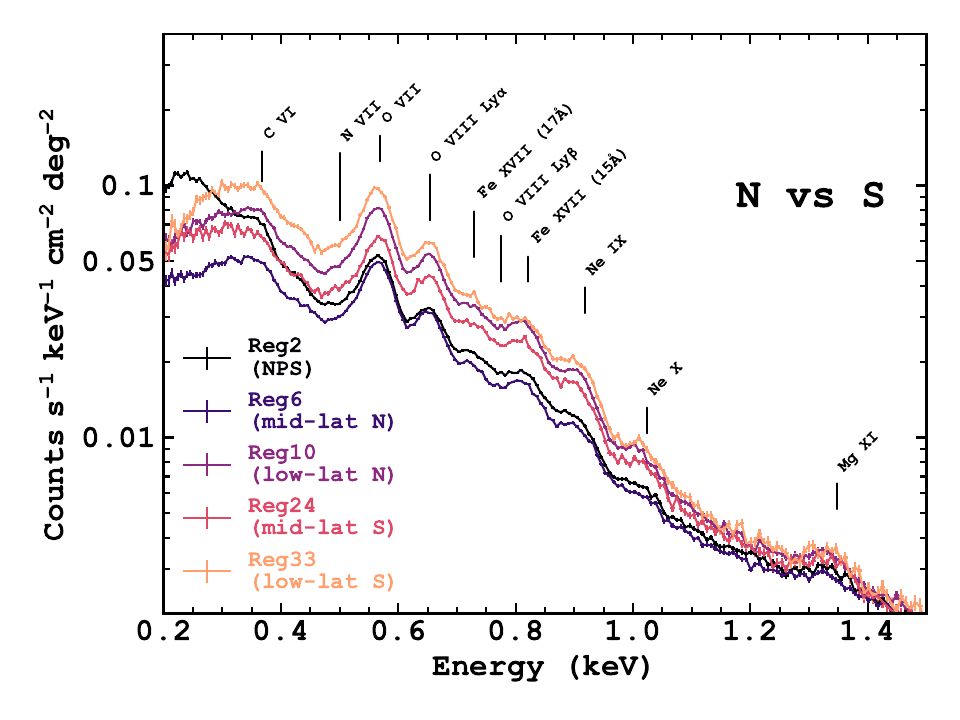}
    \caption{Comparison of spectra in the northern and southern eROSITA bubbles.}
    \label{fig:NS_com}
\end{figure}

Figure~\ref{fig:NS_com} shows a comparison of the spectra extracted from the northern and southern hemispheres. It demonstrates that the spectral features are similar regardless of hemispheres. Therefore, it is not surprising that we obtain similar temperatures in both hemispheres in the fitting.
The main differences in these spectra are foreground absorption column density and EM, both of which decrease with latitudes.

\section{Justifications of model complexities}

\subsection{The need for free C, O, Fe and Ne abundances in the CGM} \label{app:CGM_abund}
It may seem that even our spectral template in the background regions is excessively complicated compared to most emission studies of the MW CGM, with the glaring complexity of free C, O, Fe, and Ne abundances. In this Appendix, we show that the inclusion of these complexities is driven by the high S/N spectra and is a statistically significant modification. We present one example of a general trend observed during background region fitting.

\begin{figure}[htbp]
     \centering
     \includegraphics[width=\linewidth]{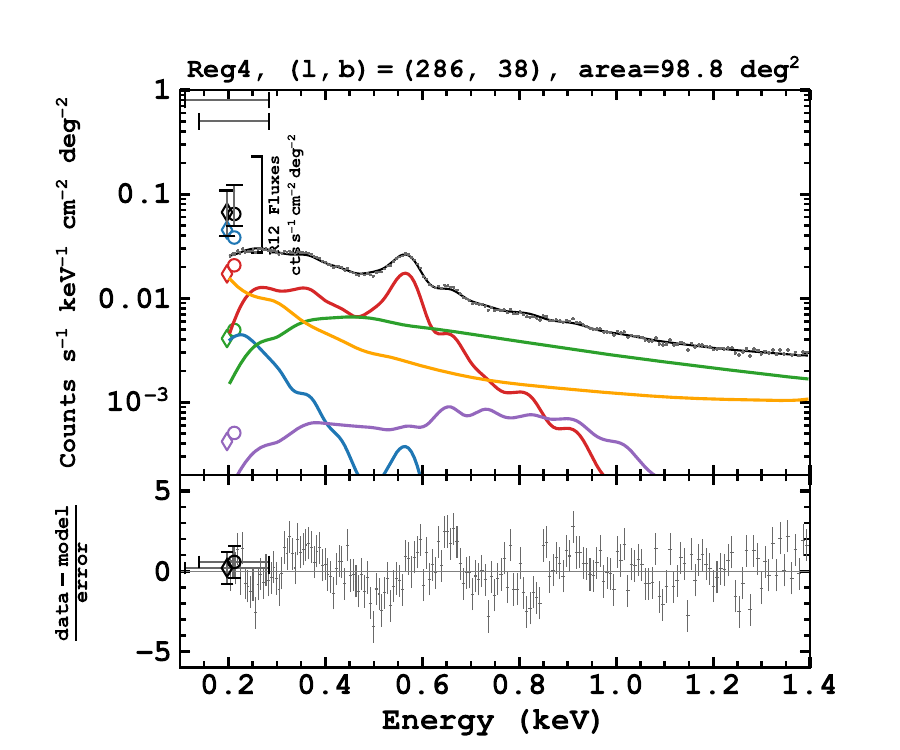}
     \includegraphics[width=\linewidth]{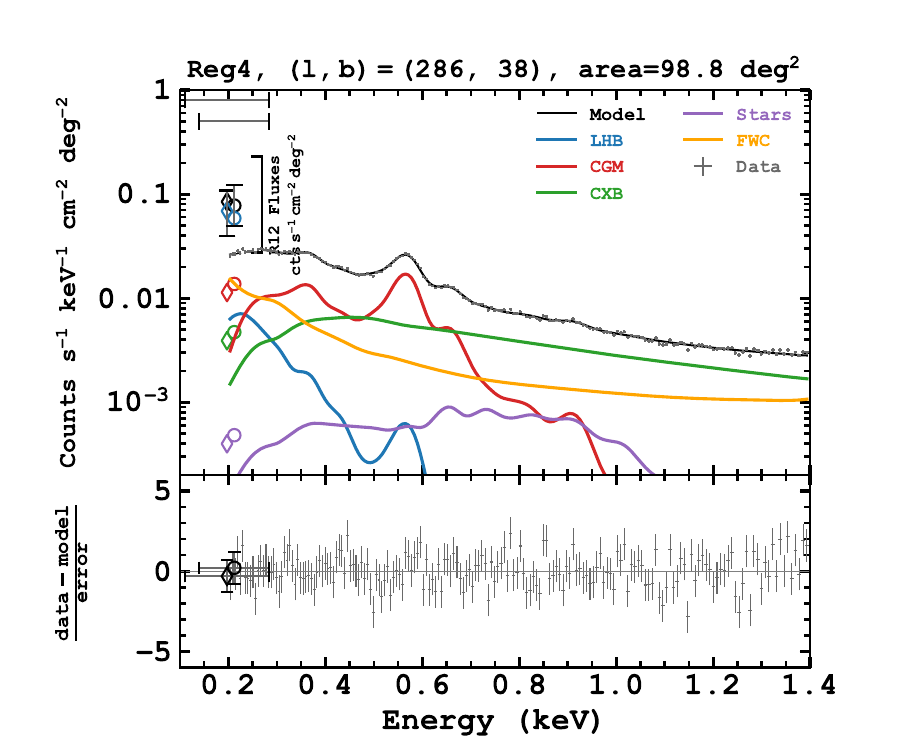}
     \caption{{\it Top:} Spectrum of background region reg~4 fitted by a model that allows the CGM abundance to vary in the ratio of solar abundance ($\chi^2/{\rm dof}=1.55$; 178 dof). {\it Bottom:} The same spectrum fitted by a model that lets CGM C, Fe and Ne abundances vary and deviate from the ratio to solar abundance that is applied to other elements heavier than He (the ratio is fitted simultaneously). The fit has a $\chi^2/{\rm dof}$ of 1.04, with ${\rm dof=175}$.}
     \label{fig:bkg_abund}
\end{figure}

We take reg~4 as an example, and begin by fitting it using an LHB+CGM+CXB+stars model, where the CGM abundance is free to vary as a ratio to solar abundance. This is common practice in fitting spectra of the SXRB. In fact, for SXRB spectra with lower S/N, the CGM is often fixed within the range 0.1--1 solar abundance.
It is not a poor fit, with $\chi^2/{\rm dof}=1.55$ (178 dof) (Fig.~\ref{fig:bkg_abund}). However, there are certainly residuals below 1.0\,keV. We found these residuals are best remedied by allowing the abundances of C, O, Fe and Ne to vary, while other elements heavier than He are linked to the O abundance\footnote{Throughout this work, `O abundance' is always linked to other unspecified elements heavier than He.}. This flexibility results in significant improvements in the fit ($\chi^2/{\rm dof}=1.04$; 174 dof). This corresponds to a F-statistic of 30.1 and a $p$-value of $\approx10^{-15}$.
The best-fit parameters of all background regions are reported in Table~{\ref{tab:bkg_param}}. All background regions consistently show high C/Fe, O/Fe, and Ne/Fe ratios that are significant considering the fitting uncertainties, suggesting this may be a real property of the MW CGM (either these elements are overabundant or Fe is depleted relative to solar). An in-depth discussion of the CGM properties is beyond the scope of this work.

\subsection{The need for free O, Fe and Ne abundances in eRObub.}\label{app:erobub_abund}
\begin{figure}
        \centering
        \includegraphics[width=\linewidth]{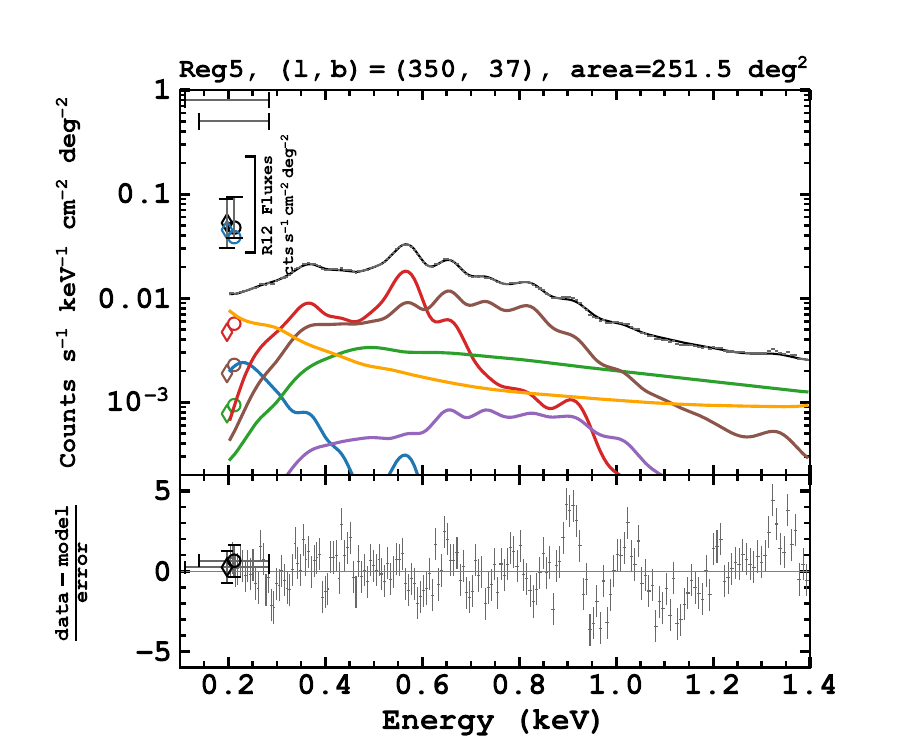}
        \includegraphics[width=\linewidth]{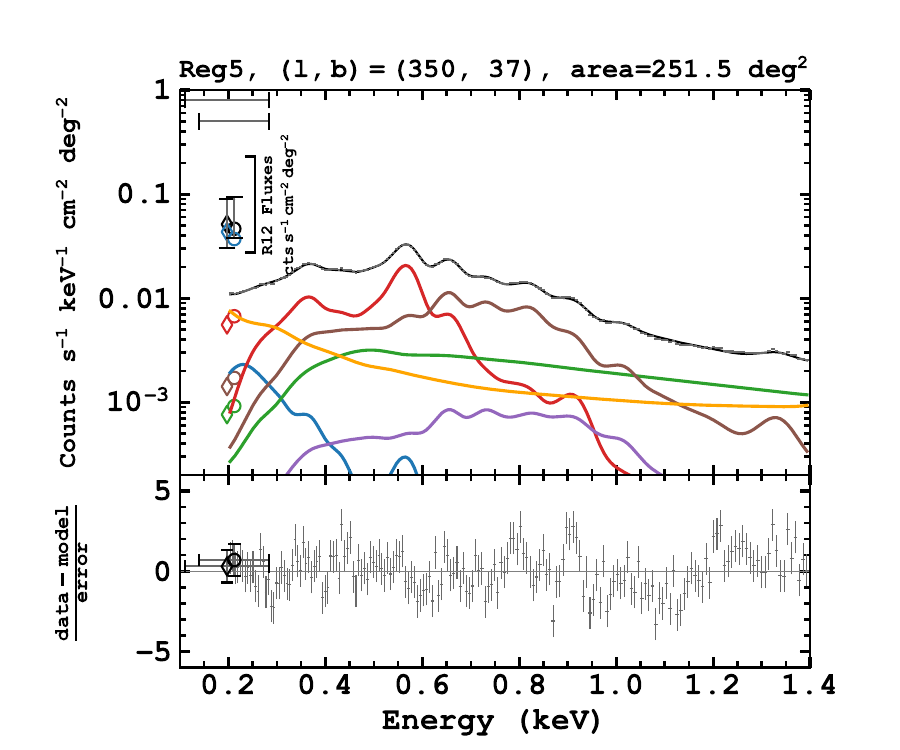}
        \caption{Comparison of the spectral fit in reg~5. {\it Top:} eRObub abundance is allowed to vary in a ratio to solar abundance only ($\chi^2/{\rm dof}=2.33$; 176 dof). {\it Bottom:} Fe and Ne abundances are allowed to deviate from the solar ratio in addition ($\chi^2/{\rm dof}=1.70$; 174 dof).}
        \label{fig:erobub_abund}
\end{figure}

Similarly to the CGM, more freedom in elemental abundance is also needed for the eRObub component(s) in most regions, especially where the eRObub is bright compared to the CGM. Reg~5 is the prime example. Figure~\ref{fig:erobub_abund} shows a comparison of the model not allowing (but the metal abundance is allowed to vary as a single ratio to solar abundance) and allowing additional freedom in Fe and Ne abundance in reg~5. In the former case, there is strong tension between the data and the model near the \ion{Ne}{XI} and \ion{Ne}{X} lines at $E\sim0.9$\,keV. Allowing flexibility in the Ne and Fe abundances alleviates much of the tension in the relevant energy range. The improvement is highly significant; the F-test returns an F-statistic of 33.89, corresponding to a $p$-value of about $10^{-13}$ ($7.6\,\sigma$). One may notice that above $E\gtrsim 1.1$\,keV, the tension between the data and the model persists. Indeed, allowing Mg abundance to vary can curb this problem ($\chi^2/{\rm dof}$ drops further to 1.52; 173 dof). The need to free C and/or Mg abundance is evident in some regions where the eRObub is bright, but it is not as critical as freeing up Fe and Ne abundances.

\subsection{Comparison of one- versus two-layer absorption} \label{app:CGM_abs}
We have employed a two-layer absorption model in reg~13, 14, 15, 30, and 31, which are the eRObub regions closest to the Galactic plane. In the two-layer absorption model, one \texttt{hot} model is applied to the CXB component ($N_{\rm H, CXB}$) and the other to the CGM, eRObub, and stellar components ($N_{\rm H,CGM}$) (see Tables~\ref{tab:bkg_param}, \ref{tab:erobub_params}). The former is fixed to the total Galactic value as the CXB is extragalactic in nature, while the latter are located within the Galaxy and should therefore be attenuated by less material along the line of sight. Statistically, because the former is fixed, our two-layer absorption model does not introduce more degrees of freedom than a one-layer model that is perhaps more commonly used.

In Table~\ref{tab:1v2abs}, we compare the two prescriptions on the relevant regions. Since the additional absorption model is multiplicative, the F-test is not appropriate for model comparison \citep[e.g.][]{Orlandini12}. Instead, we compute the Akaike Information Criterion (AIC) and the Bayesian Information Criterion (BIC) for the two models and compare their values. The more negative $\Delta$AIC or $\Delta$BIC is, the more the data prefer the two-layer absorption model.

\begin{table*}[htbp]
\caption{Comparison of the fit statistics of the one-layer versus the two-layer absorption models.}
\centering
\renewcommand{\arraystretch}{1.2}
    {\small
    \begin{tabular}{c|ccc|ccc|ccc|c}
    \hline\hline
    Reg & $l$ & $b$ & dof & \multicolumn{3}{c|}{One-layer absorption} & \multicolumn{3}{c|}{Two-layer absorption} &  $\Delta$AIC\\
    & $(^{\circ})$ & $(^{\circ})$  &  & $\chi^2/{\rm dof}$ & AIC & BIC &  $\chi^2/{\rm dof}$ & AIC & BIC & or $\Delta$BIC \tablefootmark{$\dagger$}  \\\hline
    13 & $342$ & $18$ &$173$ & $1.253$ &$242.76$ & $284.69$ & $1.245$ & $239.41$ & $278.05$ & $-3.35$\\
    14 & $327$ & $18$ &$173$ & $1.710$ &$321.89$ & $363.83$ & $1.703$ & $318.55$ & $357.20$ & $-3.34$\\
    15 & $315$ & $18$ &$173$ & $1.704$ &$324.83$ & $373.37$ & $1.680$ & $318.64$ & $363.88$ & $-6.19$\\
    30 & $352$ & $-12$ &$173$ & $1.498$ &$285.14$ & $327.07$ & $1.481$ & $280.18$ & $318.83$ & $-4.95$\\
    31 & $334$ & $-12$ &$173$ & $2.093$ &$388.12$ & $430.06$ & $2.017$ & $372.88$ & $411.52$ & $-15.25$\\\hline
    \end{tabular}
    }
    \tablefoot{
    \tablefoottext{$^{\dagger}$}{$\Delta {\rm AIC} = \Delta{\rm BIC}$ if the dof are the same.}
    }
    \label{tab:1v2abs}
    \end{table*}

\section{Parameter table for the single-temperature model} \label{app:1T}
Table~\ref{tab:1T} provides the full parameter table of the single temperature (1T) model.

\begin{table*}
\caption{Best-fit parameters of the 1T model.} \label{tab:1T}
\centering
\renewcommand{\arraystretch}{1.2}
{\small
    \begin{tabular}{c|ccccccc}
    \hline\hline
    Reg & $l$ & $b$ & ${\rm EM_{LHB}}$ & ${\rm EM_{CGM}}$ & ${\rm CXB_{norm}}$ & $\log\left(\frac{N_{\rm H,CXB}}{{\rm cm^{-2}}}\right)$ & $\log\left(\frac{N_{\rm H,CGM}}{{\rm cm^{-2}}}\right)$ \\
     & (\degr) & (\degr) & ($10^{-3}\,{\rm cm^{-6}\,pc}$) & ($10^{-2}\,{\rm cm^{-6}\,pc}$) & ($10^{-3}\,{\rm ph\,s^{-1}\,cm^{-2}\,deg^{-2}}$) &  &  \\\hline
    $1$ & $335$ & $54$ & $4.1\pm0.3$ & $9.08^{+0.40}_{-0.30}$ & $3.91^{+0.04}_{-0.03}$ & $20.53\pm0.02$ & $\ldots$ \\ 
    $2$ & $341$ & $69$ & $4.8^{+0.5}_{-0.7}$ & $12.63^{+1.81}_{-0.41}$ & $3.64^{+0.18}_{-0.48}$ & $20.47^{+0.02}_{-0.03}$ & $\ldots$ \\ 
    $5$ & $350$ & $37$ & $2.1\pm0.1$ & $5.01\pm0.06$ & $3.48^{+0.04}_{-0.05}$ & $20.76\pm0.01$ & $\ldots$ \\ 
    $6$ & $324$ & $38$ & $3.1\pm0.1$ & $0.00^{+0.08}_{0.00}$ & $4.15^{+0.01}_{-0.02}$ & $20.69\pm0.01$ & $\ldots$ \\ 
    $9$ & $352$ & $26$ & $3.0^{+0.1}_{-0.2}$ & $0.51^{+0.15}_{-0.01}$ & $3.98^{+0.07}_{-0.05}$ & $20.92\pm0.01$ & $\ldots$ \\ 
    $10$ & $331$ & $25$ & $5.4\pm0.2$ & $0.55\pm0.01$ & $3.44\pm0.08$ & $20.65\pm0.01$ & $\ldots$ \\ 
    $11$ & $314$ & $24$ & $6.9\pm0.4$ & $0.43^{+0.03}_{-0.02}$ & $3.66^{+0.05}_{-0.06}$ & $20.51\pm0.01$ & $\ldots$ \\ 
    $13$ & $342$ & $18$ & $2.8\pm0.3$ & $0.33^{+0.06}_{-0.13}$ & $4.41^{+0.37}_{-0.24}$ & $21.15$ & $20.85\pm0.01$ \\ 
    $14$ & $327$ & $18$ & $3.6\pm0.4$ & $0.43^{+0.01}_{-0.02}$ & $3.83^{+0.21}_{-0.15}$ & $21.02$ & $20.59^{+0.02}_{-0.01}$ \\ 
    $15$ & $315$ & $18$ & $4.5\pm0.3$ & $0.26\pm0.03$ & $4.07^{+0.07}_{-0.09}$ & $21.01$ & $20.63\pm0.01$ \\ 
    $17$ & $336$ & $-64$ & $6.7^{+0.1}_{-0.0}$ & $2.67^{+0.98}_{-0.02}$ & $3.89^{+0.01}_{-0.02}$ & $20.29\pm0.00$ & $\ldots$ \\ 
    $19$ & $351$ & $-40$ & $4.8^{+0.2}_{-0.1}$ & $1.36^{+0.03}_{-0.02}$ & $3.88^{+0.02}_{-0.03}$ & $20.40^{+0.01}_{-0.00}$ & $\ldots$ \\ 
    $20$ & $327$ & $-40$ & $4.6\pm0.2$ & $1.19\pm0.02$ & $3.42^{+0.04}_{-0.03}$ & $20.32\pm0.02$ & $\ldots$ \\ 
    $23$ & $348$ & $-25$ & $4.0\pm0.3$ & $3.13^{+0.08}_{-0.09}$ & $3.85\pm0.05$ & $20.54\pm0.01$ & $\ldots$ \\ 
    $24$ & $331$ & $-25$ & $4.0\pm0.2$ & $3.72\pm0.07$ & $3.47\pm0.08$ & $20.59\pm0.01$ & $\ldots$ \\ 
    $27$ & $350$ & $-17$ & $4.3\pm0.5$ & $8.47^{+0.20}_{-0.16}$ & $3.54^{+0.09}_{-0.15}$ & $20.52\pm0.01$ & $\ldots$ \\ 
    $33$ & $332$ & $-17$ & $3.2\pm0.3$ & $9.16^{+0.16}_{-0.19}$ & $3.72^{+0.08}_{-0.07}$ & $20.62\pm0.01$ & $\ldots$ \\ 
    $30$ & $352$ & $-12$ & $2.7^{+1.2}_{-1.3}$ & $2.40^{+6.20}_{-2.40}$ & $4.84^{+0.39}_{-0.69}$ & $21.17$ & $20.78\pm0.03$ \\ 
    $31$ & $334$ & $-12$ & $3.9^{+5.0}_{-3.9}$ & $1.08^{+11.34}_{-1.08}$ & $4.58^{+1.35}_{-4.58}$ & $21.25$ & $20.84^{+0.14}_{-0.15}$ \\ 
    \hline
    \end{tabular}

    \vspace{0.2cm}\begin{tabular}{c|ccccccccc}
    \hline\hline
    Reg & $kT_{\rm eRObub}$ & ${\rm EM_{eRObub}}$ & $Z_{\rm C, eRObub}$ & $Z_{\rm O, eRObub}$ & $Z_{\rm Ne, eRObub}$ & $Z_{\rm Mg, eRObub}$ & $Z_{\rm Fe, eRObub}$ & $\chi^2/{\rm dof}$ & dof \\
     & (eV) & ($10^{-3}\,{\rm cm^{-6}\,pc}$) & ($Z_{\rm C, \odot})$ & ($Z_{\rm O, \odot})$ & ($Z_{\rm Ne, \odot})$ & ($Z_{\rm Mg, \odot})$ & ($Z_{\rm Fe, \odot})$ &  &  \\\hline
    $1$ & $238^{+7}_{-6}$ & $<0.9$ & $14.29^{+58691.95}_{-3.25}$\tablefootmark{(a)} & $3.86^{+6020.07}_{-0.80}$\tablefootmark{(a)} & $7.70^{+9630.81}_{-1.60}$\tablefootmark{(a)} & $\ldots$ & $1.00$ & $2.93$ & $176$ \\ 
    $2$ & $320^{+45}_{-16}$ & $<2.3$ & $\ldots$ & $3.21^{+2458.69}_{-0.22}$\tablefootmark{(a)} & $3.98^{+17881.39}_{-0.27}$\tablefootmark{(a)} & $\ldots$ & $1.99^{+1087.41}_{-0.12}$\tablefootmark{(a)}& $3.60$ & $176$ \\ 
    $5$ & $296^{+5}_{-2}$ & $24.8^{+1.9}_{-2.0}$ & $\ldots$ & $0.28^{+0.03}_{-0.02}$ & $0.43\pm0.03$ & $0.49\pm0.04$ & $0.22\pm0.02$ & $3.03$ & $175$ \\ 
    $6$ & $187\pm0$ & $85.4^{+2.1}_{-2.6}$ & $0.42^{+0.03}_{-0.02}$ & $0.15^{+0.01}_{-0.00}$ & $0.36\pm0.01$ & $0.35\pm0.05$ & $0.18\pm0.00$ & $3.28$ & $174$ \\ 
    $9$ & $249\pm1$ & $100.1^{+3.6}_{-57.1}$ & $\ldots$ & $0.10^{+0.01}_{-0.00}$ & $0.28^{+0.02}_{-0.01}$ & $0.26\pm0.03$ & $0.15^{+0.01}_{-0.06}$ & $3.24$ & $175$ \\ 
    $10$ & $285\pm5$ & $75.2\pm2.7$ & $\ldots$ & $0.12\pm0.01$ & $0.22\pm0.01$ & $0.23\pm0.01$ & $0.12\pm0.00$ & $4.10$ & $175$ \\ 
    $11$ & $208^{+5}_{-2}$ & $155.9^{+5.3}_{-8.6}$ & $\ldots$ & $0.06\pm0.00$ & $0.16\pm0.01$ & $0.26\pm0.03$ & $0.06\pm0.00$ & $2.51$ & $175$ \\ 
    $13$ & $216^{+6}_{-10}$ & $323.6^{+47.7}_{-113.5}$ & $\ldots$ & $0.08\pm0.01$ & $0.22^{+0.03}_{-0.02}$ & $0.21\pm0.04$ & $0.13^{+0.01}_{-0.03}$ & $1.67$ & $175$ \\ 
    $14$ & $237^{+7}_{-10}$ & $205.0^{+18.5}_{-9.5}$ & $\ldots$ & $0.03\pm0.00$ & $0.11\pm0.01$ & $0.14\pm0.03$ & $0.05\pm0.00$ & $1.68$ & $175$ \\ 
    $15$ & $199^{+5}_{-3}$ & $244.6^{+11.9}_{-21.9}$ & $\ldots$ & $0.05\pm0.00$ & $0.17\pm0.01$ & $0.32\pm0.04$ & $0.06\pm0.00$ & $1.71$ & $175$ \\ 
    $17$ & $206^{+6}_{-0}$ & $16.2^{+0.1}_{-10.0}$ & $\ldots$ & $0.12^{+0.01}_{-0.00}$ & $0.33^{+0.01}_{-0.03}$ & $0.91^{+0.11}_{-0.16}$ & $0.15\pm0.00$ & $2.29$ & $175$ \\ 
    $19$ & $227^{+6}_{-0}$ & $35.1^{+0.2}_{-1.7}$ & $\ldots$ & $0.04\pm0.00$ & $0.14\pm0.01$ & $\ldots$ & $0.10^{+0.00}_{-0.01}$ & $2.20$ & $176$ \\ 
    $20$ & $372\pm2$ & $16.7^{+1.0}_{-1.1}$ & $\ldots$ & $0.18\pm0.01$ & $0.18\pm0.01$ & $\ldots$ & $0.13\pm0.01$ & $1.99$ & $176$ \\ 
    $23$ & $247^{+2}_{-4}$ & $53.5^{+3.5}_{-2.3}$ & $\ldots$ & $0.12\pm0.01$ & $0.22\pm0.01$ & $0.37\pm0.05$ & $0.18\pm0.01$ & $2.01$ & $175$ \\ 
    $24$ & $299^{+7}_{-5}$ & $37.0^{+2.8}_{-2.6}$ & $\ldots$ & $0.22\pm0.02$ & $0.31\pm0.02$ & $0.34\pm0.03$ & $0.18\pm0.01$ & $2.61$ & $175$ \\ 
    $27$ & $296^{+8}_{-5}$ & $49.1^{+4.3}_{-4.6}$ & $\ldots$ & $0.22^{+0.03}_{-0.02}$ & $0.34^{+0.04}_{-0.03}$ & $\ldots$ & $0.23\pm0.02$ & $1.71$ & $176$ \\ 
    $30$ & $198^{+53}_{-6}$ & $310.4^{+75.5}_{-162.8}$ & $\ldots$ & $0.09^{+0.02}_{-0.03}$ & $0.22^{+0.04}_{-0.05}$ & $0.24^{+0.16}_{-0.15}$ & $0.13\pm0.02$ & $2.05$ & $175$ \\ 
    $31$ & $189^{+195}_{-35}$ & $436.7^{+275.8}_{-190.5}$ & $\ldots$ & $0.07^{+0.06}_{-0.07}$ & $0.18^{+0.18}_{-0.14}$ & $0.22^{+0.95}_{-0.22}$ & $0.08^{+0.08}_{-0.06}$ & $5.13$ & $175$ \\ 
    $33$ & $247^{+2}_{-5}$ & $98.7^{+5.3}_{-4.3}$ & $\ldots$ & $0.07\pm0.00$ & $0.20\pm0.01$ & $0.25\pm0.03$ & $0.11\pm0.01$ & $1.90$ & $175$ \\ \hline
        \end{tabular}
    }
    \tablefoot{
    \tablefoottext{$a$}{Unbound abundances is caused by ${\rm EM_{eRObub}}$ not having a meaningful lower bound and providing only an upper limit.}
    }
\end{table*}

\section{Alternative explanations to the enhanced \texorpdfstring{\ion{Fe}{XVII}\,15\,\AA/17\,\AA}{Fe XVII 15A/17A} line ratio}
We considered two alternative scenarios, resonance scattering and charge exchange combined with ionised gas absorption, in producing the enhanced \ion{Fe}{XVII}\,15\,\AA/17\,\AA~line ratio in addition to the two-temperature plasma model. Despite the arguably more interesting physics involved, we found they were not as convincing as the two-temperature model.

\begin{figure}[htbp]
    \centering
    \includegraphics[width=0.5\textwidth]{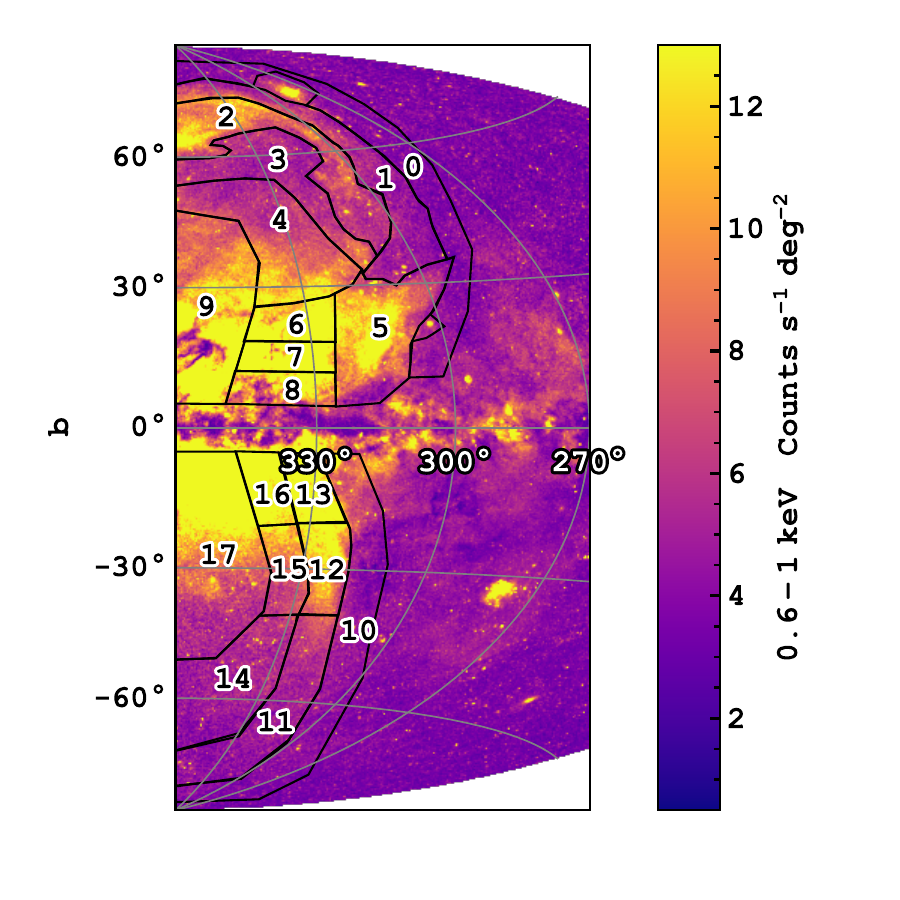}
    \caption{Region definition for Appendices~\ref{app:RS} and \ref{app:CX}.}
    \label{fig:reg}
\end{figure}

For Appendices~\ref{app:RS} and \ref{app:CX}, the spectral extraction regions were defined slightly differently from the main text as illustrated by Fig.~\ref{fig:reg}. Reg~0 and 10 in Fig.~\ref{fig:reg} are the background regions for the analyses below.

\subsection{Resonance scattering} \label{app:RS}
One possible mechanism to produce an enhanced \ion{Fe}{XVII} 15\,\AA/17\,\AA~ratio is by resonance scattering (RS). \ion{Fe}{XVII}\,$\lambda\lambda$ 15.01, 15.26\,\AA~lines have higher oscillator strengths compared to the \ion{Fe}{XVII}\,$\lambda\lambda$ 16.77, 17.05\,\AA~lines (see Table~\ref{tab:RS}), thus the former can become optically thick more easily compared to the latter \citep{Gilfanov87}. This effect has been observed in cores of galaxy clusters from the evidence that the \ion{Fe}{XVII}\,15\,\AA~lines are being suppressed due to the non-negligible optical depth \citep[e.g.][see also a review from \citet{Churazov10}]{Xu02,Sanders06,Werner09}. In fact, in the simplest case of a gaseous sphere in hydrostatic equilibrium, usually appropriate for galaxy clusters or elliptical galaxies, the expectation from RS is that the photons of the strong resonance lines will diffuse out from the core, suppressing the line intensity towards the core while enhancing it at the outskirts. If RS is responsible for the anomalous line ratio, our data suggest a scenario where the \ion{Fe}{XVII}\,15\AA~photons are scattered into the line of sight, analogous to cluster outskirts.

\begin{figure}
    \centering
    \includegraphics[width=0.49\textwidth]{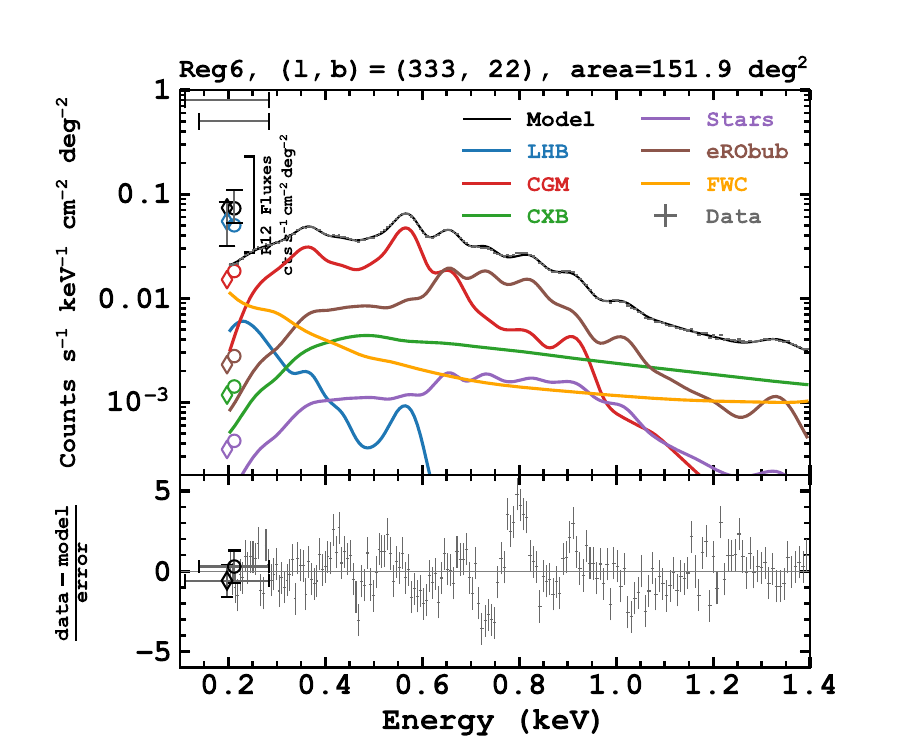}
        \includegraphics[width=0.49\textwidth]{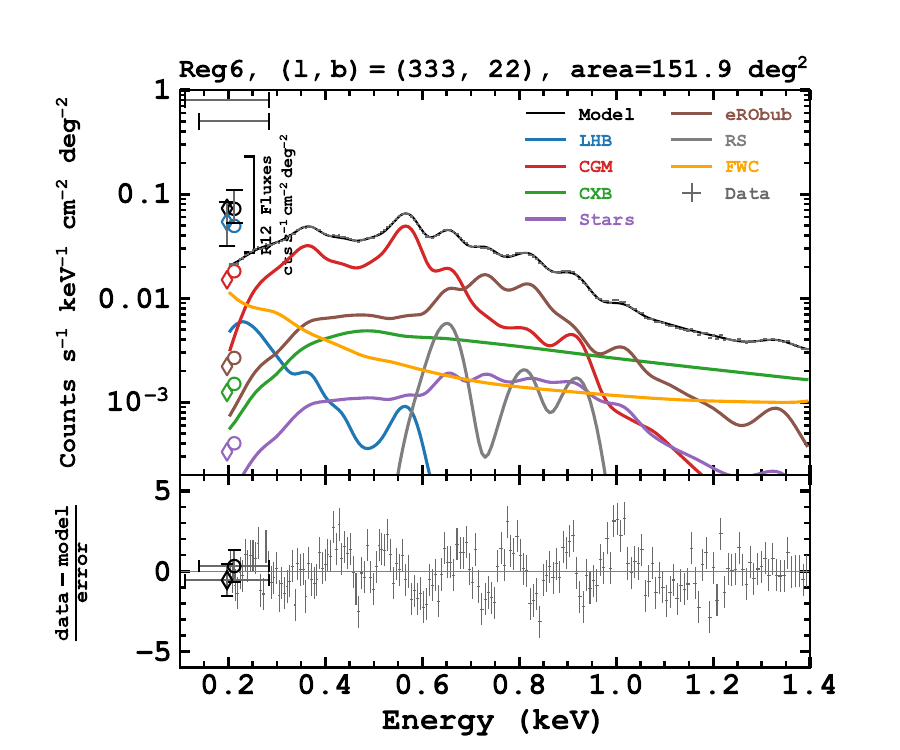} 
    \caption{Comparison of the one temperature model with the model with additional RS lines in Reg~6 defined in Fig.~\ref{fig:reg}.}
    \label{fig:RS}
\end{figure}

We added the list of resonance lines in Table~\ref{tab:RS} in addition to the single temperature models to test the RS scenario further. The listed resonance lines with oscillator strength $f>10^{-2}$ were added as Gaussian lines of free normalisations, with line centres fixed at the listed wavelengths and a fixed width of FWHM=5\,eV (negligible compared to eROSITA spectra resolution of FWHM$\sim$$60$\,eV \citep{Predehl21}, but numerically beneficial during fitting). The RS lines were also subjected to the same absorption column density as other background components. We show an example comparison of a single temperature model and the model with additional RS lines in Fig.~\ref{fig:RS}.

With the additional resonance lines, the anomalous \ion{Fe}{XVII}\,15\,\AA/17\,\AA~ratio is remedied, by the help an enhanced \ion{Fe}{XVII}\,15\,\AA~line and almost no additional \ion{Fe}{XVII}\,17\,\AA~contribution by RS. The best-fit parameters (and upper limits of resonance lines if intensity is consistent with 0) are shown in Table\,\ref{tab:RS_mod}. It is common for the low-latitude regions to show enhanced \ion{O}{VIII}\,Ly$\alpha$ and \ion{Ne}{IX} line in the RS component. eROSITA's CCD detectors do not have adequate spectral resolution to resolve and identify the individual resonance lines. However, if we look into the best-fit parameters for indications, the \ion{Fe}{XVII}\,$\lambda$\,15.26\,\AA~line is stronger than the \ion{Fe}{XVII}\,$\lambda$\,15.01\,\AA~line despite the lower $f$, not following the expectation from RS exactly. The cause of this can be seen from the residuals from the single temperature model (upper panel of Fig.~\ref{fig:RS}) that the residual peaks almost exactly at $0.80$\,keV (15.5\,\AA), centred closer to the \ion{Fe}{XVII}\,15\,\AA~lines that have smaller $f$. This is an argument against the RS hypothesis, and the main reason why it is not our preferred solution to the anomalous Fe line ratio.
Nonetheless, the $0.80$\,keV under-subtraction could also be caused by a contribution from the closely spaced, but weaker, \ion{O}{VIII}\,Ly$\beta$ below $0.8$\,keV, complicating the line identification under eROSITA $\sim$$60\,$eV spectral resolution.

\begin{table*}[htbp]
    \caption{Best-fit parameters of the models with additional resonance lines.}
    \centering
    \renewcommand{\arraystretch}{1.2}
    {\footnotesize
\begin{tabular}{c|ccccccc}
\hline\hline
Reg\tablefootmark{$\dagger\dagger$} & $\rm{EM_{LHB}}$ & $\rm{EM_{CGM}}$ & $\rm{CXB_{norm}}$ & $kT_{\rm{eRObub}}$ & $\rm{EM_{eRObub}}$ & $Z_{\rm{O,eRObub}}$ & $Z_{\rm{Ne,eRObub}}$ \\
 & (${\rm 10^{-3}\,cm^{-6}\,pc}$) & (${\rm 10^{-2}\,cm^{-6}\,pc}$) & (${\rm ph\,s^{-1}\,cm^{-2}\,keV^{-1}\,deg^{-2}}$) & (keV) & (${\rm 10^{-2}\,cm^{-6}\,pc}$) & ($Z_{\rm{O,\odot}}$) & ($Z_{\rm{Ne,\odot}}$) \\ \hline
2  & $4.28^{+0.12}_{-0.13}$ & $5.66^{+0.04}_{-0.04}$ & $4.05^{+0.11}_{-0.39}$ & $0.370^{+0.008}_{-0.006}$ & $1.24^{+0.17}_{-0.21}$ & $0.20^{+0.03}_{-0.16}$ & $0.20^{+0.03}_{-0.16}$ \\
3  & $3.68^{+0.11}_{-0.11}$ & $4.45^{+0.04}_{-0.04}$ & $3.65^{+0.03}_{-0.03}$ & $0.368^{+0.005}_{-0.005}$ & $1.64^{+0.05}_{-0.05}$ & $0.05^{+0.00}_{-0.00}$ & $0.05^{+0.00}_{-0.00}$ \\
4\tablefootmark{*}  & $2.34^{+0.06}_{-0.06}$ & $6.88^{+0.04}_{-0.06}$ & $3.54^{+0.03}_{-0.03}$ & $0.374^{+0.003}_{-0.003}$ & $2.07^{+0.09}_{-0.09}$ & $0.11^{+0.01}_{-0.01}$ & $0.11^{+0.01}_{-0.01}$ \\
5\tablefootmark{*}  & $3.15^{+0.07}_{-0.07}$ & $11.17^{+0.05}_{-0.05}$ & $3.81^{+0.02}_{-0.02}$ & $0.364^{+0.004}_{-0.003}$ & $2.22^{+0.05}_{-0.05}$ & $0.05^{+0.00}_{-0.00}$ & $0.05^{+0.00}_{-0.00}$ \\
6\tablefootmark{*}  & $2.18^{+0.11}_{-0.11}$ & $13.98^{+0.12}_{-0.12}$ & $3.38^{+0.08}_{-0.07}$ & $0.378^{+0.004}_{-0.006}$ & $4.21^{+0.23}_{-0.23}$ & $0.14^{+0.01}_{-0.01}$ & $0.14^{+0.01}_{-0.01}$ \\
7\tablefootmark{*}  & $2.75^{+0.11}_{-0.11}$ & $18.81^{+0.18}_{-0.18}$ & $3.41^{+0.11}_{-0.10}$ & $0.379^{+0.005}_{-0.005}$ & $6.86^{+0.33}_{-0.33}$ & $0.10^{+0.01}_{-0.01}$ & $0.10^{+0.01}_{-0.01}$ \\
8  & $1.99^{+0.05}_{-0.05}$ & $5.68^{+1.89}_{-1.12}$ & $4.43^{+0.03}_{-1.57}$ & $0.227^{+0.081}_{-0.018}$ & $4.10^{+0.70}_{-3.73}$ & $0.20^{+0.00}_{-0.15}$ & $0.20^{+0.00}_{-0.15}$ \\
9  & $1.78^{+0.21}_{-0.22}$ & $9.98^{+0.36}_{-0.71}$ & $4.07^{+0.18}_{-0.43}$ & $0.375^{+0.004}_{-0.010}$ & $2.12^{+0.13}_{-0.81}$ & $0.22^{+0.16}_{-0.09}$ & $0.22^{+0.16}_{-0.09}$ \\
11 & $3.72^{+0.18}_{-0.35}$ & $2.26^{+0.03}_{-0.18}$ & $3.43^{+0.06}_{-0.11}$ & $0.550^{+0.027}_{-0.037}$ & $0.79^{+0.15}_{-0.08}$ & $0.06^{+0.01}_{-0.01}$ & $0.06^{+0.01}_{-0.01}$ \\
12\tablefootmark{*} & $2.16^{+0.13}_{-0.14}$ & $6.13^{+0.07}_{-0.07}$ & $2.77^{+0.09}_{-0.09}$ & $0.499^{+0.008}_{-0.009}$ & $2.86^{+0.17}_{-0.16}$ & $0.09^{+0.01}_{-0.01}$ & $0.09^{+0.01}_{-0.01}$ \\
13\tablefootmark{*} & $2.25^{+0.10}_{-0.10}$ & $13.98^{+0.12}_{-0.12}$ & $3.91^{+0.06}_{-0.06}$ & $0.365^{+0.003}_{-0.003}$ & $6.26^{+0.16}_{-0.16}$ & $0.05^{+0.00}_{-0.00}$ & $0.05^{+0.00}_{-0.00}$ \\
14 & $5.13^{+0.17}_{-0.19}$ & $2.42^{+0.03}_{-0.94}$ & $3.63^{+0.21}_{-0.04}$ & $0.510^{+0.027}_{-0.192}$ & $0.58^{+0.76}_{-0.04}$ & $0.05^{+0.00}_{-0.00}$ & $0.05^{+0.00}_{-0.00}$ \\
15 & $2.28^{+0.17}_{-0.17}$ & $4.40^{+0.07}_{-0.07}$ & $3.06^{+0.10}_{-0.10}$ & $0.500^{+0.014}_{-0.014}$ & $1.98^{+0.18}_{-0.17}$ & $0.09^{+0.01}_{-0.01}$ & $0.09^{+0.01}_{-0.01}$ \\
16\tablefootmark{*} & $1.09^{+0.20}_{-0.25}$ & $13.62^{+0.23}_{-0.49}$ & $4.20^{+0.17}_{-0.25}$ & $0.360^{+0.017}_{-0.007}$ & $5.21^{+0.53}_{-0.75}$ & $0.07^{+0.01}_{-0.02}$ & $0.07^{+0.01}_{-0.02}$ \\
17\tablefootmark{*} & $3.11^{+0.09}_{-0.07}$ & $7.60^{+0.04}_{-0.05}$ & $3.90^{+0.03}_{-0.04}$ & $0.370^{+0.002}_{-0.002}$ & $2.61^{+0.16}_{-0.08}$ & $0.11^{+0.00}_{-0.01}$ & $0.11^{+0.00}_{-0.01}$ \\  \hline
\end{tabular}
\vspace{0.5cm}

\begin{tabular}{c|ccccccc}
\hline\hline
Reg\tablefootmark{$\dagger\dagger$} & $Z_{\rm{Mg,eRObub}}$ & $Z_{\rm{Fe,eRObub}}$ & $\log{\left(\frac{N_{\rm{H}}}{\rm{cm^{-2}}}\right)}$ & $I_{\ion{Ne}{IX}}$ & $I_{\ion{Fe}{XVII}\,\lambda\,15.01\,\text{\AA}}$ & $I_{\ion{Fe}{XVII}\,\lambda\,15.26\,\text{\AA}}$ & $I_{\ion{O}{VIII}\,{\rm Ly}\beta}$ \\
\scriptsize & ($Z_{\rm{Mg,\odot}}$) & ($Z_{\rm{Fe,\odot}}$) &  & (${\rm 10^{33}\,ph\,s^{-1}\,deg^{-2}}$) & (${\rm 10^{33}\,ph\,s^{-1}\,deg^{-2}}$) & (${\rm 10^{33}\,ph\,s^{-1}\,deg^{-2}}$) & (${\rm 10^{33}\,ph\,s^{-1}\,deg^{-2}}$) \\ \hline
2  & $0.20^{+0.03}_{-0.16}$ & $0.20^{+0.03}_{-0.16}$ & $20.441^{+0.008}_{-0.008}$ & $10.22^{+0.64}_{-0.62}$ & $6.22^{+3.68}_{-3.84}$ & $4.86^{+4.22}_{-3.97}$ & $<2.21$ \\
3  & $0.05^{+0.00}_{-0.00}$ & $0.05^{+0.00}_{-0.00}$ & $20.462^{+0.008}_{-0.008}$ & $3.26^{+0.55}_{-0.56}$ & $6.30^{+0.84}_{-3.45}$ & $<2.61$ & $<2.61$ \\
4\tablefootmark{*}  & $0.11^{+0.01}_{-0.01}$ & $0.11^{+0.01}_{-0.01}$ & $20.642^{+0.004}_{-0.004}$ & $9.33^{+0.57}_{-0.56}$ & $<1.80$ & $15.29^{+0.98}_{-0.98}$ & $<1.80$ \\
5\tablefootmark{*}  & $0.05^{+0.00}_{-0.00}$ & $0.05^{+0.00}_{-0.00}$ & $20.646^{+0.003}_{-0.003}$ & $11.56^{+0.58}_{-0.58}$ & $<1.96$ & $14.60^{+0.99}_{-0.99}$ & $<1.96$ \\
6\tablefootmark{*}  & $0.14^{+0.01}_{-0.01}$ & $0.14^{+0.01}_{-0.01}$ & $20.707^{+0.005}_{-0.005}$ & $24.83^{+1.64}_{-1.72}$ & $<6.13$ & $27.47^{+2.60}_{-2.63}$ & $<6.13$ \\
7\tablefootmark{*}  & $0.10^{+0.01}_{-0.01}$ & $0.10^{+0.01}_{-0.01}$ & $20.831^{+0.004}_{-0.004}$ & $44.34^{+2.13}_{-2.28}$ & $<7.16$ & $45.89^{+3.57}_{-3.56}$ & $<7.16$ \\
8  & $0.20^{+0.00}_{-0.15}$ & $0.20^{+0.00}_{-0.15}$ & $21.055^{+0.046}_{-0.023}$ & $42.80^{+1.53}_{-15.82}$ & $<87.35$ & $54.03^{+2.20}_{-49.36}$ & $<56.80$ \\
9  & $0.22^{+0.16}_{-0.09}$ & $0.22^{+0.16}_{-0.09}$ & $20.903^{+0.016}_{-0.017}$ & $26.47^{+4.02}_{-4.91}$ & $<29.72$ & $<40.85$ & $<4.24$ \\
11 & $0.06^{+0.01}_{-0.01}$ & $0.06^{+0.01}_{-0.01}$ & $20.116^{+0.024}_{-0.085}$ & $<2.33$ & $<2.57$ & $<2.33$ & $<2.33$ \\
12\tablefootmark{*} & $0.09^{+0.01}_{-0.01}$ & $0.09^{+0.01}_{-0.01}$ & $20.535^{+0.009}_{-0.009}$ & $11.65^{+1.59}_{-1.56}$ & $<5.34$ & $3.74^{+2.53}_{-3.01}$ & $<5.34$ \\
13\tablefootmark{*} & $0.05^{+0.00}_{-0.00}$ & $0.05^{+0.00}_{-0.00}$ & $20.748^{+0.004}_{-0.004}$ & $21.23^{+1.60}_{-1.52}$ & $<6.85$ & $31.71^{+2.89}_{-7.19}$ & $<6.85$ \\
14 & $0.05^{+0.00}_{-0.00}$ & $0.05^{+0.00}_{-0.00}$ & $20.157^{+0.078}_{-0.137}$ & $<5.68$ & $<5.70$ & $<2.15$ & $<2.15$ \\
15 & $0.09^{+0.01}_{-0.01}$ & $0.09^{+0.01}_{-0.01}$ & $20.471^{+0.014}_{-0.014}$ & $3.52^{+1.61}_{-1.72}$ & $<5.96$ & $<6.61$ & $<5.96$ \\
16\tablefootmark{*} & $0.07^{+0.01}_{-0.02}$ & $0.07^{+0.01}_{-0.02}$ & $20.820^{+0.011}_{-0.015}$ & $7.72^{+5.74}_{-4.07}$ & $<9.43$ & $39.33^{+12.75}_{-13.90}$ & $<6.93$ \\
17\tablefootmark{*} & $0.11^{+0.00}_{-0.01}$ & $0.11^{+0.00}_{-0.01}$ & $20.591^{+0.006}_{-0.004}$ & $5.82^{+0.70}_{-0.73}$ & $<4.35$ & $23.27^{+1.33}_{-5.16}$ & $<1.29$ \\\hline
\end{tabular}
\vspace{0.5cm}

\begin{tabular}{c|cccccc}
\hline\hline
Reg\tablefootmark{$\dagger\dagger$} & $I_{\ion{Fe}{XVII}\,\lambda\,16.77\,\text{\AA}}$ & $I_{\ion{Fe}{XVII}\,\lambda\,17.05\,\text{\AA}}$ & $I_{\ion{O}{VIII}\,{\rm Ly}\alpha}$ & $I_{\ion{O}{VII}}$ & $\chi^2/{\rm dof}$\tablefootmark{\textdagger} & Sky Area \\
\scriptsize & (${\rm 10^{33}\,ph\,s^{-1}\,deg^{-2}}$) & (${\rm 10^{33}\,ph\,s^{-1}\,deg^{-2}}$) & (${\rm 10^{33}\,ph\,s^{-1}\,deg^{-2}}$) & (${\rm 10^{33}\,ph\,s^{-1}\,deg^{-2}}$) &  & (${\rm deg^2}$) \\ \hline
2  & $<2.21$ & $7.12^{+1.97}_{-2.11}$ & $42.68^{+1.68}_{-1.65}$ & $<2.21$ & 1.84 & 421.59 \\
3  & $<2.61$ & $5.90^{+0.98}_{-1.84}$ & $20.91^{+1.29}_{-1.28}$ & $<2.61$ & 1.42 & 356.25 \\
4\tablefootmark{*}  & $<1.80$ & $2.79^{+1.04}_{-1.07}$ & $49.69^{+1.39}_{-1.42}$ & $<2.20$ & 2.15 & 518.70 \\
5\tablefootmark{*}  & $<1.96$ & $9.16^{+1.06}_{-1.07}$ & $81.79^{+1.43}_{-1.34}$ & $<1.96$ & 4.65 & 474.29 \\
6\tablefootmark{*}  & $<6.13$ & $9.53^{+3.09}_{-3.18}$ & $112.85^{+3.81}_{-4.07}$ & $<6.13$ & 1.60 & 151.89 \\
7\tablefootmark{*}  & $<7.16$ & $16.15^{+4.19}_{-4.32}$ & $194.97^{+5.21}_{-5.41}$ & $<7.16$ & 2.23 & 130.02 \\
8  & $<56.82$ & $<4.78$ & $15.77^{+156.17}_{-2.08}$ & $<103.87$ & 4.69 & 157.79 \\
9  & $<4.24$ & $22.19^{+3.01}_{-20.80}$ & $94.78^{+33.43}_{-13.09}$ & $<27.76$ & 3.67 & 579.68 \\
11 & $<2.33$ & $2.01^{+1.57}_{-1.69}$ & $2.43^{+3.80}_{-1.27}$ & $<12.32$ & 3.03 & 399.10 \\
12\tablefootmark{*} & $<5.34$ & $9.84^{+2.38}_{-2.43}$ & $56.34^{+2.86}_{-2.94}$ & $<5.34$ & 1.60 & 174.53 \\
13\tablefootmark{*} & $<6.85$ & $10.52^{+3.36}_{-3.20}$ & $25.91^{+4.41}_{-4.31}$ & $<6.85$ & 2.95 & 135.97 \\
14 & $<2.15$ & $<2.09$ & $1.89^{+1.37}_{-1.33}$ & $<16.99$ & 1.58 & 433.18 \\
15 & $<5.96$ & $3.83^{+2.44}_{-2.48}$ & $24.60^{+2.92}_{-2.90}$ & $<5.96$ & 1.31 & 156.16 \\
16\tablefootmark{*} & $<6.93$ & $7.68^{+12.25}_{-7.17}$ & $<22.59$ & $<6.93$ & 2.47 & 134.38 \\
17\tablefootmark{*} & $<1.29$ & $3.39^{+2.09}_{-1.48}$ & $22.87^{+1.98}_{-1.89}$ & $<1.29$ & 2.84 & 719.29 \\\hline
\end{tabular}
\tablefoot{
\tablefoottext{$\dagger\dagger$}{We note that the region definition in the table follows Fig.~\ref{fig:reg}, which is not the same as the main text.} \tablefoottext{*}{Regions where RS is necessary.} \tablefoottext{\textdagger}{dof=167.}
}
}
    \label{tab:RS_mod}
\end{table*}

The main source of resonance photons could be from the hot ISM in the Galactic plane, but it is less likely to be from the eRObub itself. It is because if the eRObub is the main source, the situation is likely similar to RS in SNRs, where the denser gas at the shell has higher optical depth and scatters the photons away from the line of sight \citep{Kaastra95,Amano20,Li24}. This would create a suppressed \ion{Fe}{XVII}\,15\,\AA/17\,\AA~ratio, contrary to our observation. Of course, the discussion above implicitly assumed a shell-like density profile, which may not necessarily apply to the eRObub. On the other hand, a more realistic model strongly depends on the geometry and densities of the hot gas, which we do not have enough observational constraints on, unlike galaxy clusters or elliptical galaxies, where one can assume spherical symmetry. More detailed work involving, for instance, Monte Carlo simulations, is helpful in quantifying the effect of RS in non-spherical geometries. Nonetheless, suppose the source of resonance photons is either the hot ISM or coronal emission from stars in the MW, then they should be, on zeroth order, distributed in a way following the disk-like MW potential. Even without the spherical symmetry, it is still natural to expect RS to cause a diffusion of resonance photons from the disk towards higher vertical distances and be scattered into our line-of-sight.

\subsection{Charge exchange and ionised absorber} \label{app:CX}
\citet{Gu16} suggests charge exchange (CX) and the presence of an ionised absorber yields a statistically significant improvement on the spectral fits of the {\it Suzuku}/XIS and {\it XMM-Newton}/EPIC data of the NPS. The evidence of CX mainly relies on the observation of line centroid shift of the \ion{O}{VII} He$\alpha$ triplet towards the forbidden line, which is evident in the {\it Suzaku}/XIS data, but less so in the {\it XMM-Newton}/EPIC data. The need for an ionised absorber originates from the anomalously high \ion{O}{VIII}\,Ly$\beta$/Ly$\alpha$ ratio.

From the eROSITA spectra, it is not obvious that there is a systematic shift in the \ion{O}{VII} He$\alpha$ centroid redwards to the forbidden line under the assumption of CIE plasma. We suspect the shift \citet{Gu16} observed could be caused by small calibration imperfections in the {\it Suzaku}/XIS at $\sim0.5$\,keV, where similar residual can be seen when the standard calibration source SNR~1E~0102.2-7219 is fitted with the IACHEC model \citep{Plucinsky17}. Otherwise, the {\it XMM-Newton}/EPIC spectra agree with the CIE assumption reasonably well at \ion{O}{VII}\,He$\alpha$, similar to eROSITA spectra. Despite the intrinsic attractiveness of CX in a Galactic outflow scenario, where there is an interface of hot ionised plasma and neutral matter for CX to occur, the eROSITA spectra are unable to confirm it.

On the other hand, the observation of the enhanced \ion{O}{VIII}\,Ly$\beta$/Ly$\alpha$ ratio in the NPS is also apparent in the EPIC spectra in \citet{Gu16}. They proposed the presence of an ionised absorber to explain the ratio enhancement, as the \ion{O}{VIII}\,Ly$\alpha$ line has a higher transition probability than the Ly$\beta$ line, hence it would be more strongly absorbed than the Ly$\beta$ line and boost the Ly$\beta$/Ly$\alpha$ ratio. We initially suspected an ionised absorber could also explain the anomalous \ion{Fe}{XVII}\,15\,\AA/17\,\AA~ratio we observed, due to the close energy proximity of the \ion{O}{VIII}\,Ly$\beta$ to the \ion{Fe}{XVII}\,15\,\AA~lines. However, after extended testing with additional CX and ionised absorber models, it was found not to be the case.

\begin{figure}[htbp]
    \centering
    \includegraphics[width=0.49\textwidth]{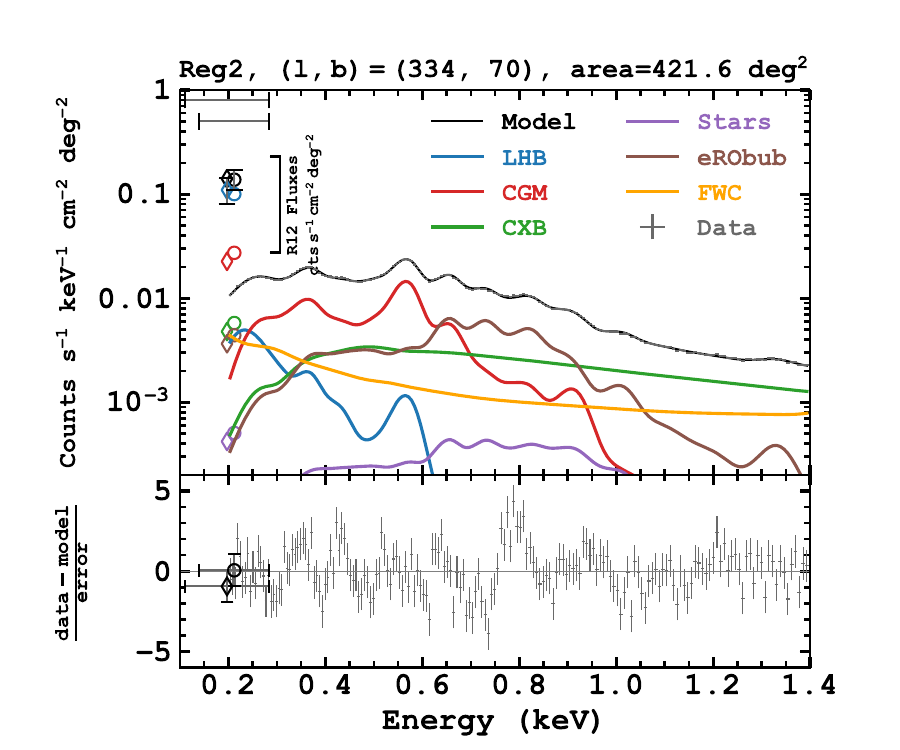}
    \includegraphics[width=0.49\textwidth]{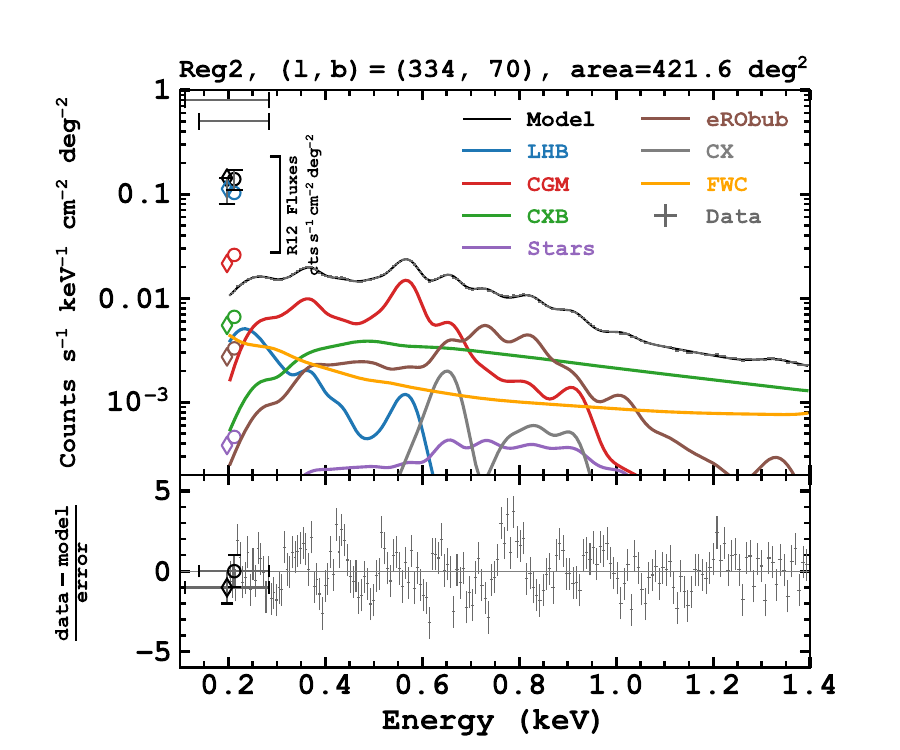}
    \caption{Comparison between the baseline model ({\it top}) and the model including charge exchange (CX) and an ionised absorber ({\it bottom}) in reg~2 defined in Fig.~\ref{fig:reg}, which is the inner part of the NPS extension into the western Galactic hemisphere. The main residual at $0.8\,$keV remains significant and does not sufficiently suppress the enhanced \ion{Fe}{XVII}\,15\,\AA/17\,\AA~ratio.}
    \label{fig:CX}
\end{figure}

Figure~\ref{fig:CX} demonstrates the lack of substantial improvement after including both CX and ionised absorber (\texttt{CX}$\times$\texttt{hot}) in the modelling, especially near $0.8\,$keV. The CX component was modelled using the \texttt{CX} model in \texttt{SPEX} \citep{Gu16_CX}, where the CX plasma component has abundance coupled to the eRObub. However, the temperature was allowed a small freedom to vary near the eRObub temperature. The ionised absorber was modelled using the \texttt{hot} model, but with a variable temperature around $0.17$\,keV and variable column density. The \texttt{CX} model has the merit of producing stronger line blend around 15\,\AA, but not near 17\,\AA, hence partially helpful in explaining the anomalous \ion{Fe}{XVII}\,15\,\AA/17\,\AA~ratio. However, it never produces a single sharp line at $0.8$\,keV, hence cannot fully explain the residual. In addition, absorption by an ionised absorber does not only boost the \ion{O}{VIII}\,Ly$\beta$/Ly$\alpha$ ratio, but also decreases the \ion{Fe}{XVII}\,15\,\AA~resonance lines more strongly than the 17\,\AA~lines due to the stronger oscillator strength, hindering the increase of the \ion{Fe}{XVII}\,15\,\AA/17\,\AA~ratio as is required by the data. A closer look and comparison of the spectral fits in \citet{Gu16} and ours reveals the residual peaks are located at slightly different energies: the {\it Suzaku} and {\it XMM-Newton} spectra in \citet{Gu16} have the residual peak closer to the \ion{O}{VIII}\,Ly$\beta$ line at 775\,eV while the eROSITA spectra are consistently at $\simeq 800$\,eV. Therefore, we concluded that invoking CX and an ionised absorber cannot satisfactorily explain the eROSTIA data.

\section{Parameter table for the plane-parallel shock model} \label{app:neij}
Table~\ref{tab:neij} provides the full parameter table of the plane-parallel shock (\texttt{neij}) model.
\begin{table*}[htbp]
    \caption{Best-fit parameters of the \texttt{neij} model.}
    \label{tab:neij}
    \centering    
    \renewcommand{\arraystretch}{1.2}
    {\small
    \begin{tabular}{c|ccccccc}
    \hline\hline

    Reg & $l$ & $b$ & ${\rm EM_{LHB}}$ & ${\rm EM_{CGM}}$ & ${\rm CXB_{norm}}$ & $\log\left(\frac{N_{\rm H,CXB}}{{\rm cm^{-2}}}\right)$ & $\log\left(\frac{N_{\rm H,CGM}}{{\rm cm^{-2}}}\right)$ \\
     & (\degr) & (\degr) & ($10^{-3}\,{\rm cm^{-6}\,pc}$) & ($10^{-2}\,{\rm cm^{-6}\,pc}$) & ($10^{-3}\,{\rm ph\,s^{-1}\,cm^{-2}\,deg^{-2}}$) &  &  \\\hline
    $1$ & $335$ & $54$ & $4.0\pm0.1$ & $9.73\pm0.11$ & $3.64^{+0.03}_{-0.02}$ & $20.53\pm0.01$ & $\ldots$ \\ 
    $2$ & $341$ & $69$ & $3.7\pm0.2$ & $10.96\pm0.18$ & $3.55^{+0.10}_{-0.04}$ & $20.40\pm0.01$ & $\ldots$ \\ 
    $5$ & $350$ & $37$ & $1.8\pm0.1$ & $4.29^{+0.11}_{-0.10}$ & $3.13\pm0.05$ & $20.72\pm0.01$ & $\ldots$ \\ 
    $6$ & $324$ & $38$ & $2.7\pm0.1$ & $3.29^{+0.08}_{-0.11}$ & $3.15^{+0.02}_{-0.06}$ & $20.65\pm0.01$ & $\ldots$ \\ 
    $9$ & $352$ & $26$ & $2.4^{+0.1}_{-0.2}$ & $0.44\pm0.01$ & $1.90^{+0.04}_{-0.10}$ & $20.85\pm0.01$ & $\ldots$ \\ 
    $10$ & $331$ & $25$ & $4.3\pm0.2$ & $0.46\pm0.01$ & $2.65^{+0.05}_{-0.06}$ & $20.59\pm0.01$ & $\ldots$ \\ 
    $11$ & $314$ & $24$ & $6.3^{+1.7}_{-0.1}$ & $0.44^{+0.11}_{-0.13}$ & $3.51^{+0.31}_{-0.22}$ & $20.47^{+0.07}_{-0.00}$ & $\ldots$ \\ 
    $13$ & $342$ & $18$ & $2.4\pm0.3$ & $0.42^{+0.15}_{-0.31}$ & $1.97^{+0.37}_{-0.28}$ & $21.15$ & $20.79\pm0.01$ \\ 
    $14$ & $327$ & $18$ & $3.5^{+0.4}_{-0.2}$ & $0.41^{+0.03}_{-0.00}$ & $4.02^{+0.04}_{-0.38}$ & $21.02$ & $20.60\pm0.01$ \\ 
    $15$ & $315$ & $18$ & $4.5^{+0.7}_{-0.6}$ & $0.26\pm0.07$ & $4.00^{+0.24}_{-0.18}$ & $21.01$ & $20.62^{+0.03}_{-0.01}$ \\ 
    $17$ & $336$ & $-64$ & $6.5^{+0.1}_{-5.3}$ & $2.80^{+1.82}_{-1.90}$ & $3.61^{+0.40}_{-0.57}$ & $20.17^{+0.26}_{-1.20}$ & $\ldots$ \\ 
    $19$ & $351$ & $-40$ & $4.3\pm0.2$ & $1.48\pm0.02$ & $2.71\pm0.08$ & $20.32\pm0.02$ & $\ldots$ \\ 
    $20$ & $327$ & $-40$ & $4.4^{+0.6}_{-0.8}$ & $1.14^{+0.07}_{-0.12}$ & $2.60^{+0.29}_{-0.20}$ & $20.30^{+0.05}_{-0.09}$ & $\ldots$ \\ 
    $23$ & $348$ & $-25$ & $2.9^{+0.3}_{-0.4}$ & $2.98^{+0.08}_{-0.07}$ & $3.01^{+0.10}_{-0.15}$ & $20.46^{+0.03}_{-0.02}$ & $\ldots$ \\ 
    $24$ & $331$ & $-25$ & $2.9\pm0.3$ & $2.60^{+0.10}_{-0.11}$ & $2.43^{+0.12}_{-0.11}$ & $20.49\pm0.01$ & $\ldots$ \\ 
    $27$ & $350$ & $-17$ & $3.3\pm0.5$ & $7.80^{+0.35}_{-0.33}$ & $2.78\pm0.11$ & $20.48\pm0.02$ & $\ldots$ \\ 
    $33$ & $332$ & $-17$ & $3.2^{+0.5}_{-0.4}$ & $9.23^{+0.40}_{-1.23}$ & $3.64^{+0.41}_{-0.15}$ & $20.62^{+0.03}_{-0.01}$ & $\ldots$ \\ 
    $30$ & $352$ & $-12$ & $2.6^{+0.3}_{-0.2}$ & $4.72^{+2.02}_{-4.72}$ & $4.39^{+0.06}_{-0.70}$ & $21.17$ & $20.75^{+0.05}_{-0.00}$ \\ 
    $31$ & $334$ & $-12$ & $3.8^{+2.3}_{-0.2}$ & $3.68^{+4.58}_{-3.68}$ & $3.85^{+1.23}_{-0.70}$ & $21.25$ & $20.80^{+0.10}_{-0.03}$ \\ \hline
    \end{tabular}

    \vspace*{0.2cm}\begin{tabular}{c|cccccccccc}
    \hline\hline

Reg & $kT_{\rm eRObub}$\tablefootmark{(a)} & ${\rm EM_{eRObub}}$ & $\log\left(\frac{u}{\rm cm^{-3}\,s}\right)$ & $Z_{\rm C, eRObub}$ & $Z_{\rm O, eRObub}$ & $Z_{\rm Ne, eRObub}$ & $Z_{\rm Mg, eRObub}$ & $Z_{\rm Fe, eRObub}$ & $\chi^2/{\rm dof}$ & dof \\
 & (eV) & ($10^{-3}\,{\rm cm^{-6}\,pc}$) &  & ($Z_{\rm C, \odot})$ & ($Z_{\rm O, \odot})$ & ($Z_{\rm Ne, \odot})$ & ($Z_{\rm Mg, \odot})$ & ($Z_{\rm Fe, \odot})$ &  &  \\\hline
$1$ & $537\pm11$ & $0.9\pm0.0$ & $11.45^{+0.05}_{-0.06}$ & $4.22^{+0.55}_{-0.49}$ & $1.24^{+0.10}_{-0.09}$ & $1.15^{+0.09}_{-0.08}$ & $\ldots$ & $1.00$ & $1.40$ & $175$ \\ 
$2$ & $374^{+3}_{-16}$ & $9.1^{+1.2}_{-1.3}$ & $12.13^{+0.28}_{-0.04}$ & $\ldots$ & $0.51^{+0.81}_{-0.07}$ & $0.55^{+0.10}_{-0.07}$ & $\ldots$ & $0.35^{+0.47}_{-0.04}$ & $2.39$ & $175$ \\ 
$5$ & $366\pm3$ & $30.1^{+1.6}_{-1.7}$ & $12.01^{+0.08}_{-0.07}$ & $\ldots$ & $0.15\pm0.02$ & $0.19^{+0.02}_{-0.01}$ & $0.24\pm0.02$ & $0.12\pm0.01$ & $2.48$ & $174$ \\ 
$6$ & $398^{+2}_{-43}$ & $29.2^{+2.5}_{-6.9}$ & $11.52^{+0.00}_{-0.01}$ & $0.26^{+0.49}_{-0.02}$ & $0.08^{+0.09}_{-0.00}$ & $0.11^{+0.08}_{-0.00}$ & $0.14\pm0.01$ & $0.08\pm0.00$ & $3.19$ & $173$ \\ 
$9$ & $420^{+5}_{-2}$ & $71.3^{+2.8}_{-2.7}$ & $11.57^{+0.03}_{-0.04}$ & $\ldots$ & $0.06\pm0.00$ & $0.11\pm0.01$ & $0.13\pm0.01$ & $0.08\pm0.00$ & $2.13$ & $174$ \\ 
$10$ & $364^{+3}_{-2}$ & $73.6\pm1.8$ & $11.87^{+0.05}_{-0.06}$ & $\ldots$ & $0.07\pm0.00$ & $0.11^{+0.00}_{-0.01}$ & $0.14\pm0.01$ & $0.08\pm0.00$ & $3.83$ & $174$ \\ 
$11$ & $227^{+14}_{-30}$ & $130.8^{+52.4}_{-15.3}$ & $12.53^{+0.07}_{-0.02}$ & $\ldots$ & $0.06\pm0.01$ & $0.14^{+0.05}_{-0.02}$ & $0.22\pm0.02$ & $0.05^{+0.02}_{-0.01}$ & $2.61$ & $174$ \\ 
$13$ & $316^{+9}_{-97}$ & $190.5^{+219.4}_{-39.1}$ & $11.89^{+6.11}_{-0.06}$ & $\ldots$ & $0.06^{+0.06}_{-0.02}$ & $0.13^{+0.00}_{-0.01}$ & $0.14^{+0.22}_{-0.01}$ & $0.08^{+0.08}_{-0.02}$ & $1.80$ & $174$ \\ 
$14$ & $227^{+19}_{-0}$ & $220.7^{+0.9}_{-27.4}$ & $13.26^{+4.74}_{-0.24}$ & $\ldots$ & $0.03\pm0.00$ & $0.12^{+0.00}_{-0.02}$ & $0.14\pm0.02$ & $0.05^{+0.00}_{-0.01}$ & $1.69$ & $174$ \\ 
$15$ & $206^{+8}_{-14}$ & $228.2^{+43.9}_{-19.8}$ & $12.87^{+5.13}_{-0.07}$ & $\ldots$ & $0.05\pm0.01$ & $0.16^{+0.03}_{-0.02}$ & $0.30^{+0.03}_{-0.07}$ & $0.06\pm0.01$ & $1.75$ & $174$ \\ 
$17$ & $548^{+595}_{-334}$ & $4.2^{+24.7}_{-4.2}$ & $10.83^{+0.74}_{-0.59}$ & $\ldots$ & $0.09^{+0.55}_{-0.06}$ & $0.12^{+10.18}_{-0.12}$ & $0.23^{+51.04}_{-0.06}$ & $0.12^{+4.51}_{-0.09}$ & $3.07$ & $174$ \\ 
$19$ & $505^{+7}_{-8}$ & $23.2\pm1.1$ & $14.91^{+3.09}_{-1.65}$ & $\ldots$ & $0.09\pm0.01$ & $0.00^{+0.00}_{0.00}$ & $\ldots$ & $0.04\pm0.00$ & $1.71$ & $175$ \\ 
$20$ & $535^{+22}_{-17}$ & $22.0^{+3.0}_{-4.1}$ & $12.36^{+0.48}_{-0.55}$ & $\ldots$ & $0.13^{+0.03}_{-0.05}$ & $0.04^{+0.03}_{-0.02}$ & $\ldots$ & $0.07^{+0.02}_{-0.01}$ & $1.59$ & $175$ \\ 
$23$ & $368^{+5}_{-14}$ & $44.5^{+1.3}_{-15.9}$ & $11.85^{+0.65}_{-0.03}$ & $\ldots$ & $0.08^{+0.14}_{-0.00}$ & $0.09^{+0.00}_{-0.01}$ & $0.17^{+0.13}_{-0.02}$ & $0.10^{+0.01}_{-0.00}$ & $1.96$ & $174$ \\ 
$24$ & $432^{+10}_{-9}$ & $47.3\pm1.8$ & $11.47\pm0.06$ & $\ldots$ & $0.06^{+0.01}_{-0.00}$ & $0.09\pm0.01$ & $0.13\pm0.01$ & $0.08\pm0.00$ & $1.96$ & $174$ \\ 
$27$ & $360\pm3$ & $54.7\pm4.1$ & $12.31^{+0.18}_{-0.14}$ & $\ldots$ & $0.17^{+0.03}_{-0.02}$ & $0.18\pm0.02$ & $\ldots$ & $0.14\pm0.01$ & $1.64$ & $175$ \\ 
$30$ & $227^{+23}_{-35}$ & $216.8^{+154.8}_{-61.3}$ & $12.43^{+0.05}_{-0.02}$ & $\ldots$ & $0.08\pm0.02$ & $0.19^{+0.00}_{-0.03}$ & $0.21^{+0.02}_{-0.03}$ & $0.12^{+0.02}_{-0.01}$ & $2.03$ & $174$ \\ 
$31$ & $227^{+20}_{-47}$ & $261.1^{+278.1}_{-58.2}$ & $12.03^{+5.97}_{-0.01}$ & $\ldots$ & $0.05^{+0.04}_{-0.02}$ & $0.13^{+0.10}_{-0.03}$ & $0.19^{+0.01}_{-0.02}$ & $0.06^{+0.04}_{-0.01}$ & $2.70$ & $174$ \\ 
$33$ & $255^{+10}_{-28}$ & $93.7^{+22.5}_{-8.7}$ & $13.36^{+4.64}_{-0.25}$ & $\ldots$ & $0.07\pm0.01$ & $0.20^{+0.04}_{-0.01}$ & $0.24^{+0.09}_{-0.04}$ & $0.11^{+0.02}_{-0.01}$ & $1.95$ & $174$ \\  \hline
    \end{tabular}
    }
    \tablefoot{
    \tablefoottext{$a$}{This refers to the post-shock temperature; the pre-shock temperature is fixed to that of the CGM, shown in Table~\ref{tab:bkg_param}.}
    }

\end{table*}

\section{NPS abundance comparison} \label{App:NPS}
\begin{figure}[htbp]
    \centering
    \includegraphics[width=0.49\textwidth]{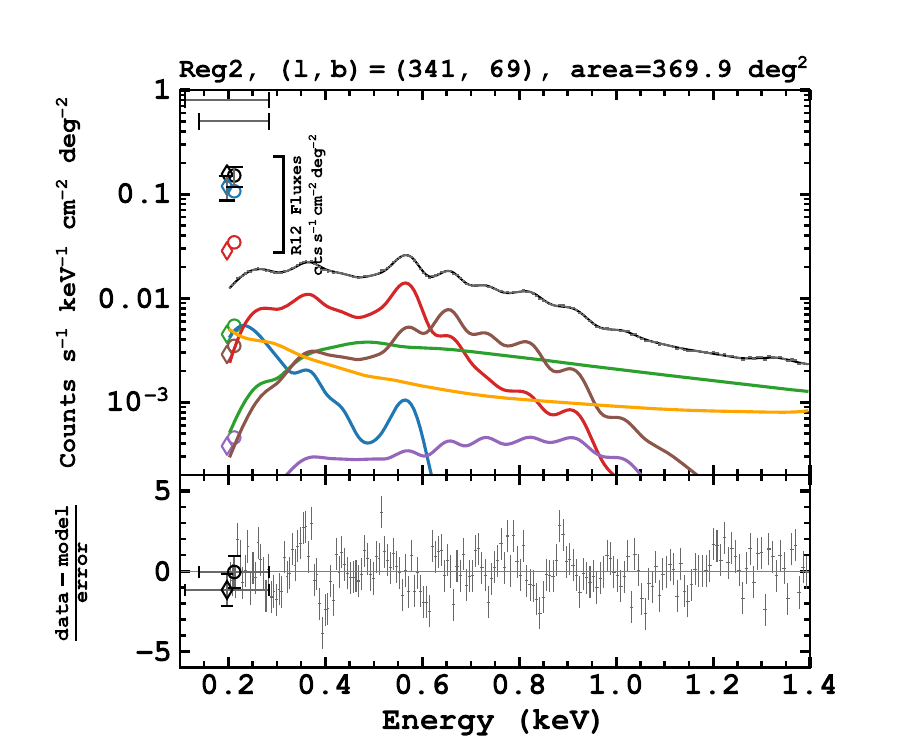}
    \includegraphics[width=0.49\textwidth]{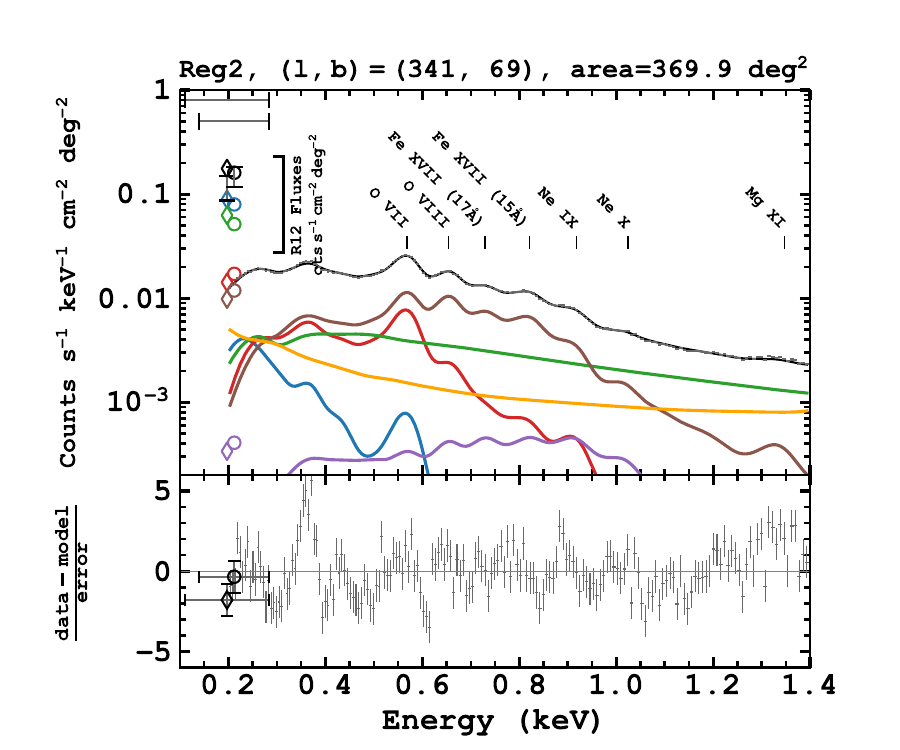}
    \caption{Comparison of NPS region fitted by free ({\it top}) and fixed ({\it bottom}) abundance. We note that in both cases, Ne and Fe abundances were left free.}
    \label{fig:NPS}
\end{figure}

We provide more information on the comparison of the free and fixed (0.2\,$Z_\odot$ except Fe and Ne) abundance model on the NPS. The spectral fits are compared in Fig.~\ref{fig:NPS}. As mentioned in Sect.~\ref{subsubsec:2T}, the free oxygen (linked to all other metals except for Fe and Ne) abundance model performs better statistically with a $p$-value of $\approx10^{-9}$ than the fixed abundance model in the F-test. Visually, improvements can be seen near the \ion{C}{VI} line near $0.36\,$\,keV and the \ion{Mg}{XI} line near $1.35\,$keV. Strong residual remains near the \ion{C}{VI} line even in the free abundance case (C was linked with O in the fit); a higher C abundance will likely deliver a better fit. The higher \ion{O}{VII} line from the eRObub component is caused by a lower temperature $\approx0.21\,{\rm keV}$ as opposed $\approx0.25\,{\rm keV}$ in the free abundance case. The lowering of the eRObub temperature in the fixed-abundance model was compensated for by a lower CGM EM.

\section{Temperature, emission measure and $\chi^2/{\rm dof}$ maps of fits on the constant S/N regions}
\begin{figure*}[htbp]
    \centering
    \includegraphics[width=0.49\linewidth]{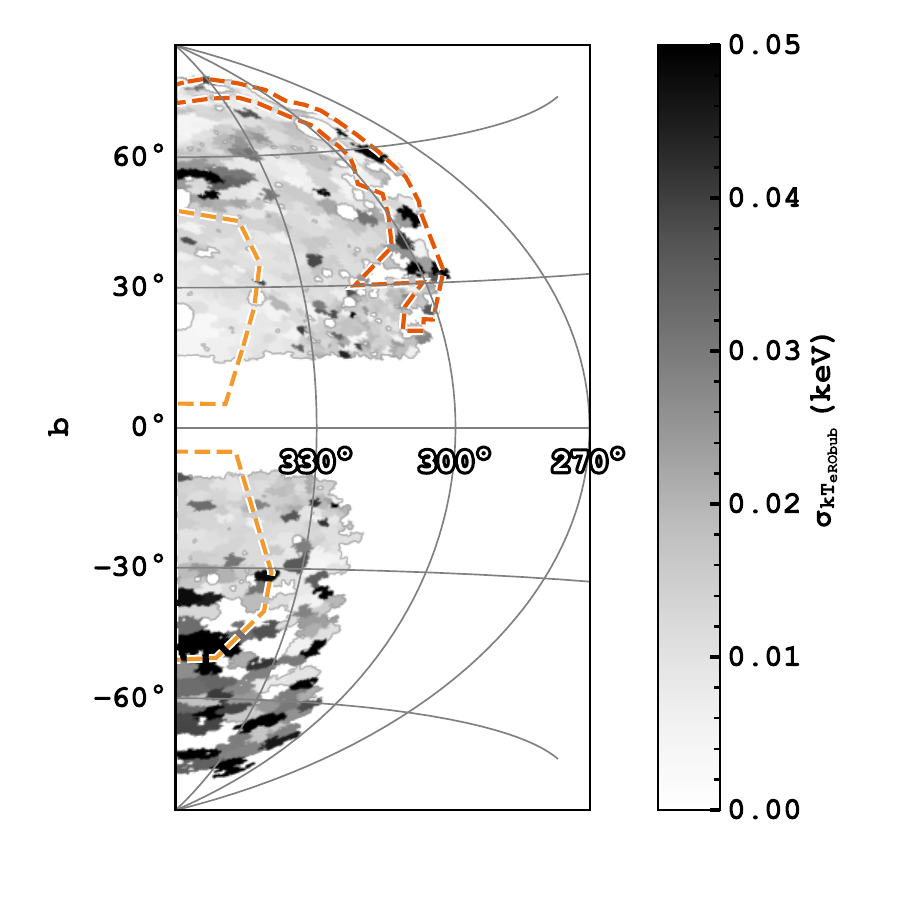}
    \includegraphics[width=0.49\linewidth]{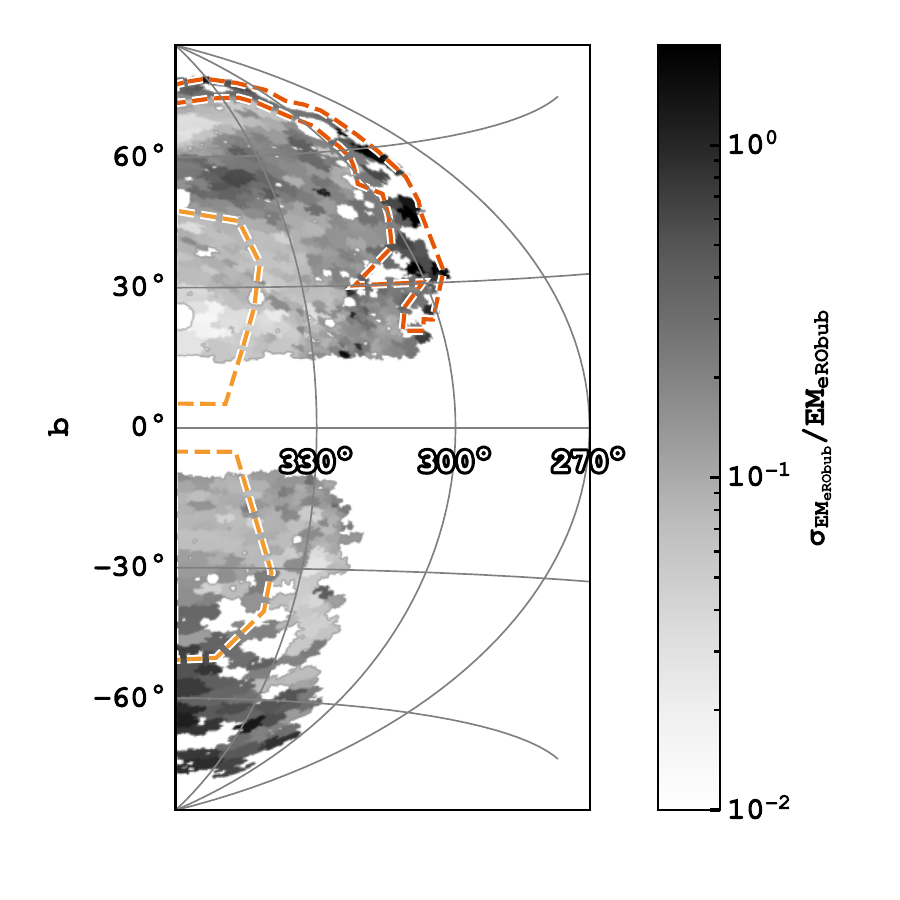}
    \includegraphics[width=0.49\linewidth]{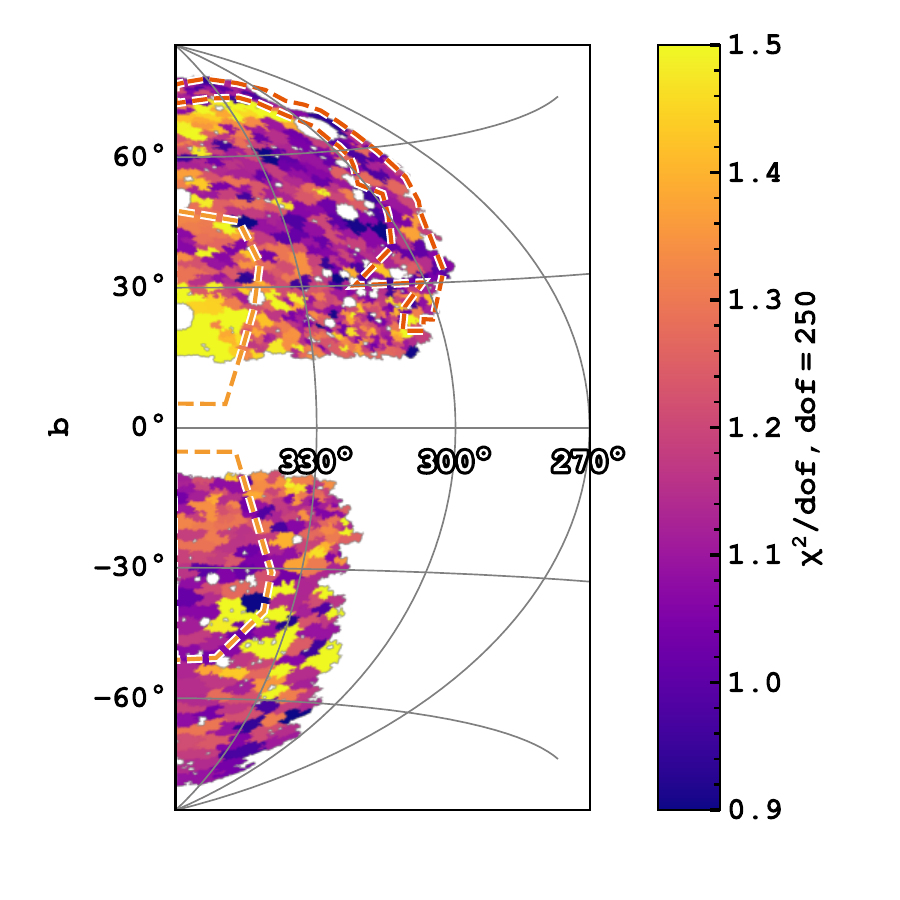}
    \caption{Temperature uncertainty, fractional emission measure uncertaity and $\chi^2/{\rm dof}$ maps of the spectral fits on the constant S/N regions.}
    \label{fig:contbin_err}
\end{figure*}

Figure~\ref{fig:contbin_err} shows the temperature uncertainty, fractional emission measure uncertainty, and $\chi^2/{\rm dof}$ maps of the spectral fits on the constant S/N regions.

\section{Supplementary information on morphological fits} \label{sec:fit_details}

\begin{figure*}[htbp]
    \centering
    \includegraphics[width=0.8\textwidth]{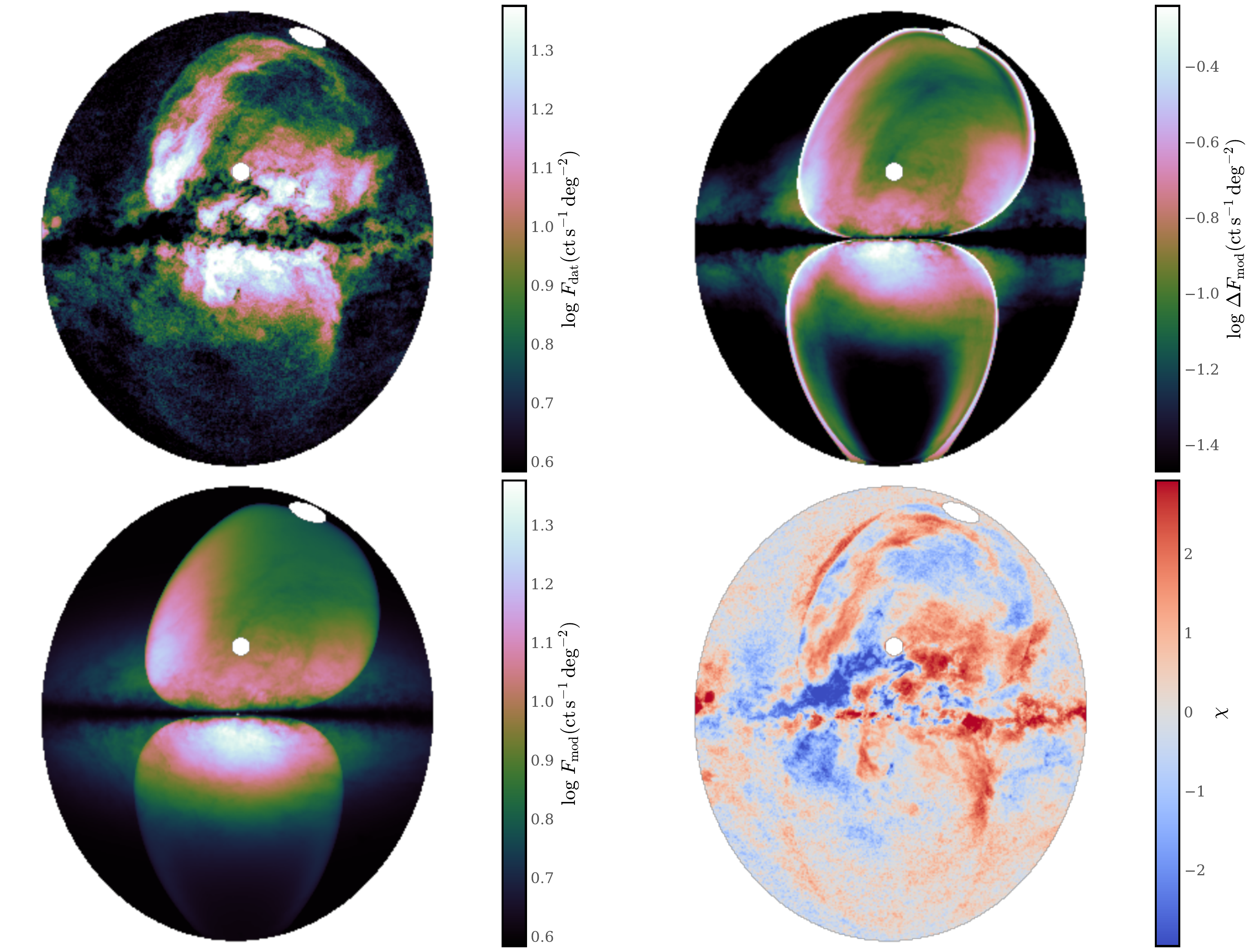}
    \caption{Visualisation of the `best fit' of the bubble morphology. The top left panel shows the fitted data, the compact-source-masked $0.6$--$1.0\,\rm keV$ count rate image of the region surrounding the eRObub. The lower left shows the corresponding morphological fit, meaning the median of the posterior at a given location. The upper right shows the corresponding error, i.e., the posterior uncertainty. Finally, the lower right shows the normalised deviation from the best fit $\chi$, taking into account both the fitted systematic scatter $s$ and the model uncertainty.}
    \label{fig:postpred}
\end{figure*}

This section provides further quantitative details on the morphological fit of the eRObub performed in Sect.~\ref{Morph}. Figure \ref{fig:postpred} displays a quantitative comparison of the observed count rate, with the posterior prediction, and associated uncertainty of our modelling, as well as the deviations of the data from the best fit (taking into account systematic scatter). This demonstrates much more complex structures in the real eRObub than our model can produce. Strong deviations are clearly present around the eastern base of the northern bubble, where absorption appears strongly underestimated, as well as along the NPS, which in reality is even thicker than predicted for a mildly supersonic shock wave shell. Furthermore, the western edge of the southern bubble and the Lotus Petal Cloud \citep{Liu24} show excess observed emission. 

\begin{figure*}[htbp]
    \centering
    \includegraphics[width=0.45\textwidth]{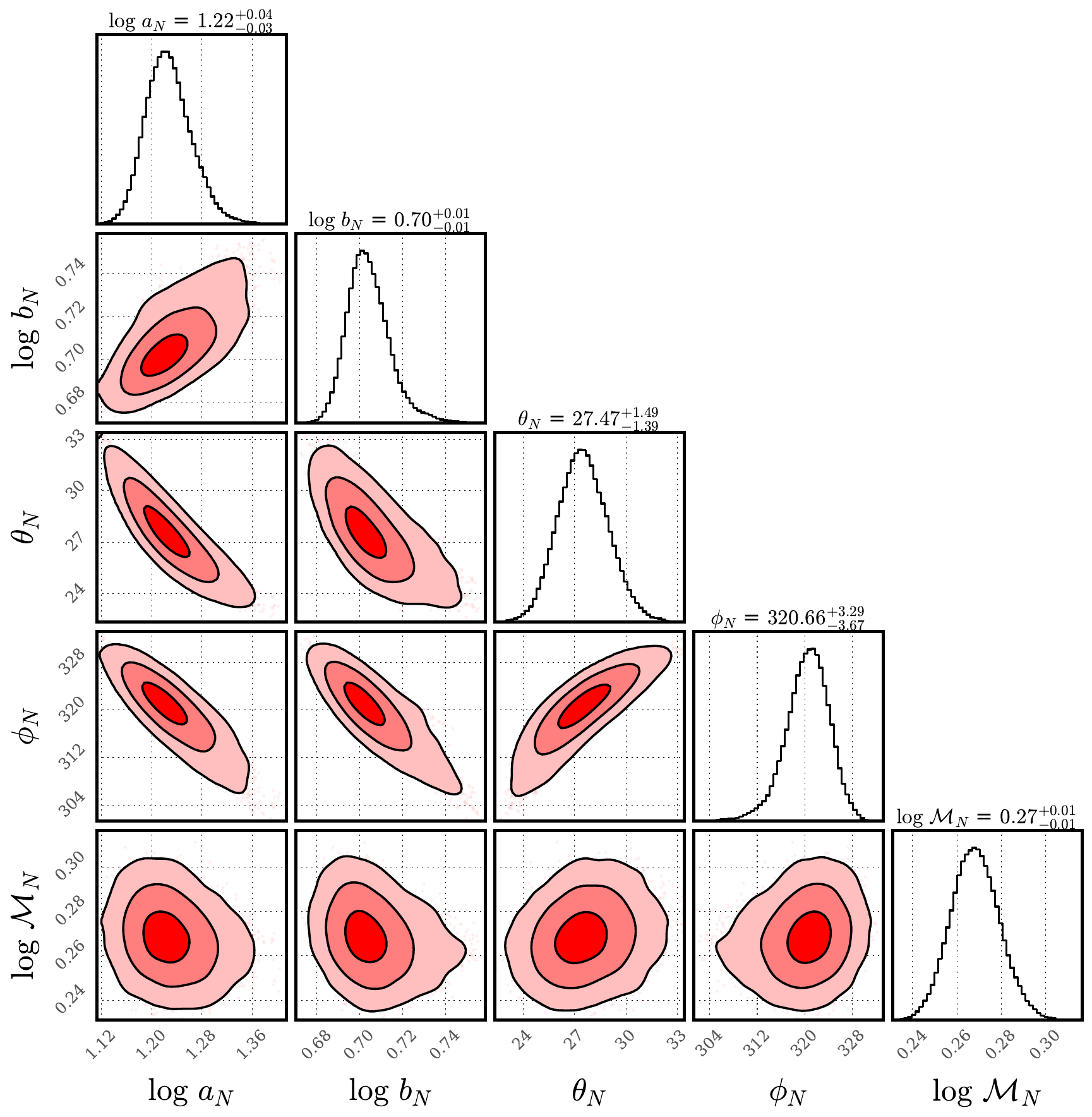}
    \includegraphics[width=0.45\textwidth]{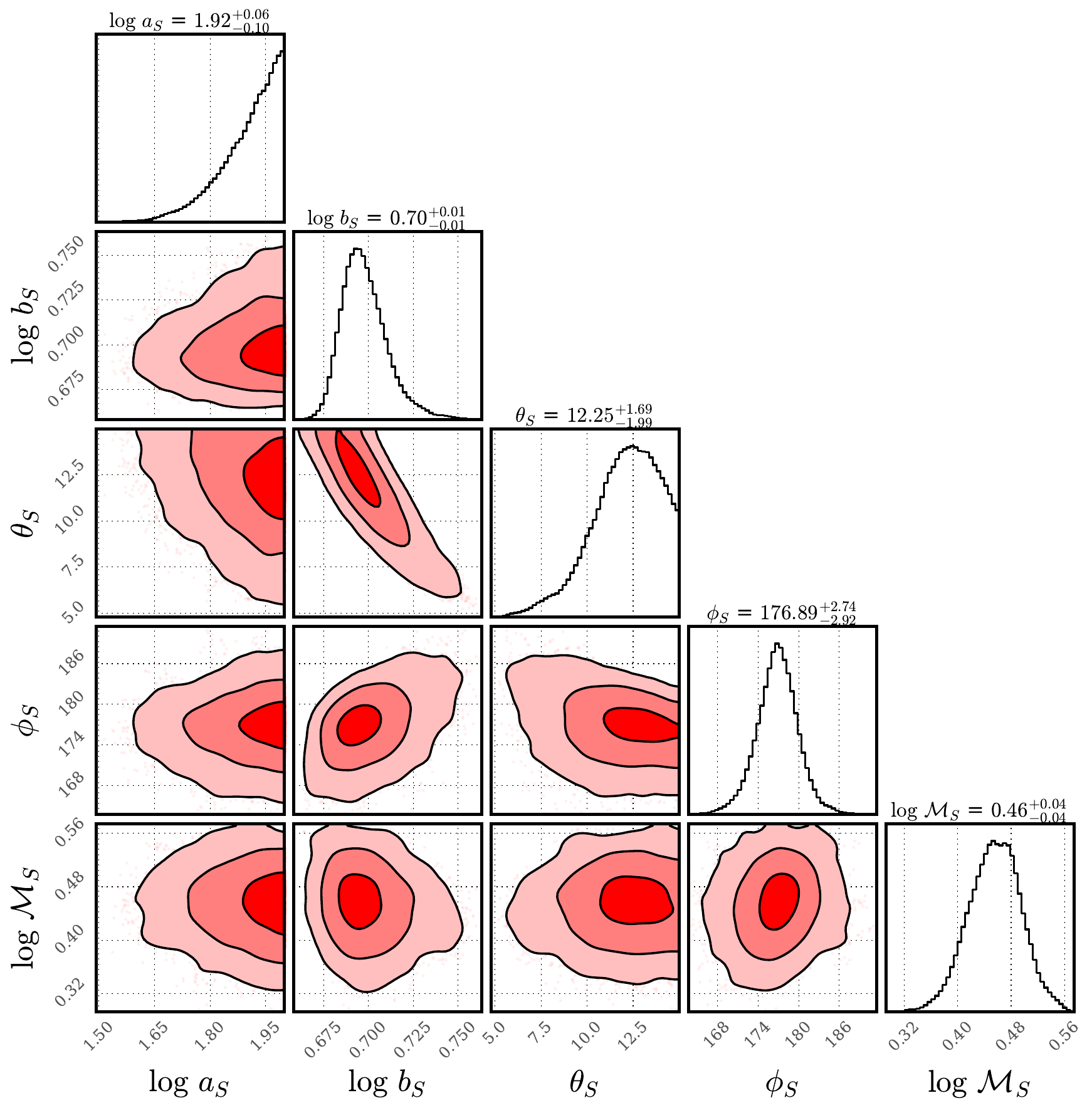}
    \caption{Geometrical parameter degeneracies. The top and bottom panels show the constraints on the parameters describing our geometrical model of the northern and southern bubbles, respectively. In each corner plot, the diagonal plots show the marginal probabilities, and the off-diagonal plots the joint probabilities (marking $1, 2,3\,\sigma$ contours), of bubble size ($a_{i}$, $b_{i}$), orientation ($\theta_{i}$, $\phi_{i}$), and shock Mach number ($\mathcal{M}_{i}$). }
    \label{fig:corners}
\end{figure*}

In Fig.~\ref{fig:corners}, we illustrate the multidimensional constraints on the model parameters of the two bubbles derived with MCMC. These plots showcase the bimodality of certain parameter constraints, for instance, the semi-minor axes ($b_{N,S}$) of the two ellipsoids, with potential peaks at $5.0$ and $5.8\,\rm kpc$. Furthermore, parameter degeneracies are observed, such as between the northern bubble vertical size $a_{N}$ and its tilt $\theta_{N}$, which are dictated mostly by the  Galactic latitude of the northern bubble edge. The same effect leads the vertical size of the southern bubble to be entirely unconstrained above a lower limit, as the top of the bubble has effectively no influence on the fit.      
Overall, these exercises reemphasise that, while the general discussion of the eRObub morphologies (Sect.~\ref{Morph}) may be illuminating, the presented tools are little more than toy models designed to separate projection effects from intrinsic morphology, and are incapable of tracing the physical complexity of the actual phenomenon known as the eRObub.

\end{appendix}
\end{document}